\documentclass[10pt]{article}
\usepackage[top=0.85in,left=1in,right=1in,footskip=0.75in]{geometry}
\usepackage{amsmath,amssymb}
\usepackage{changepage}
\usepackage{textcomp,marvosym}
\usepackage{cite}
\usepackage{nameref,hyperref}
\usepackage[right]{lineno}
\usepackage[nopatch=eqnum]{microtype}
\DisableLigatures[f]{encoding = *, family = * }
\usepackage[table]{xcolor}
\usepackage{array}
\newcolumntype{+}{!{\vrule width 2pt}}
\newlength\savedwidth

\raggedright
\usepackage[aboveskip=1pt,labelfont=bf,labelsep=period,justification=raggedright,singlelinecheck=off]{caption}

\makeatletter
\renewcommand{\@biblabel}[1]{\quad#1.}
\makeatother

\usepackage[most]{tcolorbox}
\definecolor{orange}{HTML}{ce5003}
\definecolor{lightcream}{HTML}{fff2cc}
\definecolor{darkblue}{HTML}{20124d}
\definecolor{blue}{HTML}{00709d}
\definecolor{lightblue}{HTML}{ecfaff}

\usepackage{lastpage,fancyhdr,graphicx}
\usepackage{epstopdf}
\fancyheadoffset[L]{2.25in}
\fancyfootoffset[L]{2.25in}
\usepackage{booktabs}
\usepackage{rotating}
\usepackage{bbm}
\usepackage{multirow}

\begin{document}
\vspace*{0.2in}

\begin{flushleft}
{\Large
\textbf\newline{Local US officials’ views on the impacts and governance of AI: Evidence from 2022 and 2023 survey waves} 
}
\newline
\\
Sophia Hatz\textsuperscript{1\Yinyang*},
Noemi Dreksler\textsuperscript{2\Yinyang},
Kevin Wei\textsuperscript{2,3\Yinyang},
Baobao Zhang\textsuperscript{2,4}
\\
\bigskip
\textbf{1} Department of Peace and Conflict Research, Uppsala University, Uppsala, Sweden
\\
\textbf{2} Centre for the Governance of AI, Oxford, United Kingdom
\\
\textbf{3} Harvard Law School, Harvard University, Cambridge, Massachusetts, USA
\\
\textbf{4} Political Science Department, Maxwell School of Citizenship and Public Affairs, Syracuse University, New York, USA
\\
\bigskip

\Yinyang These authors contributed equally to this work.
* Corresponding author: sophia.hatz@pcr.uu.se

\end{flushleft}
\section*{Abstract}

This paper presents a survey of local US policymakers’ views on the future impact and regulation of AI. Our survey provides insight into US policymakers’ expectations regarding the effects of AI on local communities and the nation, as well as their attitudes towards specific regulatory policies. Conducted in two waves (2022 and 2023), the survey captures changes in attitudes following the release of ChatGPT and the subsequent surge in public awareness of AI. Local policymakers express a mix of concern, optimism, and uncertainty about AI’s impacts, anticipating significant societal risks such as increased surveillance, misinformation, and political polarization, alongside potential benefits in innovation and infrastructure. Many also report feeling underprepared and inadequately informed to make AI-related decisions. On regulation, a majority of policymakers support government oversight and favor specific policies addressing issues such as data privacy, AI-related unemployment, and AI safety and fairness. Democrats show stronger and more consistent support for regulation than Republicans, but the latter experienced a notable shift towards majority support between 2022 and 2023. Our study contributes to understanding the perspectives of local policymakers—key players in shaping state and federal AI legislation—by capturing evolving attitudes, partisan dynamics, and their implications for policy formation. The findings highlight the need for capacity-building initiatives and bi-partisan coordination to mitigate policy fragmentation and build a cohesive framework for AI governance in the US.

\newpage
\renewcommand{\thefigure}{\Alph{figure}}
\section{Executive summary}

The release of ChatGPT in November 2022 marked a watershed moment that thrust Artificial Intelligence (AI) into the global spotlight, intensifying debates about both its societal implications and regulatory needs. As the adoption and capabilities of AI systems increase, understanding how policymakers view and plan to govern this technology becomes increasingly critical. 

To explore local US policymakers' perspectives on AI's future impact and their attitudes toward specific regulatory policies, we conducted two waves of surveys targeting local elected officials -- one in April-May 2022 (\textit{n} = 524) and another in May-June 2023 (\textit{n} = 504). This timing meant we captured local officials' attitudes both before and after ChatGPT's release and the increasing attention AI received in the following period, allowing us to explore shifts in regulatory stance and risk perception. The sample included  county-level, municipal-level, and township-level elected officials from across the country (not state- or federal-level officials).

\begin{tcolorbox}[colback=lightblue, colframe=blue, title=\textbf{Key findings}, fonttitle=\large, breakable]
In regard to the impacts of AI over 2025--2050 in their local community and the broader country we found that:
\begin{itemize}
    \item \textbf{Most local policymakers anticipate AI will create significant societal risks over the next decades.} The majority of officials expect increased surveillance (83\%), misinformation (69\%), and political polarization (59\%), while expressing concerns about decreased data security (64\%) and threats to US democracy (53\%).
    \item \textbf{While local officials think AI will be positive for US innovation (62\%) and transportation/infrastructure (51\%) in the coming decades, they have more measured views about its economic impacts.} Local officials are more likely to think jobs (17\% increase vs. 48\% decrease), income levels (22\% increase vs. 41\% decrease), and inequality  (38\% increase, 10\% decrease, 32\% no effect) will be negatively rather than positively affected in their local community. Regarding broader economic impacts, there are mixed expectations -- 38\% expect the US economy to grow thanks to AI, 24\% expect it to decline, 16\% think it will have no effect, and 22\% say they don't know.
    \item \textbf{Local officials are pessimistic about AI's impact on personal well-being.} At least a plurality expects negative effects on mental health (63\% worsen vs. 16\% improve), physical health (55\% worsen vs. 25\% improve), and quality of life (48\% worsen vs. 29\% improve) because of AI. 
    \item \textbf{Respondents expressed uncertainty regarding some of AI’s impacts, particularly on international issues.} The highest rates of ``don't know'' responses were for questions about international conflicts (26\%) and the probability of great power war (24\%). But around a fifth of respondents also responded ``don't know'' in regard to the local impacts of AI on inequality (20\%) and bias and discrimination (21\%), as well as the broader impact of AI on the US economy (22\%). 
    \item \textbf{Local officials became more pessimistic about the long-term effects of AI until 2100.} In 2022, 37\% of respondents thought the overall impact of AI would be positive or very positive, while 35\% thought it would be negative or very negative. In 2023, negative views (55\%) outweighed positive ones (22\%) more starkly.
\end{itemize}

Concerning the governance of AI, we asked respondents whether they thought AI should be regulated, and whether they thought a range of AI policies would be beneficial for the country in response to AI in the years 2025--2050:
\begin{itemize}
    \item \textbf{Local officials are increasingly in favor of AI regulation and oversight.} A majority of Democrats and Republicans think AI should be regulated, and support jumped significantly from 56\% in 2022 to 74\% in 2023, with increases across party lines. Democrats are more positive about AI regulation, but Republican support for AI regulation increased more strongly from 2022 (43\%) to 2023 (68\%), than for Democrats (75\% vs 84\%).
    \item \textbf{A number of specific AI policies have majority support from local US policymakers.} These included: stricter data privacy regulations (80\%), re-training opportunities (76\%), regulation that ensures deployed AI systems are safe, robust, and fair (72\%), stronger antitrust regulations (58\%), stricter requirements for AI use in judicial decisions (55\%), and auditing AI systems used in hiring for bias (52\%). 
    \item \textbf{Democrats show greater support for a broad range of AI policies.} Democrats indicate higher support than Republicans for AI policies generally, including higher corporate taxes, stronger anti-trust regulations, stronger social safety nets, universal basic income, wage subsidies to counter wage declines, re-training, federal legislation on how local governments use AI systems,  regulation that ensures deployed AI systems are safe, robust, and fair, hiring and judicial decision auditing and requirements, and subsidies for US manufacturing of semiconductors and high-end AI hardware systems.    
    \item \textbf{Universal Basic Income (UBI) stands out as the only AI policy proposal facing strong opposition.} While most AI policies receive majority support, UBI faces majority rejection, with 58\% opposing (45\% strongly disagree, 13\% somewhat disagree) versus only 24\% agreeing it would be beneficial (13\% somewhat agree, 11\% strongly agree), with 17\% remaining neutral.
    \item \textbf{Local officials feel under-prepared for AI governance, with limited expectations of near-term involvement.} Most local policymakers (57\%) believe they are unlikely to make AI-related decisions in the next few years, with a third (30\%) considering it likely. But 52\% feel inadequately informed to make AI-related decisions today, compared to 30\% who feel adequately informed. 
\end{itemize}

Other key group differences in attitudes towards the impacts of AI and its regulation include:
\begin{itemize}
    \item \textbf{Party affiliation shapes policy preferences more strongly than expectations of impacts.} While Democrats and Republicans share similar views on most AI risk and benefit items, Democrats consistently show greater support for regulatory policies.
    \item \textbf{Views on AI policy and impacts vary to some extent across demographic subgroups.} Men are less concerned about the mental health and quality of life effects of AI, more likely to expect greater benefits from AI in the US, including the economy, and are less supportive of a robot tax than women. Meanwhile, higher education obtainment correlates with stronger support for privacy regulations, optimism about AI-driven innovation, heightened concerns about surveillance, and less concern about AI’s impact on the probability of a great power war.
\end{itemize}
\end{tcolorbox}

\begin{figure}[p]
    \centering
    \includegraphics[width=\textwidth,height=\textheight,keepaspectratio]{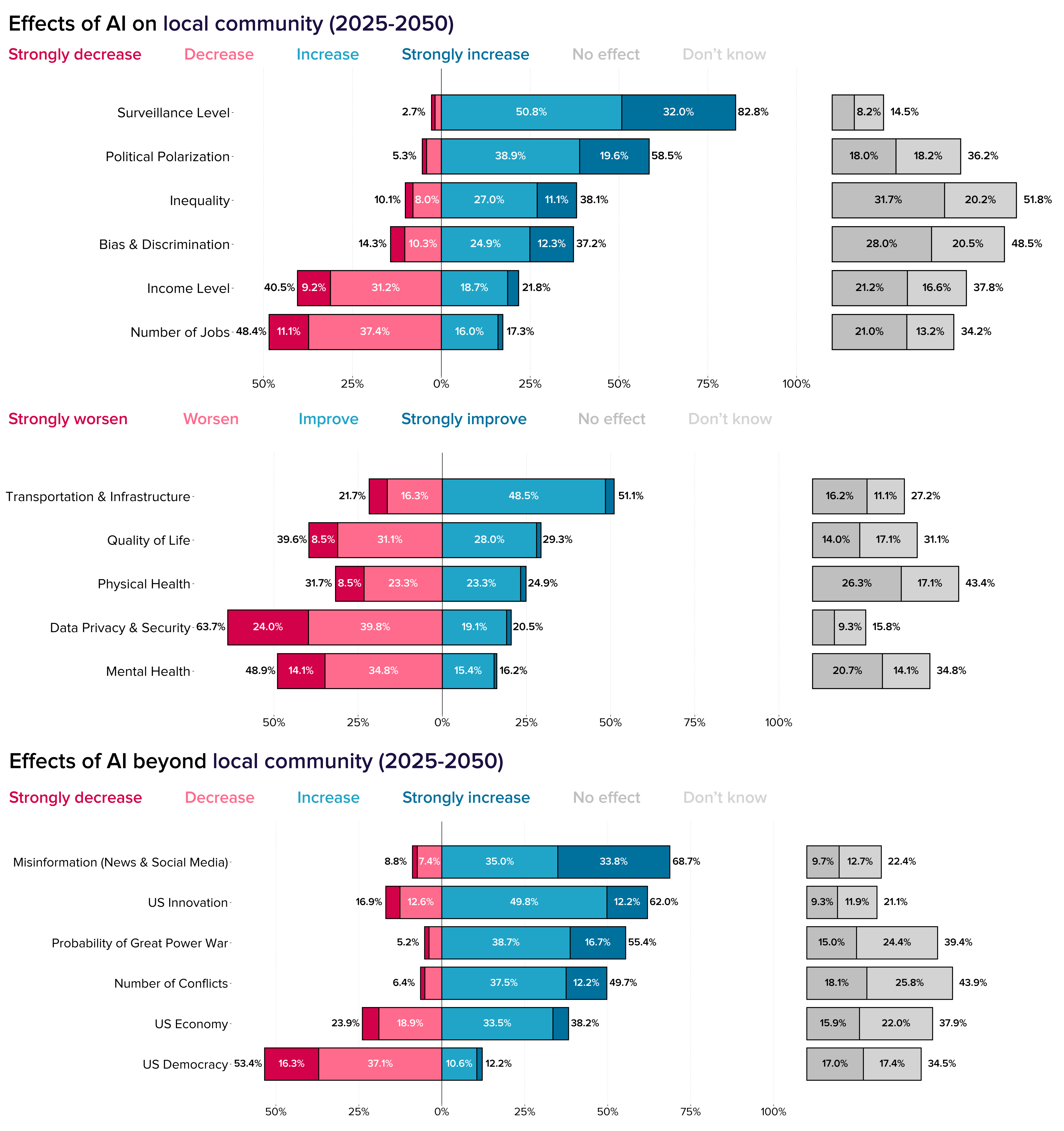}
    \caption{\textbf{Local US officials' expectations of the impacts of AI between 2025 and 2050.} The figure shows unweighted relative frequencies for QS1--3 across both survey waves.}
\end{figure}

\clearpage

\begin{figure}[p]
    \centering
    \includegraphics[width=\textwidth,height=\textheight,keepaspectratio]{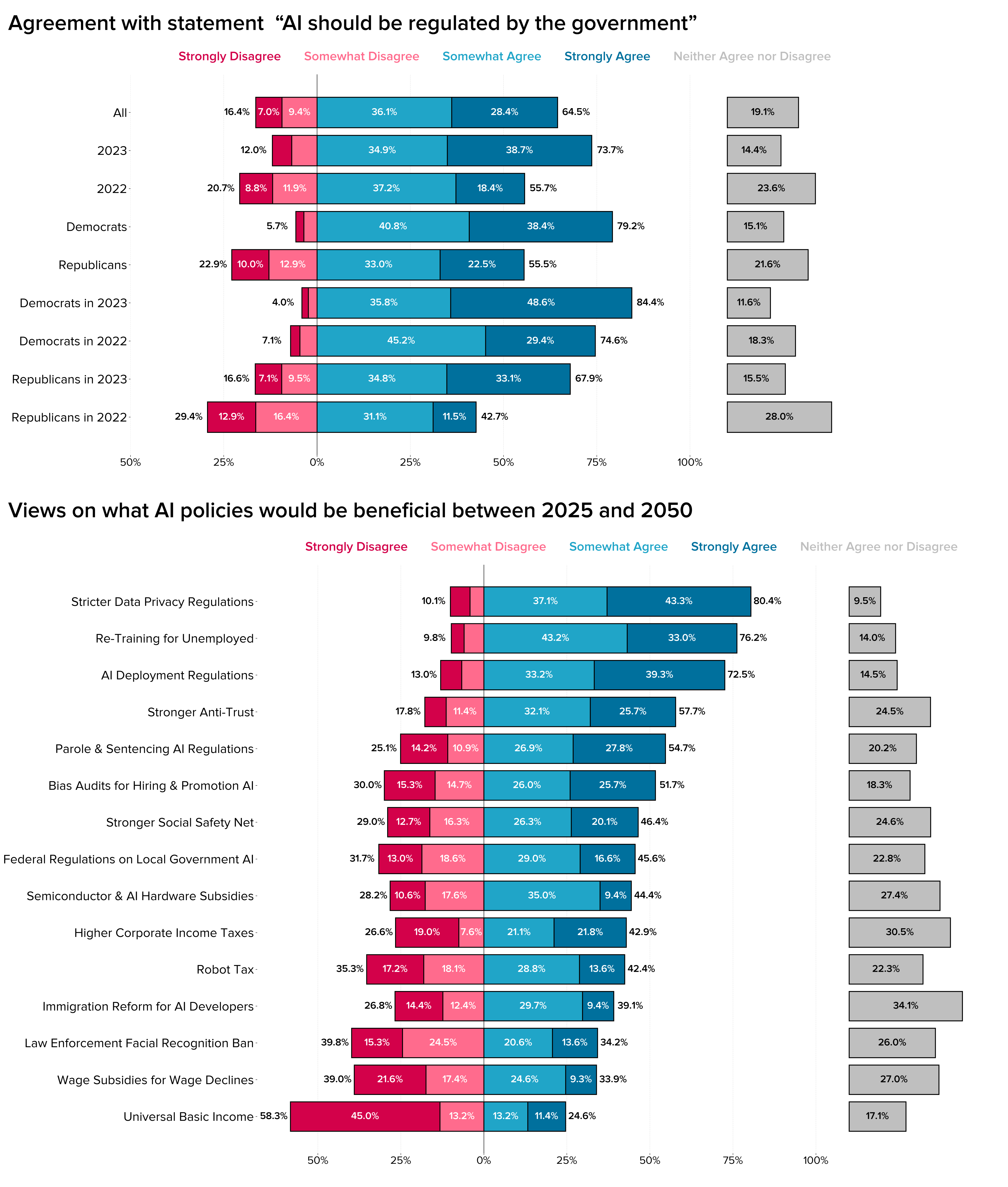}
    \caption{\textbf{Local US officials' views on what government AI regulation and which federal AI policies would be beneficial between 2025 and 2050.} The figure shows unweighted relative frequencies for QS4 and Q4.1 across both survey waves.}
\end{figure}

\begin{figure}[p]
    \centering
    \includegraphics[width=\textwidth,height=\textheight,keepaspectratio]{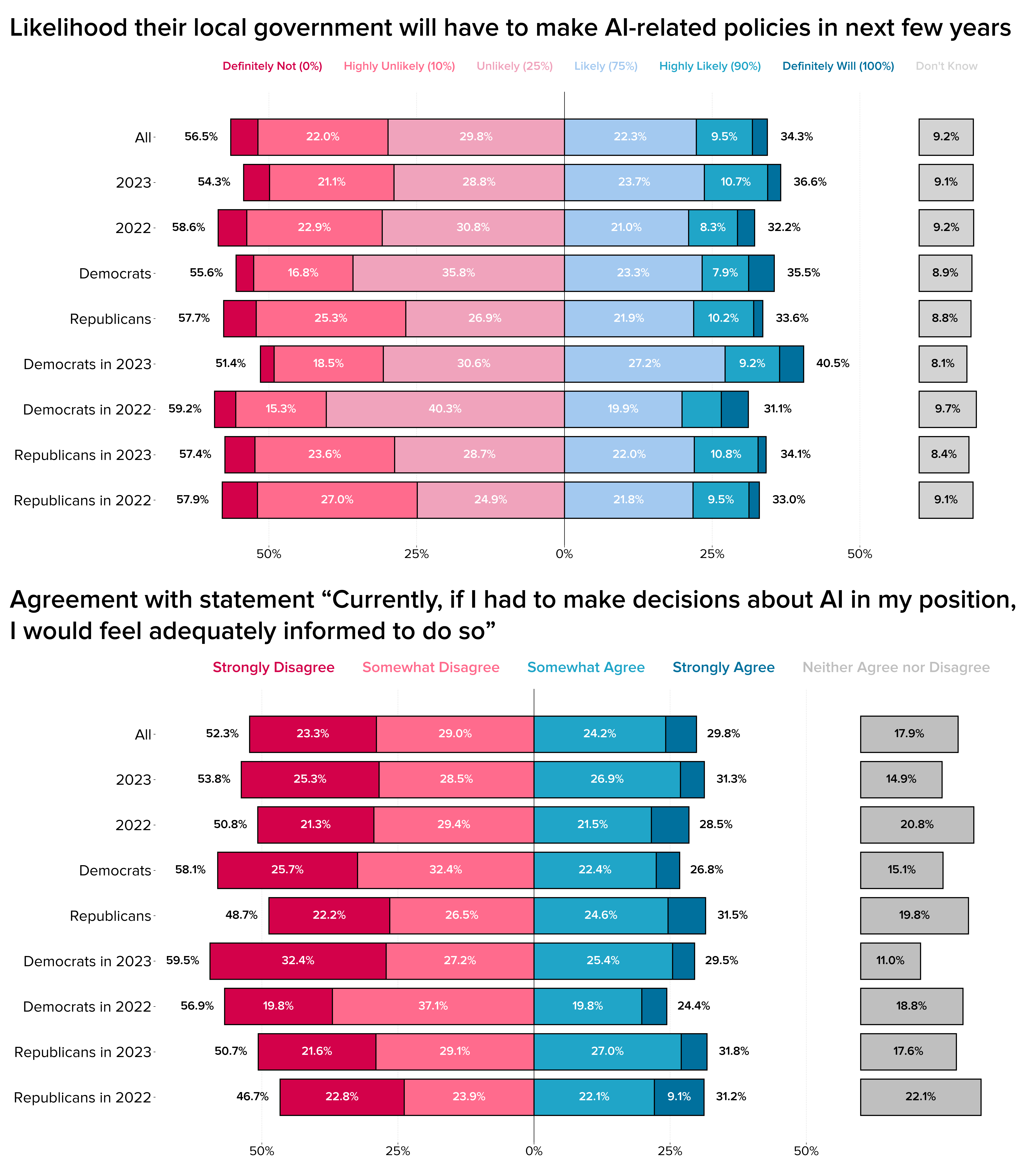}
    \caption{\textbf{Local US officials' expectations about whether their local government will need to make AI-related decisions and whether they would personally feel adequately informed to make these.} The figure shows unweighted relative frequencies for QS5 across both survey waves.}
\end{figure}

\clearpage
\renewcommand{\thefigure}{\arabic{figure}}
\setcounter{figure}{0}

\newpage

\section*{Introduction} \label{sec:Introduction}

The release of ChatGPT in November 2022 drew the world’s attention to Artificial Intelligence (AI), demonstrating the potential of AI to impact nearly all aspects of society. Both expectations of vast benefits across areas such as the economy, healthcare, education, and national security, and concerns about risks such as discrimination, job loss, and authoritarianism, have come to the forefront of public discourse. 

As the far-reaching impacts of AI materialize, governments are increasingly considering regulating AI. In the years following the launch of ChatGPT, several concrete AI regulations emerged \cite{HCAI:2024aa}. The European Union advanced landmark policy action with the EU AI Act: a legislative framework that governs AI development, marketing, and use within the EU, approved by the European Parliament in 2024 \cite{eu_ai_act,eu_ai_act_final}.  In the US, there has been a surge in regulatory policy on AI at the national, state, and local level. In October 2023, US President Biden issued an Executive Order on the ``Safe, Secure, and Trustworthy Development and Use of AI,'' including new standards for safety and security and provisions to protect citizens and workers. AI became a focal topic for state legislatures in 2023, with a massive increase in the number of proposed AI-related bills. City and county governments have also moved forward, adopting AI technologies in city services and enacting their own AI policies. As AI continues to advance rapidly, understanding how policymakers view and plan to govern this technology is increasingly critical. 

This paper examines US policymakers' perspectives on AI's future impact and regulation, focusing specifically on local elected officials. Our survey provides insight into the expectations of local US policymakers about the future effects of AI on local communities and the nation, as well as their attitudes toward specific regulatory policies. We analyze how views on AI vary between Democrats and Republicans, as well as by gender and level of education. In addition, we conducted our survey in two waves, May/June 2022 ($n = 524$) and May/June 2023 ($n = 504$), allowing us to assess how attitudes changed after the release of ChatGPT \cite{openai_chatgpt} and the substantial increase in public awareness of AI \cite{HCAI:2024aa,Heikkila:2024aa}. 

In general, local policymakers express a mixture of concern, optimism, and uncertainty about the impacts of AI over 2025--2050 in their local community and the broader country. A majority of local policymakers anticipate significant societal risks from AI over the next decades, including increased surveillance, misinformation, political polarization, decreased data security, and threats to US democracy. Some also foresee positive impacts, particularly in innovation and transportation/infrastructure. Views on AI's future economic impacts are more measured, with most expecting negative impacts on jobs and inequality, but mixed expectations about the US economy. Notably, policymakers also express a great deal of uncertainty, particularly around broader impacts such as international conflicts, the probability of great power war, bias and discrimination, and the impact on the US economy. Also reflecting this uncertainty, local policymakers feel under-prepared for AI governance, believing they are unlikely to make AI-related decisions in the next few years, and reporting they feel inadequately informed to make such decisions. 

In regard to the governance of AI, we asked respondents whether they
thought AI should be regulated, and whether they thought a range of AI policies would be beneficial for the country in the years 2025--2050. A majority of local policymakers support government regulation and oversight. Further, policymakers are in favor of a number of policies that address specific AI issues such as data privacy, AI-related unemployment, and AI safety and fairness. UBI stands out as the only AI policy proposal facing strong opposition.

However, in examining support for AI regulation between Democrats and Republicans, we find substantial differences. Democrats are generally more in favor of government regulation than Republicans, and consistently show greater support for policies addressing specific AI use cases and issues. This includes policies related to AI-driven unemployment, wage decline, regulation that ensures deployed AI systems are safe, robust, and fair, AI-use in judicial decisions, AI use in hiring, immigration reform for AI developers, and company use of robots. 

Tracking changes in attitudes towards AI over the 2022--2023 time period, we observe a significant increase in support for government regulation of AI. Although support increased among both Democrats and Republicans, the change was much greater among Republicans, who shifted to majority support in 2023. 

Understanding local officials’ views on AI is important for several reasons. First, due to the decentralized nature of US policymaking, local officials on the county-, municipal- and township-level play important roles in shaping state legislation. For example, AI guidelines introduced by the city of Seattle were influential in shaping Washington state law \cite{BSA:2023aa}. Second, in the absence of federal-level comprehensive AI law, US AI policy is largely being shaped by states, particularly since many states are controlled by one political party \cite[p.7]{Brennen:2023aa}. Our survey of local US officials thus provides insight into expectations and attitudes that will shape how AI is regulated in the US.

Our study expands on earlier survey research in a number of ways. First, our survey of local US policymakers is thematically broad, measuring expectations of future risks and benefits, as well as attitudes towards government regulation and a battery of potential policies. While existing survey research provides comprehensive understanding of US public opinion on AI \cite{Zhang:2019aa}, and prior surveys of US officials provide insights into attitudes on specific issues such as autonomous vehicles \cite{Horowitz:2021aa}, we lack a broad mapping of the views of local US political elite.

Second, our two survey waves -- fielded in 2022 and 2023 -- repeated the same questions, allowing us to capture how policymakers’ views changed over this time period.  This time period spans sudden and significant changes in AI, including the public releases of OpenAI’s ChatGPT and GPT-4, Midjourney, Stable Diffusion, Microsoft’s Bing AI Chat, and Google Bard \cite{openai_chatgpt, openai_gpt4_system_card, midjourney_2022, stabilityai_stable_diffusion_announcement, mehdi2023bing_edge_ai, pichai2023ai_journey}. Public awareness of AI and concern about its risks also increased significantly over this period \cite{Faverio:2023aa}. Third, we focus on the differences in views on AI across Democrats and Republicans, as well as changes in partisan differences over the 2022--2023 time period.  

Four key findings have clear implications for the trajectory of AI regulation in the US. First, we observed a significant increase in local policymakers' support for AI regulation in 2023. While we cannot rule out other simultaneous influences during 2022--23, many analysts attribute the surge in AI regulation globally, nationally, and in the US to the release of ChatGPT in 2022 \cite[p.17]{HCAI:2024aa}\cite{Heikkila:2024aa,Lee:2024aa}. Given that advances in AI technology tend to be sudden and significant, it’s likely that big changes such as those in 2022--23 will occur again. It seems reasonable to expect substantial jumps in support for AI regulation as AI advances.

Second, we observed polarization in views on AI regulation across party lines, with Republicans consistently showing less support for government regulation, as well as for policies addressing specific AI challenges like unemployment, safety, and fairness. Interestingly, however, this polarization was stronger in regard to policy preferences; Democrats and Republicans express more similar views on the impacts of AI. We believe this has important implications both for public opinion formation and for policy adoption. Some research suggests that on polarized issues, the public tends to form opinions in line with the political elite who share their partisanship \cite{Guisinger:2017aa,Baldassarri:2008aa}, relying less on substantive arguments about the issue \cite{Druckman:2013aa}. Further, polarization along party lines can have implications for state policy. Divergent state-level policies may result in ``patchwork'' AI legislation across the US, which could generate uncertainty and complications for companies, workers and consumers who operate across state lines \cite{Lee:2024aa}. 

Third, we also observed some indication that partisan differences are decreasing, as a majority of Republican local policymakers now support government regulation. This is likely to support emerging bi-partisan efforts towards comprehensive AI legislation, such as the SAFE Innovation Framework proposed by Senate Majority Leader Chuck Schumer \cite{Covington:2023aa}.

Finally, our finding that local officials feel uncertain about AI’s impacts and are unprepared to make AI-related policy decisions highlights the need for both capacity-building and further research. State and local legislators have begun tackling this issue through capacity-building laws establishing working groups tasked with studying AI \cite[p.12-13]{Brennen:2023aa}. Future survey research could support these efforts by evaluating the specific areas where policymakers may feel least informed or prepared, such as the technical aspects of AI, its economic implications, or its potential role in international relations, and tracking changes in competencies over time.

\section*{Literature Review} \label{sec:Literature_Review}

\subsection*{US AI policy during 2022--2023}

Over the past decade, governments have grown increasingly aware of the need to regulate AI in order to address risks while maximizing the benefits. Concrete regulatory policies have emerged in the past five years, with significant policy advances in 2023, in the year following the launch of ChatGPT \cite{HCAI:2024aa}. 

In particular, the European Union advanced landmark policy action with the EU AI Act: a legislative framework which governs AI development, marketing and use within the EU \cite{eu_ai_act_final}. This is the world’s most comprehensive legislation on AI. Other notable policy actions in 2023 include the UK’s AI Safety Summit and release of a government white paper on AI regulation \cite{UK_AI_Regulation_2023,AI_Safety_Summit_2023}, as well as China’s announcement of an ``Artificial Intelligence Law'' on its legislative agenda \cite{Heikkila:2024aa}. 

In the US, President Biden issued in 2023 an Executive Order on the ``Safe, Secure, and Trustworthy Development and Use of AI,'' including new standards for safety and security and provisions to protect citizens and workers. AI also became a high-priority issue in Congress, with a substantial jump in the number of proposed AI-related bills, rising from 88 in 2022 to 181 in 2023 \cite[p.17]{HCAI:2024aa}. Some bills –primarily focused on government uses of AI– have been voted on and have passed through committee stages \cite{Covington:2023aa}. Yet, the US does not yet have ``comprehensive'' legislation on AI: law which imposes broad new consumer protections and company requirements for AI \cite[p.13]{Brennen:2023aa}. The biggest obstacle to comprehensive legislation is Congress; where a lack of consensus makes passing legislation unlikely \cite{Lewis:2023aa,Covington:2023aa}.  

In the absence of congressional legislation on AI, there has been a surge in activity at the state-level. State legislators introduced 191 AI-related bills in 2023; a 440\% increase since 2022 \cite{BSA:2023aa}. 14 of these bills were enacted into law in 2023. Many of these laws are issue-specific, targeting specific AI use cases and concerns. The range of issues varies widely, including government use, elections, privacy and pornography \cite[p.12-13]{Brennen:2023aa}. For example, Arizona, Minnesota and Texas passed laws concerning AI in law enforcement. Another six states passed privacy reforms limiting the use of AI in profiling. Three states passed laws targeting generative AI and deepfakes in political advertising. States also enacted capacity-building laws which set up working groups or task forces to study AI, conduct inventories or develop regulatory frameworks \cite[p.12-13]{Brennen:2023aa};\cite{BSA:2023aa,Lee:2023aa}. 

At the local level, municipalities and county governments have also moved forward, enacting their own policies on AI. Most of these policies, such as in Boston, Santa Cruz County and Grove City, focus on government uses of AI \cite{Edinger:2024aa,Hurley:2024aa}. These cities have incorporated AI technologies in city services, serving as pioneers in the anticipation of risks, ethical considerations and necessary regulation \cite{Lee:2023aa}. 

Overall, there has been an enormous increase in federal, state and local government attention to AI regulation in 2023; it has become the "hot topic" in technology policy. There is also great heterogeneity in the AI laws passed, with laws targeting concerns ranging from profiling in automated decisions to deep fakes in political advertising. This highlights the diversity of perspectives on AI in terms of perceived impacts, risks and policy priorities.

\subsection*{Survey research on AI impacts and policy}

Survey research provides insight into the drivers of AI policy by examining attitudes such as expectations of risks and benefits and preferences for regulation. Prior survey research on AI tends to target one of several different audiences: the general public, AI experts, or elite audiences such as CEOs, academics and policymakers. 

Surveys of the general US public measure a wide variety of themes, including general awareness and use of AI, trust in AI, support for specific applications, views on AI labs, attitudes towards regulation, expectations of achieving intelligence milestones and concern about risks. Citizens tend to be optimistic about certain uses of AI, particularly criminal justice applications \cite{Rainie:2022aa,GovUK:2023aa} and some medical uses \cite{GovUK:2023aa,Faverio:2023aa}. They also express concern about risks such as economic harm/job loss \cite{Rainie:2022aa,Ipsos:2023aa,Marken:2023aa} and existential threats 
\cite{Ipsos:2023ab,Heath:2023ab,Gruetzemacher:2024aa}. A number of surveys asked whether citizens are overall more excited or more concerned about AI, and the results consistently show that concern outweighs excitement  \cite{GovUK:2023aa,Tyson:2023aa,Rainie:2022aa,Murray:2023aa,Penn:2023aa}. Among the surveys that include a category corresponding to being equally excited and concerned, however, a substantial proportion of respondents fall into this middle ground, reflecting ambivalence or a balanced view of AI's potential benefits and risks  \cite{Rainie:2022aa,Murray:2023aa,Heath:2023aa}. From a few public opinion surveys which repeated the same questions in subsequent waves, concerns about risks appear to have increased over the past few years \cite{Pauketat:2023aa,Tyson:2023aa}. When it comes to the regulation of AI, polls show the majority of the public is in favor \cite{MITRE:2023aa,Dreksler:2023aa,Ipsos:2023ac,GovUK:2023aa,Faverio:2023aa,Gruetzemacher:2024aa}. 

A second category is surveys of AI experts such as scientists, researchers and conference participants. In the most recent and largest survey, AI researchers assign non-negligible probabilities to both extremely good outcomes and extremely bad outcomes \cite{Grace:2024aa}. Specific risks of high concern include misinformation, increasing inequality and authoritarian control \cite[p.12]{Grace:2024aa};\cite[p. 5]{Gruetzemacher:2024aa}. When it comes to the regulation of AI and the implementation of AI safety measures, AI experts express broad support \cite{Schuett:2023aa,Heath:2023ab,Gruetzemacher:2024aa}. One study comparing US voters and AI experts finds that while both groups favor AI regulation, the public considers societal-scale risks both more likely and more impactful \cite{Gruetzemacher:2024aa}.

A final category of surveys are those that target elite audiences such as corporate managers, CEOs, entrepreneurs and academics. These surveys tend to focus on risks and opportunities, particularly in reference to business or research. Elite respondents are optimistic about potential benefits of AI, such as advances in health care and education \cite{Anderson:2023aa}, improvements in research processes \cite{Noorden:2023aa} and gains in efficiency and profitability \cite{Thomson-Reuters:2023aa,PwC:2024aa}. The risks they express most concern about include misinformation, inaccuracy and cybersecurity \cite{Anderson:2023aa,Noorden:2023aa,WEF:2024aa,PwC:2024aa,Chui:2023aa}.

While surveys of the public, AI experts and elite audiences are becoming regular, there have been relatively fewer studies of political elites. To our knowledge, the only prior study of US policymaker opinion on AI is Horowitz and Khan's survey of 690 US local officials \cite{Horowitz:2021aa}, focusing on attitudes towards the adoption of AI technology such as autonomous vehicles and autonomous surgery. As in surveys of the public, local US officials’ express both optimism and pessimism, with approval of AI varying widely depending on the specific use case.

\section*{Materials and Methods} \label{sec:Methods}

In this section, we present methodological details about our survey sample, survey questionnaire, and analysis. A pre-analysis plan was filed prior to analyzing survey results \cite{dreksler_2022-2023_2023}. The pre-analysis plan (\href{https://osf.io/k2efj}{https://osf.io/k2efj}), and the deviations from the pre-analysis plan (\href{https://osf.io/5kb23}{https://osf.io/5kb23}) can be found on OSF.

\subsection*{Sample} \label{subsec:Methods_Sample}

We conducted two national surveys of local elected officials in the United States. Data was collected in two independent waves by CivicPulse \cite{noauthor_civicpulse_nodate}, a non-profit organization that surveys local and state elected officials for academic and public policy research. CivicPulse collected random samples of US elected officials serving counties, municipalities, and townships with populations of over 1,000 residents. The first survey wave was conducted in May--June 2022 and the second in May--June 2023. CivicPulse also collected incomplete responses in 2022 but not in 2023 ($n = 46$); these incomplete responses are not reported below and were excluded from our analysis. 

\begin{figure}[htbp]
    \centering
    \includegraphics[width=\textwidth]{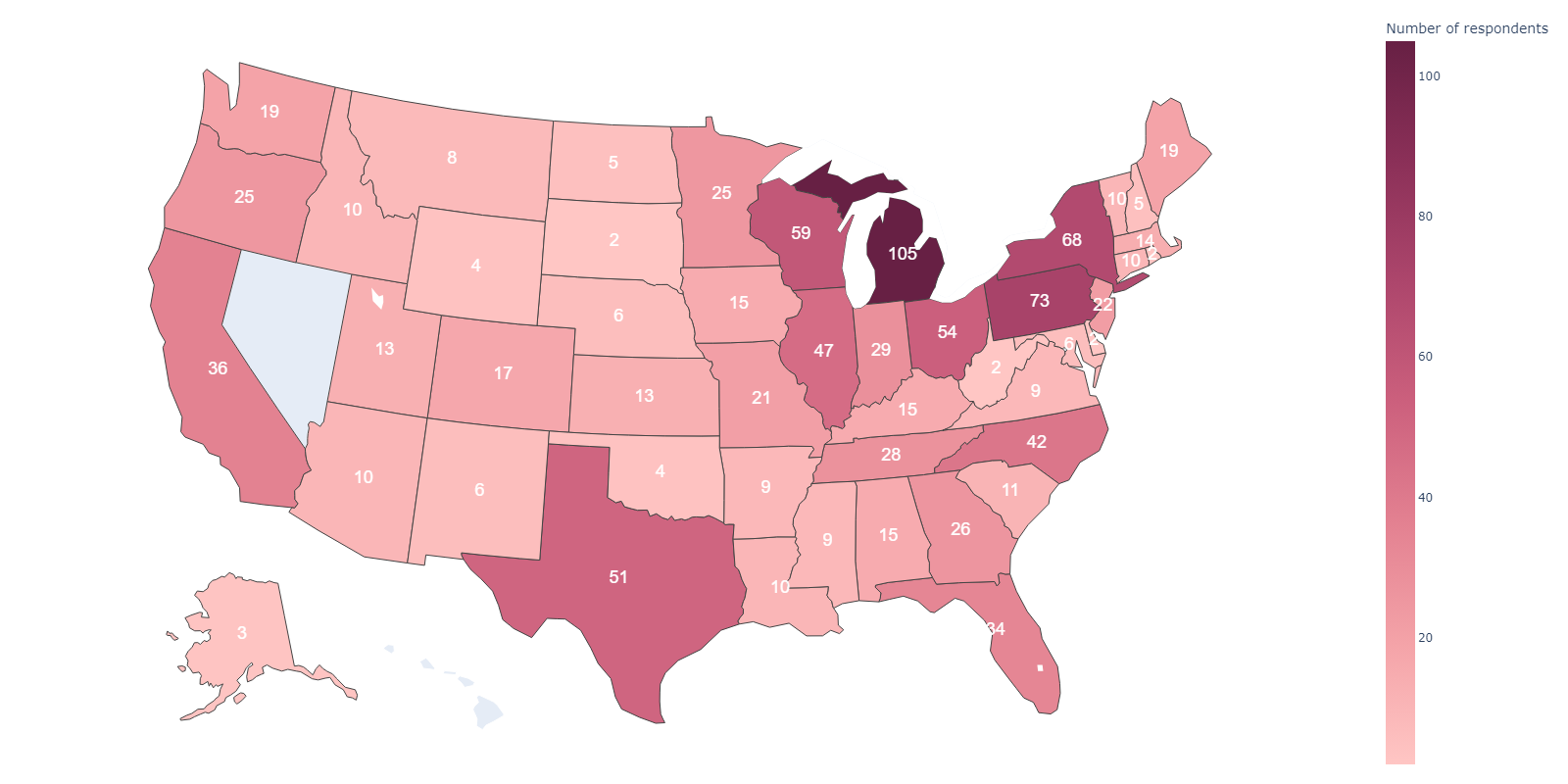}
    \caption{\textbf{Number of respondents per US state.}$^{\text{a}}$}
    \par
    \raggedright \footnotesize
    $^{\text{a}}$ The figure shows unweighted absolute frequencies across both survey waves; gray regions represent states from which we received no survey responses. \\   
    \label{fig:Methods_Chronopleth}
\end{figure}

We obtained a total of 1028 complete survey responses across both waves, with 524 and 504 responses from 2022 and 2023, respectively. The sample consisted of county-level officials, municipal-level officials, and township-level officials from across the country, with respondents' government levels contained in Table \ref{tab:Methods_Gov_Level}. Respondents' party identification are contained in Table \ref{tab:Methods_Party_Demographics}. Personal demographics of the respondents (gender, race, age, and education level) are contained in Table \ref{tab:Methods_Personal_Demographics}. Number of respondents per state are visualized in Fig \ref{fig:Methods_Chronopleth}. Sample representativeness statistics for each sample are included in \nameref{supp:samplerepresentativeness}. All tables and figures in this section are for unweighted data, using the pooled sample (i.e., combining the 2022 and 2023 samples).

\begin{table}[htbp]
    \centering
    
    \begin{tabular}{l*{6}{r}}
    \toprule
    & \multicolumn{3}{c}{\textbf{Count}} & \multicolumn{3}{c}{\textbf{Percentage}} \\
    & 2022 & 2023 & Total & 2022 & 2023 & Total \\
    \midrule
    \multicolumn{7}{l}{\textbf{Government Level}} \\
    
    Township & 125 & 111 & 236 & 23.85\% & 22.02\% & 22.96\% \\
    Municipality & 321 & 324 & 645 & 61.26\% & 64.29\% & 62.74\% \\
    County & 78 & 69 & 147 & 14.89\% & 13.69\% & 14.30\% \\
    
    \bottomrule
    \end{tabular}
    \caption{\textbf{Frequency table of respondents' level of government.} The table shows unweighted absolute and relative frequencies across both survey waves.} 
    \label{tab:Methods_Gov_Level}
\end{table}

\begin{table}[htbp]
    \centering
    
    \begin{tabular}{l*{6}{r}}
    \toprule
    & \multicolumn{3}{c}{\textbf{Count}} & \multicolumn{3}{c}{\textbf{Percentage}} \\
    & 2022 & 2023 & Total & 2022 & 2023 & Total \\
    \midrule
    Democrat & 197 & 174 & 371 & 37.81\% & 34.73\% & 36.30\% \\
    Independent & 38 & 31 & 69 & 7.29\% & 6.19\% & 6.75\% \\
    Republican & 286 & 296 & 582 & 54.89\% & 59.08\% & 56.95\% \\
    \bottomrule
    \end{tabular}
    \caption{\textbf{Frequency table of respondents' political party identification.}$^{\text{a}}$}
    \par 
    \raggedright \footnotesize
    $^{\text{a}}$ The table shows unweighted absolute and relative frequencies across both survey waves. We report transformed party data that is used for our regression model. Definitions of this variable and the transformation are described in \nameref{supp:Variable_Defs}. 
    \label{tab:Methods_Party_Demographics}
\end{table}

\begin{table}[htbp]
    \centering
    \begin{tabular}{l*{6}{r}}
    \toprule
    & \multicolumn{3}{c}{\textbf{Count}} & \multicolumn{3}{c}{\textbf{Percentage}} \\
    & 2022 & 2023 & Total & 2022 & 2023 & Total \\
    \midrule
        \multicolumn{7}{l}{\textbf{Gender}} \\  \\
        Women & 149 & 89 & 238 & 29.33\% & 24.86\% & 27.48\% \\
        Men & 359 & 262 & 621 & 70.67\% & 73.18\% & 71.71\% \\
        Other (Self-Described) & 0 & 7 & 7 & 0.00\% & 1.96\% & 0.81\% \\
        \midrule
        \multicolumn{7}{l}{\textbf{Race}$^{\text{a}}$} \\
        White & 443 & 302 & 745 & 86.02\% & 87.79\% & 86.73\% \\
        Non-White & 72 & 42 & 114 & 13.98\% & 12.21\% & 13.27\% \\
        \midrule
        \multicolumn{7}{l}{\textbf{Median Age (Birth Year Range)$^{\text{b}}$}} \\
        21 (2001 - 2005) & 1 & 0 & 1 & 0.19\% & 0.00\% & 0.12\% \\
        26 (1996 - 2000) & 0 & 1 & 1 & 0.00\% & 0.30\% & 0.12\% \\
        31 (1991 - 1995) & 2 & 5 & 7 & 0.39\% & 1.50\% & 0.83\% \\
        36 (1986 - 1990) & 12 & 9 & 21 & 2.33\% & 2.70\% & 2.48\% \\
        41 (1981 - 1985) & 19 & 8 & 27 & 3.69\% & 2.40\% & 3.18\% \\
        46 (1976 - 1980) & 36 & 23 & 59 & 6.99\% & 6.91\% & 6.96\% \\
        51 (1971 - 1975) & 32 & 31 & 63 & 6.21\% & 9.31\% & 7.43\% \\
        56 (1966 - 1970) & 43 & 33 & 76 & 8.35\% & 9.91\% & 8.96\% \\
        61 (1961 - 1965) & 44 & 49 & 93 & 8.54\% & 14.71\% & 10.97\% \\
        66 (1956 - 1960) & 89 & 57 & 146 & 17.28\% & 17.12\% & 17.22\% \\
        71 (1951 - 1955) & 107 & 57 & 164 & 20.78\% & 17.12\% & 19.34\% \\
        76 (1946 - 1950) & 78 & 41 & 119 & 15.15\% & 12.31\% & 14.03\% \\
        81 (1941 - 1945) & 35 & 13 & 48 & 6.80\% & 3.90\% & 5.66\% \\
        86 (1936 - 1940) & 15 & 3 & 18 & 2.91\% & 0.90\% & 2.12\% \\
        91 (1931 - 1935) & 2 & 1 & 3 & 0.39\% & 0.30\% & 0.35\% \\
        96 (1926 - 1930) & 0 & 1 & 1 & 0.00\% & 0.30\% & 0.12\% \\
        104 (1920 or earlier) & 0 & 1 & 1 & 0.00\% & 0.30\% & 0.12\% \\
        \midrule
        \multicolumn{7}{l}{\textbf{Education Level}} \\
        Less than high school & 2 & 1 & 3 & 0.38\% & 0.20\% & 0.29\% \\
        High school graduate & 40 & 45 & 85 & 7.68\% & 9.04\% & 8.34\% \\
        Technical/trade school & 26 & 33 & 59 & 4.99\% & 6.63\% & 5.79\% \\
        Some college & 97 & 89 & 186 & 18.62\% & 17.87\% & 18.25\% \\
        College graduate & 131 & 140 & 271 & 25.14\% & 28.11\% & 26.59\% \\
        Some graduate school & 38 & 38 & 76 & 7.29\% & 7.63\% & 7.46\% \\
        Graduate degree & 187 & 152 & 339 & 35.89\% & 30.52\% & 33.27\% \\
    \bottomrule
    \end{tabular}
    \caption{\textbf{Absolute and relative frequencies of respondents' personal demographics, totals and segmented by year.}$^{\text{b}}$ }
    
    \par
    \raggedright \footnotesize
    $^{\text{a}}$ CivicPulse collected data for sub-groups beyond `non-white' in 2022 but not in 2023, so we report only white/non-white. \\
    $^{\text{b}}$ CivicPulse collected only birth year ranges, and we impute ages by subtracting median birth year in each bin from 2024. These imputed ages are used in the regression analysis below and are treated as a continuous variable. \\
    $^{\text{c}}$ The table shows unweighted absolute and relative frequencies across both survey waves. \\
    \label{tab:Methods_Personal_Demographics}
\end{table}

\subsection*{Survey questions} \label{subsec:Methods_Survey_Questions}

The main survey questions consisted of five sets of Likert-scale questions about AI impacts and governance; questions were identical in both waves of the survey. 

Question Sets 1 through 3 (QS1--QS3) measured expectations of future AI impacts: QS1 and QS2 consisted of questions about the effects of AI on respondents' local communities, and QS3 asked about the effects of AI beyond respondents' local communities. Responses for QS1 and QS3 were measured on a 5-point Likert response scale representing respondents' expectations of the expectations of AI, from strongly decrease (-2) to strongly increase (2); the 5-point Likert response scale for QS2 was similar but presented in terms of strongly worsen (-2) to strongly improve (2). Respondents were permitted to select ``I don't know'' (``IDK'') for QS1--2 but not for QS3.

Question 4.1 (Q4.1) and Question Set 4 (QS4) measured attitudes towards AI governance. Q4.1 measured respondents' support for government regulation of AI in general (Q4.1). QS4 measured respondents' beliefs about whether specific federal policies on AI would benefit the United States. The 5-point Likert response scale in Q4.1 and QS4 represented respondents' agreement with particular policy positions or statements, from strongly disagree (-2) to strongly agree (2). Respondents were not permitted to select ``I don't know'' (``IDK'') for Q4.1 or QS4.

Q4.2 measured respondents' respondents' beliefs on long-term AI impacts until 2021, asking whether they expected overall positive or negative benefits. This question was also the outcome variable in a survey experiment, but as the experiment is not central to this paper, we describe the survey experiment design and results in \nameref{supp:Survey_Experiment}. 

The survey questions in QS1--QS4 (excluding Q4.1 and Q4.2) were presented to the respondents in grid format, and the survey incorporated a planned missing design \cite{zhang_planned_2022, pokropek_missing_2011}. Respondents were randomly presented with three of six questions in QS1, three of five questions in QS2, four of six questions in QS3, and five of fifteen questions in QS4. In practice, the proportion of missing data from the survey design fell in the range of 39--46\% for QS1, 32--38\% for QS2, 25--30\% for QS3, and 66--69\% for QS4. 

Question Set 5 (QS5) consisted of two questions (Q5.1, Q5.2) related to policymakers' preparedness for AI-related policy decisions. QS5.1 asked respondents to indicate the likelihood that their local government would make decisions about AI ``in the next few years,'' and QS5.2 asked respondents to assess whether they felt ``adequately prepared to'' make decisions about AI. The Likert scale for Q5.1 contained six options, asking respondents to rate probabilities from 0\%, 10\%, 25\%, 75\%, 90\%, and 100\%; the response scale for Q5.2 was the same as for QS4 and Q4.1. Respondents were permitted to select ``I don't know'' (``IDK'') for Q5.1 but not for Q5.2.

Finally, the survey included collected demographic information on gender, age, education level, race, political ideology, and political party. CivicPulse then joined the data on respondents’ level of government, geographical location (US state), and US census data.

The dataset also included survey weights: both pooled and unpooled weights were calculated by CivicPulse using a post-stratification raking procedure per \cite{debell_computing_2009}. Weights were calculated using Census data for the population of the local government, the proportion of the local population 25 and older with a 4-year college degree, and the county’s proportional votes for President Biden in 2020 (see \nameref{supp:Survey_Text}). 

Aside from variables constructed from individual survey questions, we constructed several indices which aggregate responses across multiple survey questions. In particular, we create a policy agreement index (average across QS4 and Q4.1); positive impacts index (average of all impacts in QS1--3 which correspond to societal benefits); positive impacts for local community index (average of local community impacts in QS1--3 which correspond to societal benefits); personal well-being and community index (average of impacts on mental health, physical health and quality of life in QS1--3); and progress and innovation index (average of transportation and infrastructure items in QS1--3). A full list of constructed indices and their component items is included in \nameref{supp:Indices_Defs}.

\nameref{supp:Survey_Text} contains the complete survey text. 

\subsection*{Analysis} \label{subsec:Methods_Analysis}

We conducted our analysis of the survey data according to the filed pre-analysis plan \cite{dreksler_2022-2023_2023}. As specified in the pre-analysis plan, we first generate summary statistics describing respondent demographics, expectations of AI impacts and attitudes towards AI governance, pooling the two survey waves. To examine how expectations of AI impacts and attitudes towards AI governance vary across respondent subgroups and over the 2022--2023 time period, we specify in equation \ref{eq:Linear_Model} a survey-weighted linear regression model \cite{lumley_fitting_2017}:

\begin{equation*} \label{eq:Linear_Model}
    \begin{aligned}
        y_i = gender + age + edu + race + party + year_{2023} + party * year_{2023} \\
        + gov_{municipality} + gov_{county} + college + pop + Biden \\
        + \mathbbm{1}_{y_i \in \{Indices \backslash Index_{Policy \ Agreement}\}} * policy_{regai}
    \end{aligned}
\end{equation*}

where $y_i \in \{QS1-4, Q4.1, Indices\}$ represents the outcome variable, which includes the questions on impacts and governance as well as the constructed indices.  

To assess group differences we include the respondent-level demographic variables $gender$, $age$, $education$, $race$ and $party$. $Party$ indicates political party identification, constructed by combining questions on ideology and self-identified political party affiliations (details in \nameref{supp:Variable_Defs}). To assess changes over time, we include the dummy variable $Year_{2023}$, which indicates the survey wave (2022 vs. 2023). We also include an interaction term for the combined effect of year and party identification: $party * year_{2023}$.

We include five control variables. $Gov_{municipality}$ and $gov_{county}$ are dummy variables for the level of government of the respondent. $College$ is the proportion of residents in each respondent's geographic unit who are 25 years or older and who have completed a 4-year, post-secondary degree, based on 2015--2019 US Census data and binned in terciles. $Pop$ is the total number of residents in each respondent's geographic unit, based on 2015--2019 US Census data and binned in terciles. $Biden$ is the proportion of votes by county for Joe Biden in the 2020 Presidential election, binned in terciles. 

$\mathbbm{1}_{y_i \in \{Indices \backslash Index_{Policy \ Agreement}\}}$ is an indicator variable for whether $y_i$ is any index except the policy agreement index, and $policy_{regai}$ is Q4.1 (which is excluded from the regression equation for the policy agreement index as it is an element of that index). 

Missing data for variables in the regression analysis was imputed using multiple imputation by chained equations (MICE) \cite{van_buuren_mice_2011, van_buuren_flexible_2018, van_ginkel_rebutting_2020}, which is appropriate for planned missingness designs \cite{kaplan_imputation_2018, little_joys_2014, baraldi_introduction_2010}. The MICE model specification contained all questions in QS1--4, and all demographic information as well as survey weights (see below). 120 imputed datasets were generated with predictive mean matching (PMM) and run for 200 iterations each. All IDK responses in QS1--4 were recoded to 0 (neutral). Running the regressions after coding the IDK responses as missing (and generating them using MICE) increased the number of statistically-adjusted regression coefficients that were significant at the $p < 0.05$ level, but all significant coefficients under the original coding scheme remained significant at the $p < 0.05$ level. Analysis on the dataset where IDKs were treated as missing, and were then imputed, are presented in \nameref{supp:Alternative_Regression_Results}. We also generated indices based on the survey questions, which are described in \nameref{supp:Indices_Defs} and on which we fit the regression model above. Note that imputed data and survey weights are only used for regression analysis, and all descriptive statistics ( relative frequency charts) do not contain imputed data and are unweighted; the weights are described in \nameref{subsec:Methods_Survey_Questions}.

Detailed definitions and coding for all variables are contained in \nameref{supp:Variable_Defs}. Statistical corrections using the Benjamini-Hochberg method were applied to all regression coefficients. The statistical analysis was conducted in R and Python.

\section*{Results} \label{sec:Results}

In this section, we present summary statistics and the results of our regression analyses. We begin by describing the distribution of responses in three categories of questions: 1) expectations of the impacts of AI on local communities and the broader country; 2) attitudes towards specific AI policies and AI regulation; and 3) preparedness for AI-related policy decisions. We then examine group differences and changes over time, presenting select regression results per the model specified in \nameref{subsec:Methods_Analysis}. 
Full regression results, descriptive statistics, and additional figures are included in \nameref{supp:Full_Regression_Results} and \nameref{supp:Additional_Figures}.

\subsection*{Expectations of the impacts of AI} \label{subsec:Results_Summary_Impacts}

\begin{figure}[htbp]
    \centering
    \includegraphics[width=\textwidth,height=\textheight,keepaspectratio]{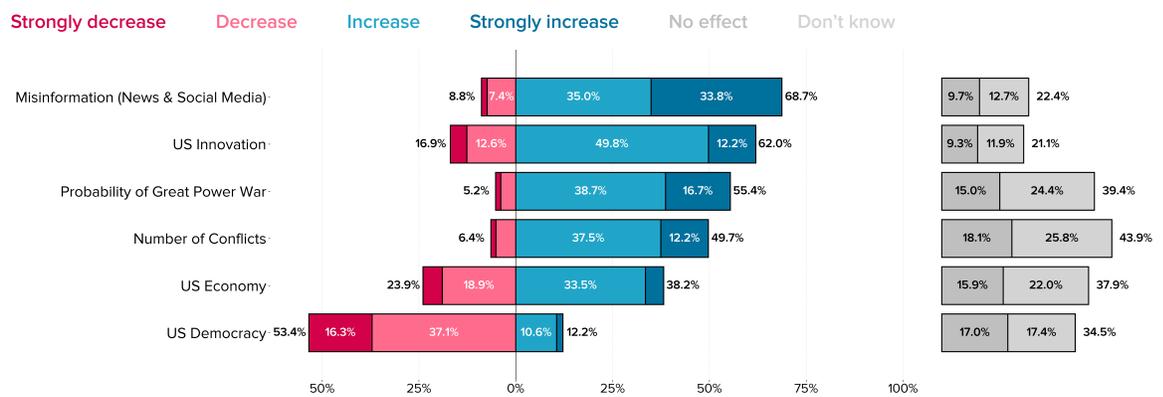}
    \caption{\textbf{Local US officials' expectations of the impacts of AI between 2025 and 2050.}$^{\text{a}}$}

    \par\raggedright \footnotesize
    $^{\text{a}}$ The figure shows unweighted relative frequencies across both survey waves. In all figures throughout, labels for bars containing relative frequencies of less than 7.0\% are hidden for readability. \\
    
    \label{fig:allimpacts}
\end{figure}

In regard to the impacts of AI over 2025--2050 in their local community and the broader country, we found that most local policymakers anticipate AI will create significant societal risks over the next decades (Fig \ref{fig:allimpacts}). A majority of respondents indicated that they somewhat or strongly agreed that AI will increase surveillance levels (83\%), increase misinformation (69\%), decrease data security (64\%), increase political polarization (59\%), increase the probability of great power war (56\%), and decrease the strength of US democracy (53\%). A plurality of respondents also expect other negative effects such as worsening mental health outcomes (49\%), increased international conflicts (49\%), and decreased numbers of jobs (48\%). 

Although these results show that respondents generally indicate a bleak view of AI, Fig \ref{fig:allimpacts} also shows that they expect positive effects in two areas: rates of innovation in the US (62\%) and local transportation/infrastructure (51\%).

Local policymakers express measured and mixed views on the economic impacts of AI in their local community. Along several economic indicators, greater proportions of local officials expect decreases rather than increases, including jobs (48\% vs. 17\%), income levels (41\% vs. 22\%), and inequality (38\% vs. 10\%). In terms of broader economic impacts, there are mixed expectations: 38\% expect the US economy to grow thanks to AI, 24\% expect it to decline, and 16\% believe it will have no effect. 

Respondents also express a great deal of uncertainty about the future impacts of AI, with high proportions of ``I don’t know'' (``IDK'') or ``neither agree nor disagree'' (``neutral'') responses. The highest rates of ``I don’t
know'' responses are for questions about international conflicts (26\%) and the probability of great power war (24\%). Around a fifth of respondents indicated ``I don’t know'' with regard to the local impacts of AI on inequality (20\%), bias and discrimination (21\%) and about the broader impact of AI on the US economy (22\%). In addition, a few questions have high proportions of neutral responses, most notably inequality (32\%), bias and discrimination (28\%), and physical health (26\%). 

\subsection*{Attitudes towards governance of AI} \label{subsubsec:Results_Summary_Governance}

To measure attitudes toward the governance of AI, we asked respondents whether they thought AI should be regulated, and whether they thought a range of AI policies
would be beneficial for the country in response to AI in the years 2025--2050.

Overall, a majority of local policymakers express support for government regulation: 64\% strongly or somewhat agree with the statement that AI should be regulated by the government, with 19\% IDK and only 16\% disagreeing. 

\begin{figure}[htbp] \label{fig:Results_All_Policies}
    \centering
    \includegraphics[width=\textwidth,height=\textheight,keepaspectratio]{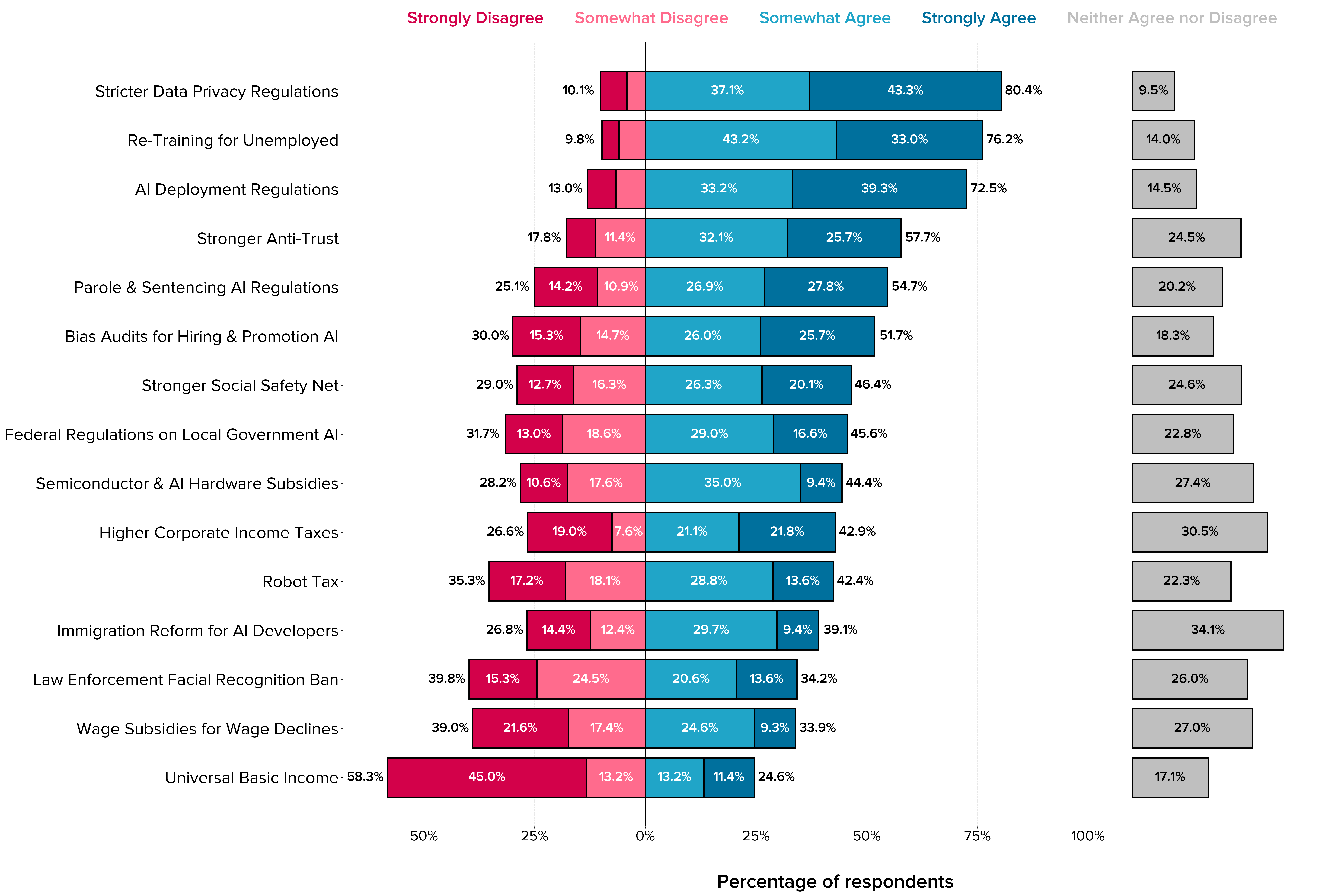}
    \caption{\textbf{Local US officials' views on what AI policies would be beneficial between 2025 and 2050.}$^{\text{a}}$}

    \par\raggedright \footnotesize
    $^{\text{a}}$ The figure shows unweighted relative frequencies across both survey waves. \\

    \label{fig:allpolicy}
\end{figure}

Further, local policymakers express support for many specific AI policies addressing particular use cases and issues. Among the policies listed in Fig \ref{fig:allpolicy}, a majority of respondents express support for stricter data privacy regulations (80\%), re-training programs for those at risk of automation-driven unemployment (76\%), regulation that ensures deployed AI systems are safe, robust, and fair (72\%), stronger antitrust policies (58\%), stricter requirements for AI use in judicial decisions (55\%), and bias audits for AI used in employment decisions (52\%). UBI stands out as the only AI policy proposal
facing strong opposition (58\% disagree). Two policies face plurality disagreement: wage subsidies for wage declines and a ban on law enforcement usage of facial recognition technologies.

Uncertainty is again evident in this category of questions, with 20\% or more of respondents expressing neutral positions on 10 of 16 questions (respondents did not have the option to select IDK for QS3--4 questions). As with the questions on AI impacts, respondents express most uncertainty with respect to foreign policy and economic questions --- i.e., immigration reform for AI developers (34\%), higher corporate income tax (31\%), semiconductor subsidies (27\%), and wage subsidies for wage declines (27\%). Note, however, that uncertainty is generally substantially lower here compared to questions on future AI impacts.

\subsection*{Policymakers’ AI preparedness and decision-making} \label{subsubsec:Results_Summary_Local}

A final category of questions relates to how likely local policymakers think they will make decisions related to AI regulation in the coming years, as well as how informed or prepared they feel to make such decisions.

\begin{figure}[htbp] 
    \centering
    \includegraphics[width=\textwidth,height=\textheight,keepaspectratio]{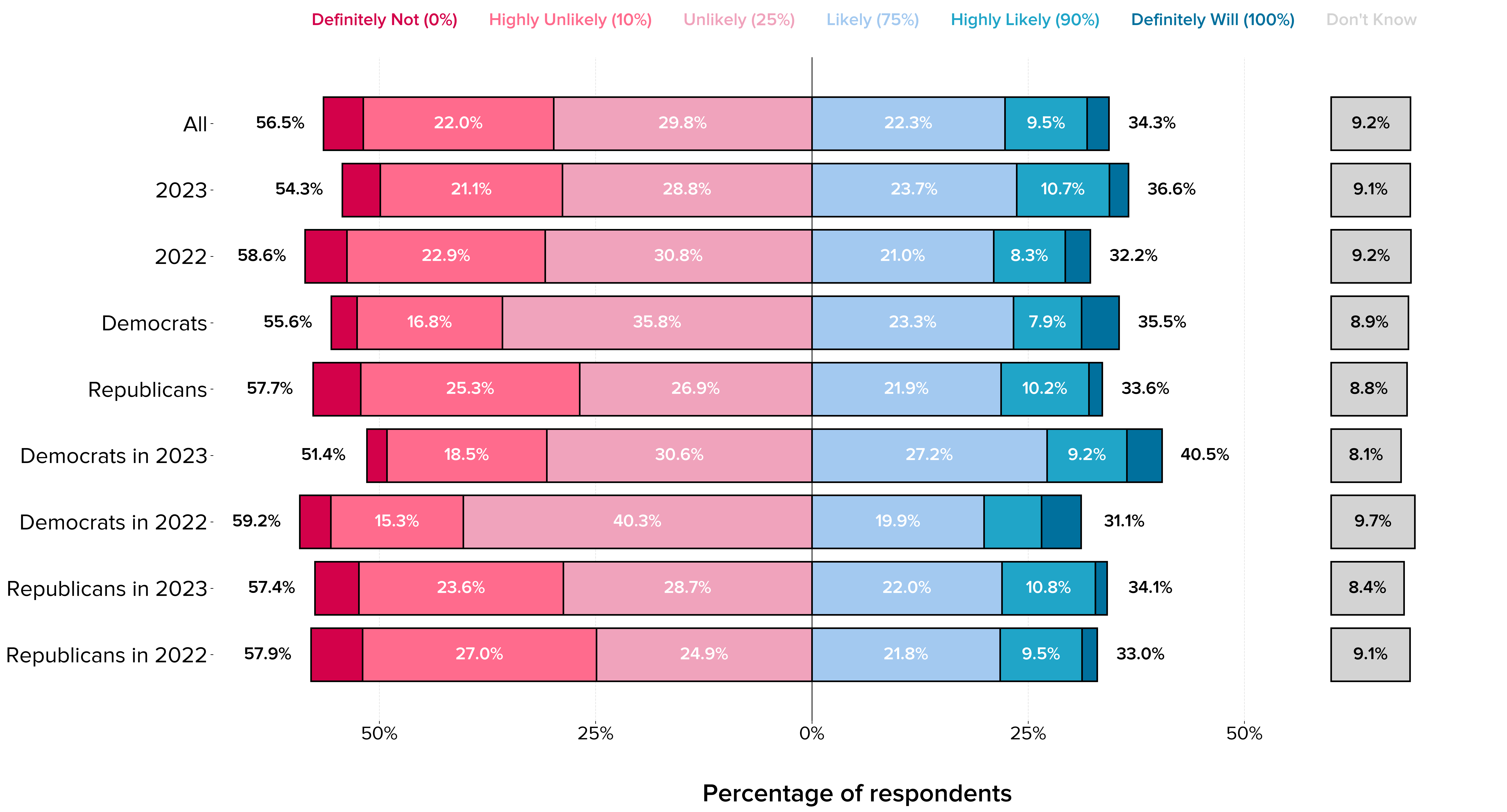}
    \caption{\textbf{Local US officials' responses to the question ``How likely is it that your local government will have to make decisions about AI-related policies and questions in the next few years?'', with overall support and segments by party and year.}$^{\text{a}}$}

    \par\raggedright \footnotesize
    $^{\text{a}}$ The figure shows unweighted relative frequencies across both survey waves. \\
    
    \label{fig:Results_Q5.1}
\end{figure}

\begin{figure}[htbp] 
    \centering
    \includegraphics[width=\textwidth,height=\textheight,keepaspectratio]{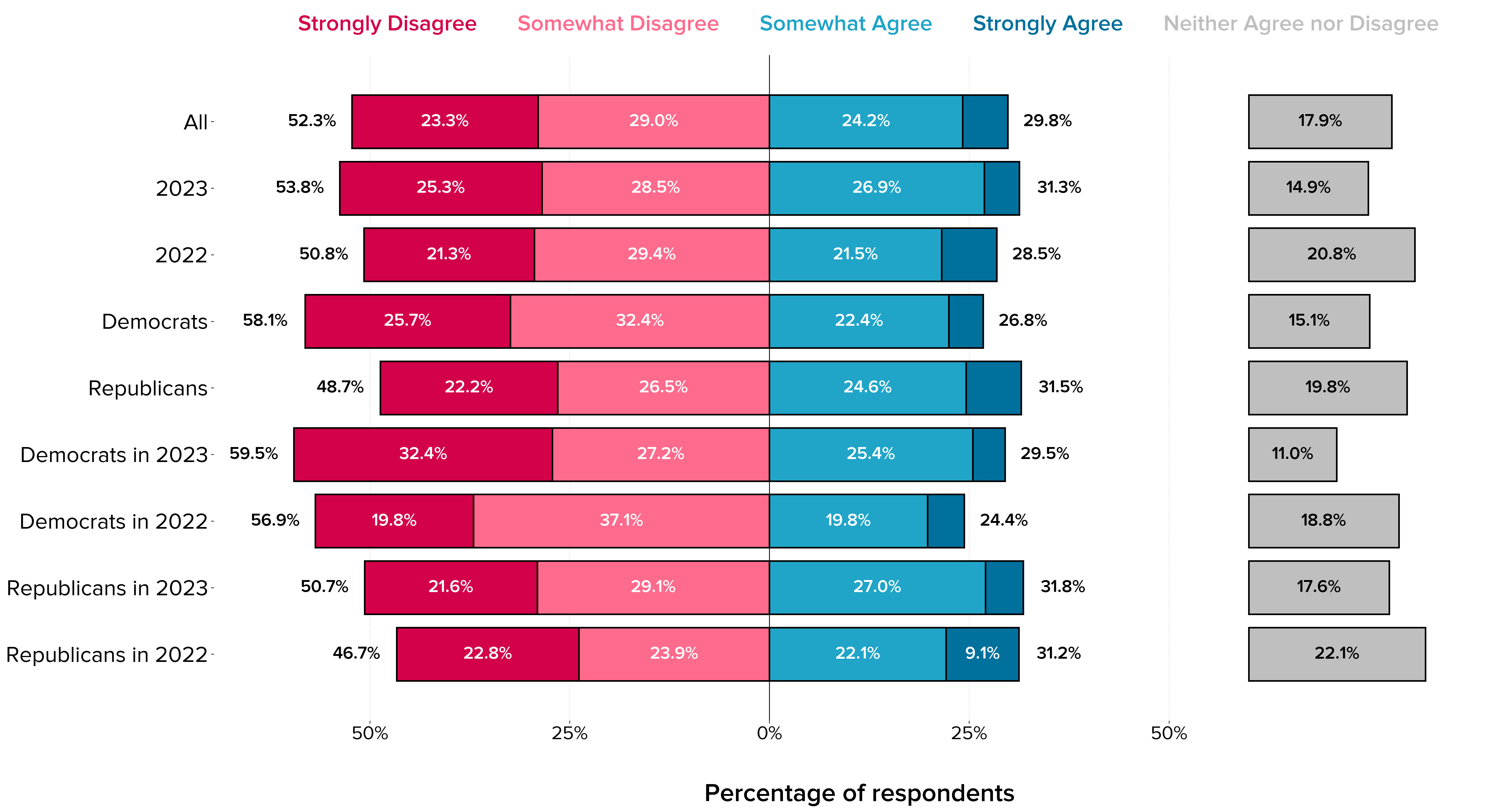}
    \caption{\textbf{Local US officials' agreement with the statement ``Currently, if I had to make decisions about AI in my position, I would feel adequately informed to do so,'' with overall support and segments by party and year.}$^{\text{a}}$}

    \par\raggedright \footnotesize
    $^{\text{a}}$ The figure shows unweighted relative frequencies across both survey waves. \\
    
    \label{fig:Results_Q5.2}
\end{figure}

Overall, local policymakers responded that they are unlikely to make decisions about AI over the next few years (57\% unlikely vs. 34\% likely) (Fig \ref{fig:Results_Q5.1}). The majority (54\%) of respondents also indicated that they did not feel adequately informed to make decisions about AI at the time of the survey (Fig \ref{fig:Results_Q5.2}).

\subsection*{Differences by party, education, and gender} \label{subsubsec:Results_Regression_Demographic}

We estimate the linear regression model specified in \nameref{sec:Methods} to assess differences in attitudes across respondent subgroups and over the 2022--2023 time period spanned by the two survey waves. 

Table \ref{tab:Results_regression_significant_1-3} shows that expectations of future AI impacts vary primarily by education level and gender, with smaller effects linked to age and political party affiliation. Local policymakers with higher education levels are more likely to anticipate improvements in transportation and infrastructure ($\beta = 0.121$, $ p < 0.001 $), express greater concern about surveillance ($\beta = 0.095$, $ p < 0.001$), and less concern about AI’s impact on the probability of great power war ($\beta = -0.071$, $ p = 0.010$). Gender differences reveal that compared to their female counterparts, male policymakers are more optimistic about AI’s impact on mental health  ($\beta = 0.313$, $ p = 0.003$), impact on quality of life ($\beta = 0.263 $, $ p = 0.025 $) and impact on the US economy ($\beta = 0.246 $, $ p = 0.044$). Age also has an influence, with older policymakers showing slightly lower concern about surveillance ($\beta = -0.010$, $p = 0.001$), but the difference is very close to zero. 

Political affiliation also shapes expectations, as Republicans are less likely than Democrats to predict positive effects of AI on physical health ($\beta = -0.181$, $ p = 0.004$) and quality of life ($\beta = -0.161$, $ p = 0.023$), though these differences remain modest.

\begin{table}[phtb]
    \centering
    \small
    
    \resizebox{\textwidth}{!}{
    \begin{tabular}{lccccccc}
        \toprule
        & \multicolumn{7}{c}{\textbf{Covariate}} \\
        \textbf{Question}$^{\text{a}}$ & $Age$ & $Education$ & $Gender$$^{\text{b}}$ & $Race$$^{\text{c}}$ & $Party$$^{\text{d}}$ & $Year$$^{\text{e}}$ & $Party * Year$ \\
        
        \midrule
        \multicolumn{7}{l}{\textbf{QS1: Local Effects of AI}} \\
        
        \hspace{1em} Surveillance Level & \textbf{-0.010\textsuperscript{**}} & \textbf{0.095\textsuperscript{***}} &  &  &  &  &  \\ 
        & \textbf{(-0.016, -0.005)} & \textbf{(0.051, 0.138)} &  &  &  &  &  \\ 
         
        \midrule
        \multicolumn{7}{l}{\textbf{QS2: Local Effects of AI}} \\

        \hspace{1em} Quality of Life &  &  & \textbf{0.263\textsuperscript{*}} &  & \textbf{-0.161\textsuperscript{*}} &  &  \\ 
         &  &  & \textbf{(0.094, 0.432)} &  & \textbf{(-0.263, -0.058)} &  &  \\ 
        \hspace{1em} Mental Health &  &  & \textbf{0.313\textsuperscript{**}} &  &  &  &  \\ 
         &  &  & \textbf{(0.149, 0.478)} &  &  &  &  \\ 
        \hspace{1em} Physical Health &  &  &  &  & \textbf{-0.181\textsuperscript{**}} &  &  \\ 
         &  &  &  &  & \textbf{(-0.278, -0.085)} &  &  \\ 
        \hspace{1em} Transportation \& Infrastructure &  & \textbf{0.121\textsuperscript{***}} &  &  &  &  &  \\ 
         &  & \textbf{(0.075, 0.167)} &  &  &  &  &  \\ 
         
        \midrule
        \multicolumn{7}{l}{\textbf{QS3: Broad Effects of AI}} \\
        \hspace{1em} US Economy &  &  & \textbf{0.246\textsuperscript{*}} &  &  &  &  \\ 
         &  &  & \textbf{(0.077, 0.414)} &  &  &  &  \\ 
        \hspace{1em} Probability of Great Power War &  & \textbf{-0.071\textsuperscript{*}} &  &  &  &  &  \\ 
         &  & \textbf{(-0.113, -0.030)} &  &  &  &  &  \\ 
        \bottomrule
    \end{tabular}
    }

    \caption{\textbf{Significant Covariates from Regression Analysis for QS1--3.}$^{\text{f}}$}

    \par 
    \raggedright \footnotesize
    * = $p < 0.05$, ** = $p < 0.01$, *** = $p < 0.001$. All statistically significant results in bold. \\

    $^{\text{a}}$ For each question, higher values represent belief that outcomes would increase. \\
    $^{\text{b}}$ 0 = woman and 1 = man. \\
    $^{\text{c}}$ 0 = white, 1 = non-white. \\
    $^{\text{d}}$ 0 = Democrat, 1 = independent or other party, and 2 = Republican. \\
    $^{\text{e}}$ 0 = 2022 and 1 = 2023. \\
    $^{\text{f}}$ The table includes only statistically significant covariates for the survey-weighted linear regression model as specified in \nameref{sec:Methods}, across both survey waves; results are hidden where adjusted p-values are above 0.05. P-values are Benjamini-Hochberg adjusted. IDK responses are re-coded as neutral (0) for QS1--2; respondents did not have the IDK option for QS3. Full results are in \nameref{supp:Full_Regression_Results}. \\
    \label{tab:Results_regression_significant_1-3}
\end{table}

We found no statistically significant differences on any questions across race, year, or the interaction between year and party. These results imply that expectations of future risks and benefits from AI are similar across white and non-white populations. Furthermore, these expectations do not appear to have shifted following the launch of ChatGPT and the accompanying increase in public awareness of AI.

Turning to attitudes towards AI governance, Table \ref{tab:Results_regression_significant_4} shows that political party affiliation strongly correlates with AI policy preferences. Republicans express significantly less support (and Democrats express significantly more support) for a number of policies addressing specific AI use cases and issues, including retraining for AI-driven unemployment ($\beta = -0.281$, $ p = 0.002$), wage subsidies for wage declines ($\beta = 0.642$, $ p < 0.001$), regulation that ensures deployed AI systems are safe, robust, and fair ($\beta = -0.307 $, $ p < 0.001$), parole and sentencing AI regulations ($\beta = -0.397 $, $ p < 0.001 $), bias audits for hiring ($\beta = -0.386$, $ p < 0.001 $), immigration reform for AI developers ($\beta = -0.317$, $ p = 0.003$), and taxes on company use of robots ($\beta = -0.298$, $ p = 0.011$). Further, Republican local policymakers express less support for government AI regulation generally ($\beta = -0.363$, $ p < 0.001 $). 

\begin{table}
    \centering
    \small
    \resizebox{\textwidth}{!}{
    \begin{tabular}{lccccccc}
        \toprule
        & \multicolumn{7}{c}{\textbf{Covariate}} \\
        \textbf{Question}$^{\text{a}}$ & $Age$ & $Education$ & $Gender$$^{\text{b}}$ & $Race$$^{\text{c}}$ & $Party$$^{\text{d}}$ & $Year$$^{\text{e}}$ & $Party * Year$ \\

        \midrule
        \multicolumn{7}{l}{\textbf{QS4.1: General Support for AI Regulation}} \\
        \hspace{1em} Support AI Regulation &  &  &  &  & \textbf{-0.363\textsuperscript{***}} & \textbf{0.374\textsuperscript{**}} &  \\ 
         &  &  &  &  & \textbf{(-0.473, -0.252)} & \textbf{(0.178, 0.569)} &  \\ 
        
        \midrule
        \multicolumn{7}{l}{\textbf{QS4: Policy Support}} \\
        \hspace{1em} Stronger Anti-Trust &  &  &  &  & \textbf{-0.359\textsuperscript{***}} &  &  \\ 
         &  &  &  &  & \textbf{(-0.490, -0.228)} &  &  \\ 
        \hspace{1em} Robot Tax &  &  & \textbf{-0.552\textsuperscript{*}} &  & \textbf{-0.298\textsuperscript{*}} &  &  \\ 
         &  &  & \textbf{(-0.875, -0.229)} &  & \textbf{(-0.472, -0.123)} &  &  \\ 
        \hspace{1em} Higher Corporate &  &  &  &  & \textbf{-0.659\textsuperscript{***}} &  &  \\ 
         \hspace{2em} Income Taxes &  &  &  &  & \textbf{(-0.821, -0.497)} &  &  \\ 
        \hspace{1em} Stronger Social Safety Net &  &  &  &  & \textbf{-0.555\textsuperscript{***}} &  &  \\ 
         &  &  &  &  & \textbf{(-0.699, -0.412)} &  &  \\ 
        \hspace{1em} Universal Basic Income &  &  &  &  & \textbf{-0.859\textsuperscript{***}} &  &  \\ 
         &  &  &  &  & \textbf{(-1.015, -0.703)} &  &  \\ 
        \hspace{1em} Immigration Reform for &  &  &  &  & \textbf{-0.317\textsuperscript{**}} &  &  \\ 
         \hspace{2em} for AI Developers &  &  &  &  & \textbf{(-0.482, -0.151)} &  &  \\ 
        \hspace{1em} Wage Subsidies for &  &  &  &  & \textbf{-0.642\textsuperscript{***}} &  &  \\ 
         \hspace{2em} for Wage Declines &  &  &  &  & \textbf{(-0.790, -0.494)} &  &  \\ 
        \hspace{1em} Re-Training for Unemployed &  &  &  &  & \textbf{-0.281\textsuperscript{**}} &  &  \\ 
         &  &  &  &  & \textbf{(-0.422, -0.139)} &  &  \\ 
        \hspace{1em} Stricter Data Privacy &  & \textbf{0.116\textsuperscript{**}} &  &  &  &  &  \\ 
         \hspace{2em} Regulations &  & \textbf{(0.049, 0.184)} &  &  &  &  &  \\ 
        \hspace{1em} AI Deployment Regulations &  &  &  &  & \textbf{-0.307\textsuperscript{***}} &  &  \\ 
         &  &  &  &  & \textbf{(-0.446, -0.168)} &  &  \\ 
        \hspace{1em} Federal Regulations on &  &  &  &  & \textbf{-0.320\textsuperscript{**}} &  &  \\ 
         \hspace{2em} Local Government AI &  &  &  &  & \textbf{(-0.491, -0.149)} &  &  \\ 
        \hspace{1em} Semiconductor \& AI &  &  &  &  & \textbf{-0.260\textsuperscript{*}} &  &  \\ 
         \hspace{2em} Hardware Subsidies &  &  &  &  & \textbf{(-0.419, -0.101)} &  &  \\ 
        \hspace{1em} Bias Audits for Hiring \& &  &  &  &  & \textbf{-0.386\textsuperscript{***}} &  &  \\ 
         \hspace{2em} Promotion AI &  &  &  &  & \textbf{(-0.561, -0.211)} &  &  \\ 
        \hspace{1em} Parole \& Sentencing &  &  &  &  & \textbf{-0.397\textsuperscript{***}} &  &  \\ 
         \hspace{2em} AI Regulations &  &  &  &  & \textbf{(-0.576, -0.219)} &  &  \\ 
         
        \bottomrule
    \end{tabular}
    }
    \caption{\textbf{Significant Covariates from Regression Analysis for Q4.1--QS4.}$^{\text{f}}$}

    \par 
    \raggedright \footnotesize
    * = $p < 0.05$, ** = $p < 0.01$, *** = $p < 0.001$. All statistically significant results in bold. \\
    
    $^{\text{a}}$ For each question, higher values represent belief that outcomes would increase. \\
    $^{\text{b}}$ 0 = woman and 1 = man. \\
    $^{\text{c}}$ 0 = white, 1 = non-white. \\
    $^{\text{d}}$ 0 = Democrat, 1 = independent or other party, and 2 = Republican. \\
    $^{\text{e}}$ 0 = 2022 and 1 = 2023. \\
    $^{\text{f}}$ The table includes only statistically significant covariates for the survey-weighted linear regression model as specified in \nameref{sec:Methods}, across both survey waves; results are hidden where adjusted p-values are above 0.05. P-values are Benjamini-Hochberg adjusted. Respondents did not have the IDK option for Q4.1--QS4. Full results are in \nameref{supp:Full_Regression_Results}. \\
    \label{tab:Results_regression_significant_4}
\end{table}

In addition to differences by political party, there are a few differences across demographic subgroups. Higher education levels correlate with stronger support for privacy regulations ($\beta = 0.117$, $ p = 0.009$), and men are less supportive of a robot tax compared to women ($\beta = -0.552$, $ p = 0.011$). 

Table \ref{tab:Results_regression_significant_5} shows regression results using the constructed indices. These confirm some general patterns we observed in Tables \ref{tab:Results_regression_significant_1-3} and \ref{tab:Results_regression_significant_4}. Republicans score significantly lower on the index aggregating personal well-being and community health questions ($\beta = -0.174$, $ p < 0.001$), and Republicans' average response across all questions is also significantly lower than Democrats ($\beta$ = -0.400, $p < 0.001$). Higher education levels correlate with a higher score on the progress and innovation index ($\beta = 0.091$, $ p < 0.001$). Male policymakers are more optimistic about AI's impacts on personal well-being and community health ($\beta = 0.266$, $ p = 0.004$).

\begin{table}[phtb]
    \centering
    
    \resizebox{\textwidth}{!}{
    \begin{tabular}{lcccccccc}
        \toprule
        & \multicolumn{8}{c}{\textbf{Covariate}} \\
        \textbf{Question}$^{\text{a}}$ & $Age$ & $Education$ & $Gender$$^{\text{b}}$ & $Race$$^{\text{c}}$ & $Party$$^{\text{d}}$ & $Year$$^{\text{e}}$ & $Party * Year$ & $Policy_{RegAI}$ \\
        
        \midrule
        \multicolumn{9}{l}{\textbf{Constructed Indices}} \\
        \hspace{1em} Policy Agreement &  &  &  &  & \textbf{-0.400\textsuperscript{***}} &  &  & \\ 
        &  &  &  &  & \textbf{(-0.469, -0.330)} &  &  & \\ 
        \hspace{1em} Positive Impacts (All) &  &  & \textbf{0.208\textsuperscript{**}} &  &  &  &  \\ 
         &  &  & \textbf{(0.100, 0.317)} &  &  &  &  \\ 
        \hspace{1em} Personal Well-Being \& &  &  & \textbf{0.266\textsuperscript{**}} &  & \textbf{-0.174\textsuperscript{***}} &  &  \\ 
         \hspace{2em} Community Health Impacts &  &  & \textbf{(0.122, 0.411)} &  & \textbf{(-0.256, -0.092)} &  &  \\ 
        \hspace{1em} Progress \& &  & \textbf{0.091\textsuperscript{***}} &  &  &  &  &  \\ 
         \hspace{2em} Innovation Impacts &  & \textbf{(0.052, 0.130)} &  &  &  &  &  \\ 
        \hspace{1em} Positive Impacts &  &  & \textbf{0.204\textsuperscript{*}} &  &  &  &  \\ 
         \hspace{2em} (Local Effects) &  &  & \textbf{(0.083, 0.324)} &  &  &  &  \\ 
        \hspace{1em} Positive Impacts &  &  & \textbf{0.195\textsuperscript{*}} &  &  &  &  \\ 
         \hspace{2em} (Broad Effects) &  &  & \textbf{(0.059, 0.332)} &  &  &  &  \\ 
        \bottomrule
    \end{tabular}
    }
    \caption{\textbf{Significant Covariates from Regression Analysis for Constructed Indices.}.$^{\text{f}}$}

    \par 
    \raggedright \footnotesize
    * = $p < 0.05$, ** = $p < 0.01$, *** = $p < 0.001$. All statistically significant results in bold. \\

    $^{\text{a}}$ For each question, higher values represent belief that outcomes would increase. \\
    $^{\text{b}}$ 0 = woman and 1 = man. \\
    $^{\text{c}}$ 0 = white, 1 = non-white. \\
    $^{\text{d}}$ 0 = Democrat, 1 = independent or other party, and 2 = Republican. \\
    $^{\text{e}}$ 0 = 2022 and 1 = 2023. \\
    $^{\text{f}}$ The table includes only statistically significant covariates for the survey-weighted linear regression model as specified in \nameref{sec:Methods}, across both survey waves; results are hidden where adjusted p-values are above 0.05. P-values are Benjamini-Hochberg adjusted. IDK responses are re-coded as neutral (0). Full results are in \nameref{supp:Full_Regression_Results}. Index definitions are in \nameref{supp:Indices_Defs}. \\
    \label{tab:Results_regression_significant_5}
\end{table}

In addition, the indices relating to positive AI future impacts reveal a new pattern. These two indices were constructed by aggregating all survey questions relating to positive AI impacts, i.e., potential societal benefits rather than societal risks. These indices include societal benefits such as increases in jobs, quality of life, democracy, and innovation but exclude risks such as increasing political polarization, surveillance, and international conflicts. Examining the positive impacts indices, we see that male policymakers are more likely to expect greater benefits from AI on the local community level ($\beta = 0.204$, $ p = 0.011$), more broadly in the country ($\beta = 0.195$, $ p = 0.048$), and overall across all societal benefits ($\beta = 0.208$, $ p = 0.003$).

\subsection*{Differences from 2022 to 2023}
\label{subsubsec:Results_Regression_Year}

Generally, both expectations of AI impacts and attitudes towards specific AI policies remained stable over 2022--2023. We found no statistically significant differences across the two years or the interaction between year and party in any of the questions relating to AI impacts or to specific AI policies such as stronger anti-trust laws, a robot tax, or UBI. 

A notable exception, however, is support for government regulation, which increased significantly from 2022 to 2023 ($\beta = 0.374$, $p = 0.003$). The magnitude of the change in support over 2022-2023, alongside the difference in support by party affiliation, are the two largest group effects we observe in the study  (see Table \ref{tab:Results_regression_significant_4}). 

\begin{figure}[htbp] 
    \centering
    \includegraphics[width=\textwidth,height=\textheight,keepaspectratio]{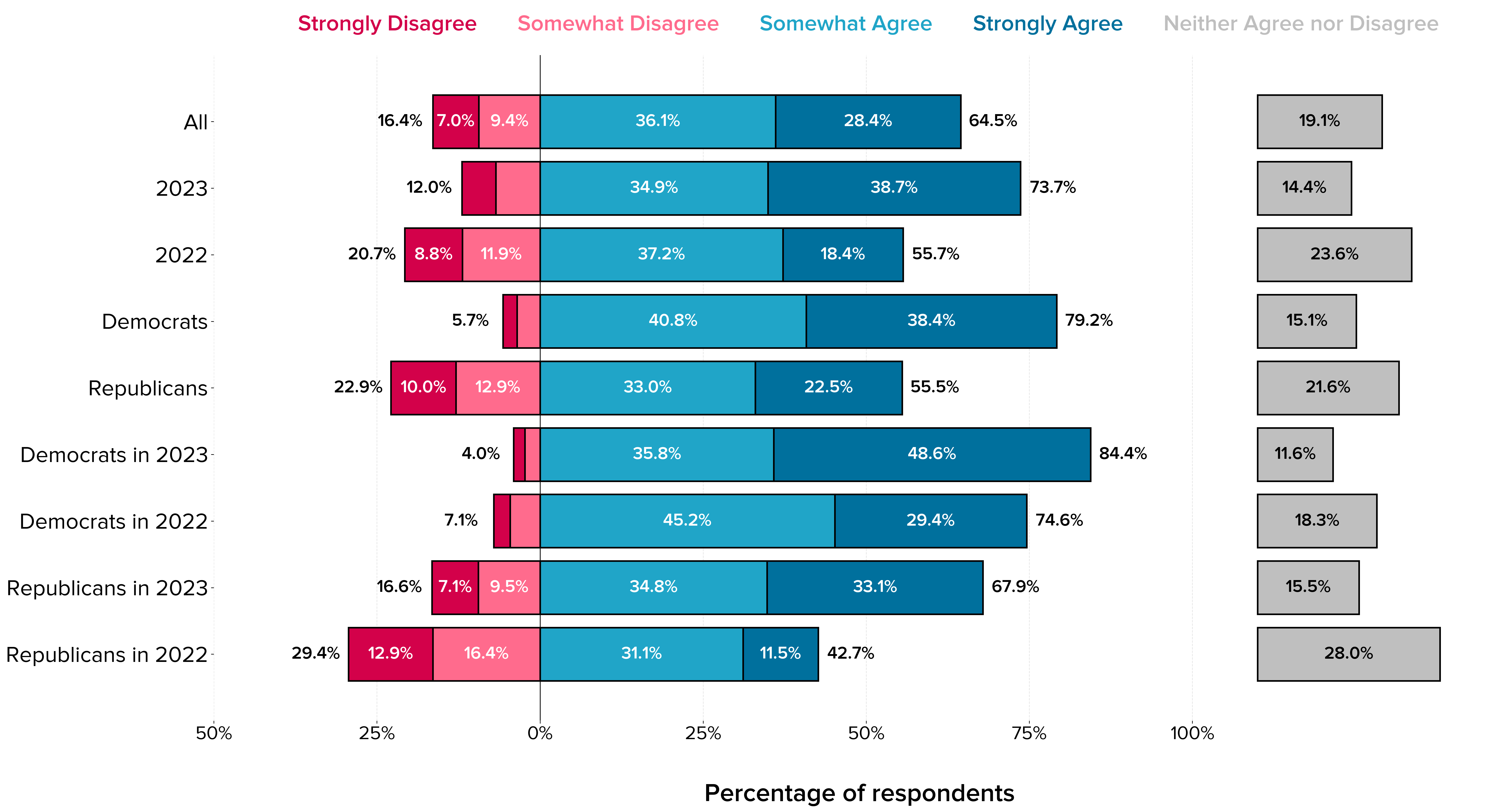}
    \caption{\textbf{Local US officials' agreement with the statement ``AI should be regulated by the government,'' with overall support and segments by party and year.}$^{\text{a}}$}

    \par\raggedright \footnotesize
    $^{\text{a}}$ The figure shows unweighted relative frequencies across both survey waves.
    
    \label{fig:Results_Q4.1}
\end{figure}

Although the interaction between year and party is not statistically significant, it is worth noting that the increase in support was much greater among Republicans compared to Democrats. As Fig \ref{fig:Results_Q4.1} shows, support increased across the board from 56\% agree or strongly agree in 2022 to 74\% agree or strongly agree in 2023. Republicans, however, shifted from minority agreement in 2022 (43\%) to majority agreement in 2023 (68\%). The shift among Democrats, who already exhibited high levels of support in 2022, was more modest, rising from 75\% agreement in 2022 to 84\% in 2023.

\section*{Discussion} \label{sec:Discussion}

The results of our study offer valuable insights into how local US policymakers view the societal impacts and governance of AI. 
Local officials in our study conveyed both optimism about AI's potential and significant concern about its risks, consistent with prior surveys of the US public and AI experts. Local officials are optimistic about AI's impacts on innovation, infrastructure, and transportation. But local officials also share concerns about specific risks also highlighted in previous research, including increasing inequality, job loss, and data security threats \cite{public_first_2023a_what_nodate,public_first_2023b_what_nodate,policy_elections_and_representation_lab_global_2024,department_for_science_innovation_and_technology_and_centre_for_data_ethics_and_innovation_2023b_public_2023,Zhang:2019aa,Gruetzemacher:2024aa}). 

A noteworthy feature of this study was its timing, spanning a period of heightened attention to AI governance, including the release of ChatGPT and subsequent federal and state-level regulatory activity in 2023. Bipartisan support for AI regulation increased markedly between the 2022 and 2023 surveys, mirroring broader national developments, such as President Biden’s Executive Order on AI governance and the surge in AI-related legislation at both the state and federal levels during 2023. The convergence in support for government regulation, with both Republicans and Democrats in favor in 2023, suggests a growing recognition across party lines of the need for oversight in the face of rapid advancements in AI. However, we also see marked partisan differences in support for specific AI policies, with Democrat support generally being significantly higher.

Despite concerns over risks and growing support for regulation, we also found high levels of uncertainty, reported lack of preparedness among policymakers, and limited expectations of near-term involvement in policymaking. These results are particularly striking given the surge in AI-related policy initiatives during this period. They suggest that increased attention to AI at the national and state levels has not yet translated into greater confidence or preparedness among local officials. This is also notable in light of the decentralized nature of US policymaking, which often places local officials at the forefront of implementing state and federal policies.

\subsection*{Limitations} \label{subsec:Discussion_Limitations}

While the survey results provide important insight into local policymakers' attitudes and expectations, which help us to understand the trajectory of AI policy-making, our study also has some limitations.

First, we asked policymakers to consider potential future impacts of AI during the 2025--2050 time period. Yet, most policymakers are not trained in forecasting; they may lack expertise and experience to make accurate or confident predictions about the trajectory of AI. It is challenging even for experts to make predictions about the costs and impacts of emerging technologies \cite{Savage:2021aa}. This means that we should expect substantial variation in reported expectations of impact, as well as notable rates of non-response. We believe some of this variation and non-response reflects the uncertain trajectory of AI development and is consistent with the finding that most policymakers' feel inadequately informed about AI. Nevertheless, some of the uncertainty could be due to the difficulty of forecasting in general, as well as the difficulty of forecasting in the context of emerging technologies in particular. 

Second, any survey has to content with selection effects and other considerations related to the survey sample. For example, it may be the case that local US officials who chose to participate differ systematically from those who did not. Details on the representativeness of the sample in terms of area-level characteristics can be found in the Supporting Information (\nameref{supp:samplerepresentativeness}), and these generally suggest that, at least in regard to most of these characteristics, the sample is somewhat representative. To improve sample representativeness in our regression analysis, we also used weights computed based on the gender of the respondent and the area-level voting and population characteristics. In turn, it is important to note that the results also do not necessarily generalize to policymakers in other countries or to state and federal-level policymakers. 

Third, the study design and choice of questions introduce further limitations. The questions we asked and specific items chosen reflect our subjective prioritization of topics and may not fully capture the range of concerns that policymakers hold. The phrasing of items likely also has an effect on the expressed beliefs and opinions of individuals in regard to the impacts they expect from AI and the specific policies they support. There may also be effects of positive and negative frames of impacts on expressed beliefs. Additionally, it is important to note that the design was not a longitudinal panel but completed in separate survey waves with different samples. Therefore, we cannot assess individual-level changes, and there may be cohort effects across the waves.

Fourth, an important consideration is the high frequency of ``I don't know'' responses to many questions. Likert-scale responses are treated as continuous variables in the linear regressions, which approach does not adequately address the unique nature of IDK responses. In our main analysis, we re-coded IDK responses as neutral (0) to ensure these responses were included in the models. However, this method has its limitations, as it assumes that uncertainty or lack of knowledge aligns with neutrality, which may not accurately reflect respondents' perspectives. To address this, we also conducted supplementary analyses using imputed IDK responses, which are presented in the \nameref{supp:Alternative_Regression_Results}. This analysis found that all coefficients that were statistically significant at $p < 0.05$ when IDK are re-coded as neutral are also significant when IDK responses are imputed. In addition, the model fit on imputed data sees 21 more statistically significant coefficients.

Fifth, our sample contained high levels of missing data, in part due to our planned missingness design. Rates of missingness were as high as 69\% for some questions in QS4; MICE procedures have been shown in simulations to be robust even with much higher rates of missingness \cite{lee_evaluation_2021, madley-dowd_proportion_2019}, but high rates of missingness may still bias the imputations. To improve the robustness of our imputations, we impute a large number of datasets, and we also conduct robustness checks (see \nameref{supp:Full_Regression_Results}) on the imputation procedure. 

Finally, our statistical analyses have inherent limitations. Given the breadth of regressions run, we adjusted significance thresholds using the Benjamini-Hochberg procedure to control for false discovery rates. While this reduces the risk of false positives, it also lowers the power of the regressions, increasing the likelihood of false negatives. Moreover, the relatively small sample size, when broken into subgroups or conditions, limits the precision of some estimates.  

\subsection*{Future directions} \label{subsec:Discussion_Future_Directions}

The findings of our study point to several promising avenues for future research. First, there is a pressing need for further capacity-building initiatives to equip local, state, and national policymakers with the knowledge and tools necessary to navigate AI’s complexities. Research could play a critical role in identifying the specific informational needs of policymakers, including areas where they feel least prepared, such as the technical aspects of AI, its economic implications, and its potential impacts on international relations.

Longitudinal studies could be particularly valuable for tracking how policymakers’ attitudes and competencies evolve over time, especially in response to rapid advancements in AI technology and policy interventions. Such research would provide a clearer picture of how exposure to AI technologies and governance frameworks shapes perceptions and decision-making capabilities. Future studies could also broaden the scope to include other key populations, such as state-level policymakers, federal agencies, and private sector leaders. 

Cross-national comparisons offer another avenue for exploration. Understanding how US local policymakers’ attitudes align or diverge from those in other countries—particularly in regions with distinct regulatory frameworks such as the EU or Asia—could provide valuable insights into how differing political, economic, and cultural contexts shape AI governance approaches and inform strategies for international coordination.

Finally, future research should delve deeper into the drivers of uncertainty among policymakers. Investigating the sources of their knowledge gaps, as well as the factors that contribute to shifts in confidence and preparedness, would help design targeted interventions to address these challenges. This could include studies on how exposure to training programs, expert advisory panels, and public debates influences policymaker readiness to engage with AI-related issues. 

\section*{Conclusion} \label{sec:Conclusion}

This study highlights several important policy implications for AI governance in the US, particularly at the local level. First, there may be growing bipartisan support for AI regulation among local US officials, rising from 55\% in 2022 to 74\% in 2023. While Democrats consistently show greater support for regulatory policies addressing AI challenges such as unemployment and AI safety and fairness, Republican support for government oversight has increased significantly, narrowing partisan divides.

Of course, in the future, AI discourse may also become more politicized and polarized as the political landscape evolves. Indeed, the persistence of polarization in specific policy preferences already raises important concerns. While Democrats and Republicans share similar views on AI risks, their divergent support for targeted regulatory measures suggests that public opinion on AI-related issues may increasingly align with partisan cues. This dynamic has the potential to exacerbate state-level policy fragmentation. Such disparities could create uncertainty and inefficiencies for businesses, workers, and consumers operating across jurisdictions.

Finally, substantial levels of uncertainty and reported lack of preparedness among local officials highlight a potential need for capacity-building initiatives. Policymakers require targeted training and resources to navigate the complexities of AI governance, particularly in addressing its broader economic, social, and international implications. Bridging these knowledge gaps could empower local governments to play a more active role in shaping AI policy and ensure that governance efforts reflect both local priorities and broader national and global objectives.

\newpage
\section*{Supporting information} \label{sec:Supporting_Info}

\paragraph*{S1 Additional figures}
\label{supp:Additional_Figures}

\begin{figure*}[hbtp]
    \centering
    \includegraphics[width=\textwidth,height=\textheight,keepaspectratio]{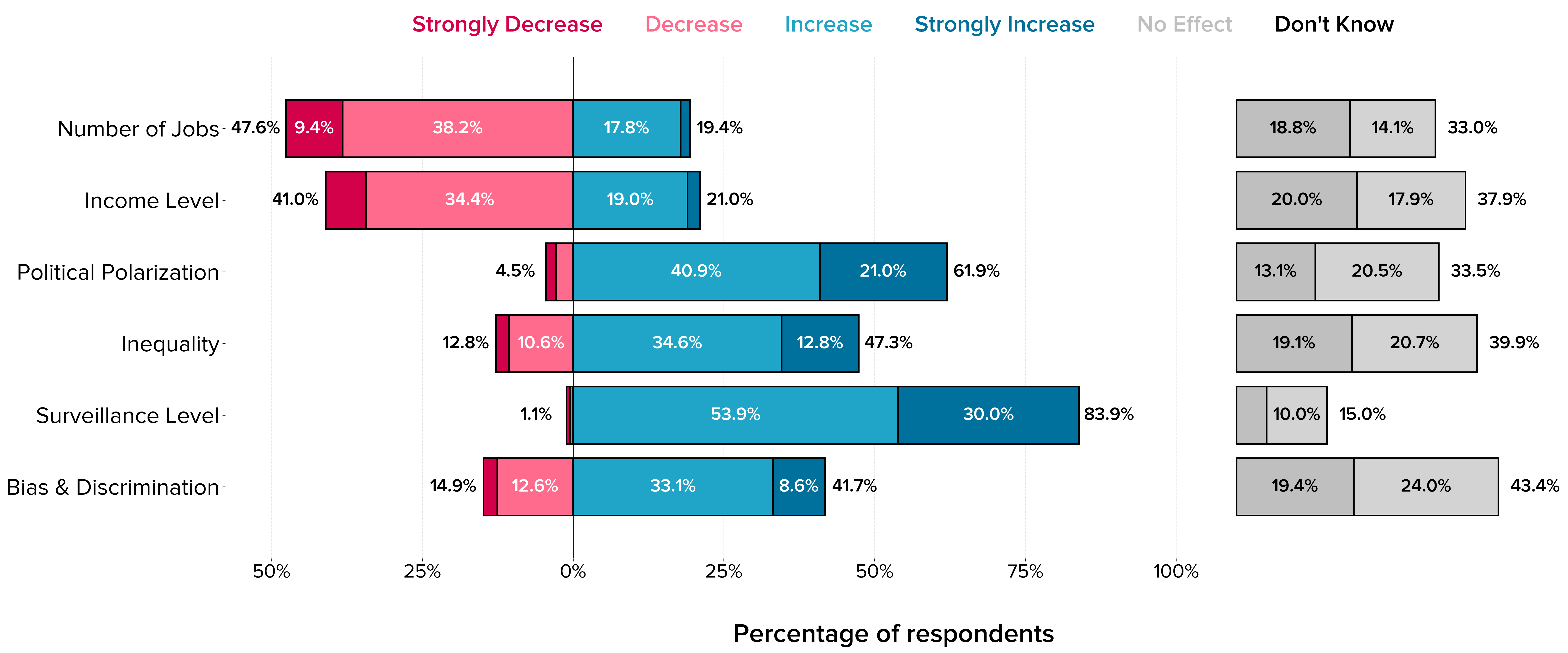}
    
    \caption{\textbf{Democratic US officials' expectations of the local impacts of AI between 2025 and 2050.} The figure shows unweighted relative frequencies for QS1 across both survey waves.}
    
    \label{fig:QS1_Dems}
\end{figure*}

\begin{figure*}
    \centering
    \includegraphics[width=\textwidth,height=\textheight,keepaspectratio]{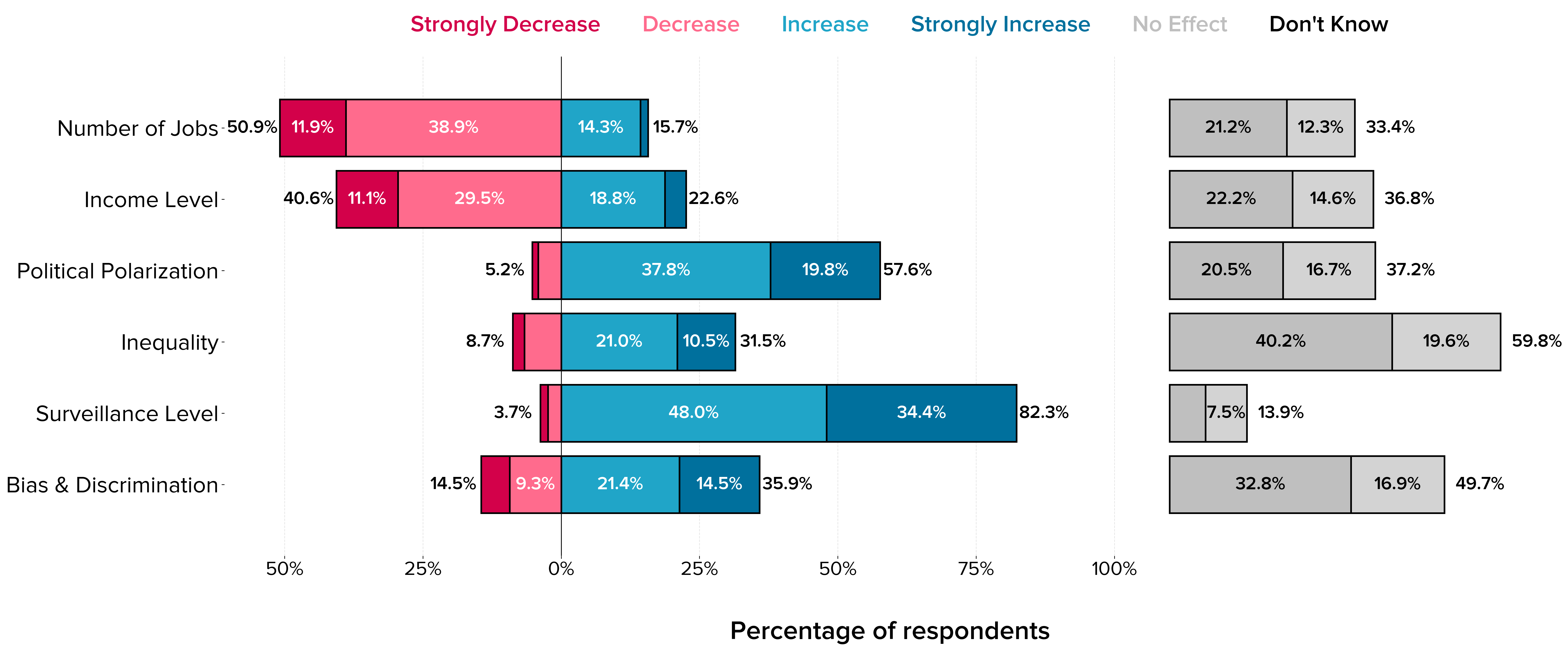}
    
    \caption{\textbf{Republican US officials' expectations of the local impacts of AI between 2025 and 2050.} The figure shows unweighted relative frequencies for QS1 across both survey waves.}
    
    \label{fig:QS1_Reps}
\end{figure*}

\begin{figure*}
    \centering
    \includegraphics[width=\textwidth,height=\textheight,keepaspectratio]{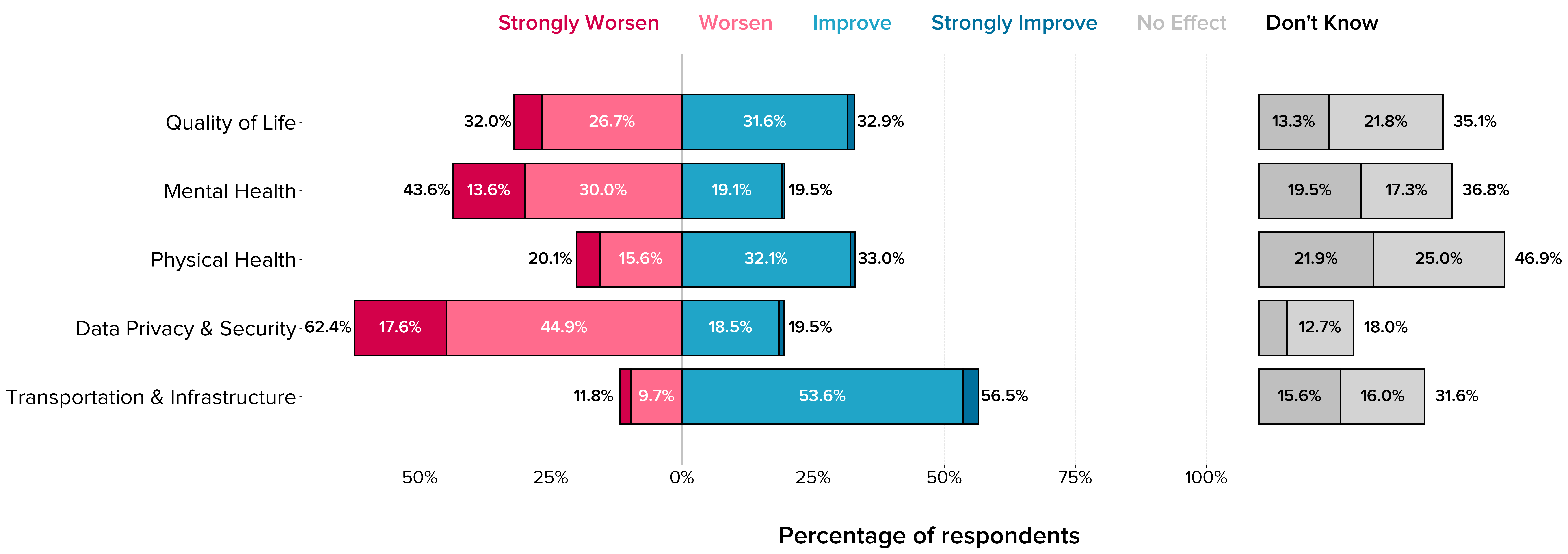}

    \caption{\textbf{Democratic US officials' expectations of the local impacts of AI between 2025 and 2050.} The figure shows unweighted relative frequencies for QS2 across both survey waves.}
    
    \label{fig:QS2_Dems}
\end{figure*}

\begin{figure*}
    \centering
    \includegraphics[width=\textwidth,height=\textheight,keepaspectratio]{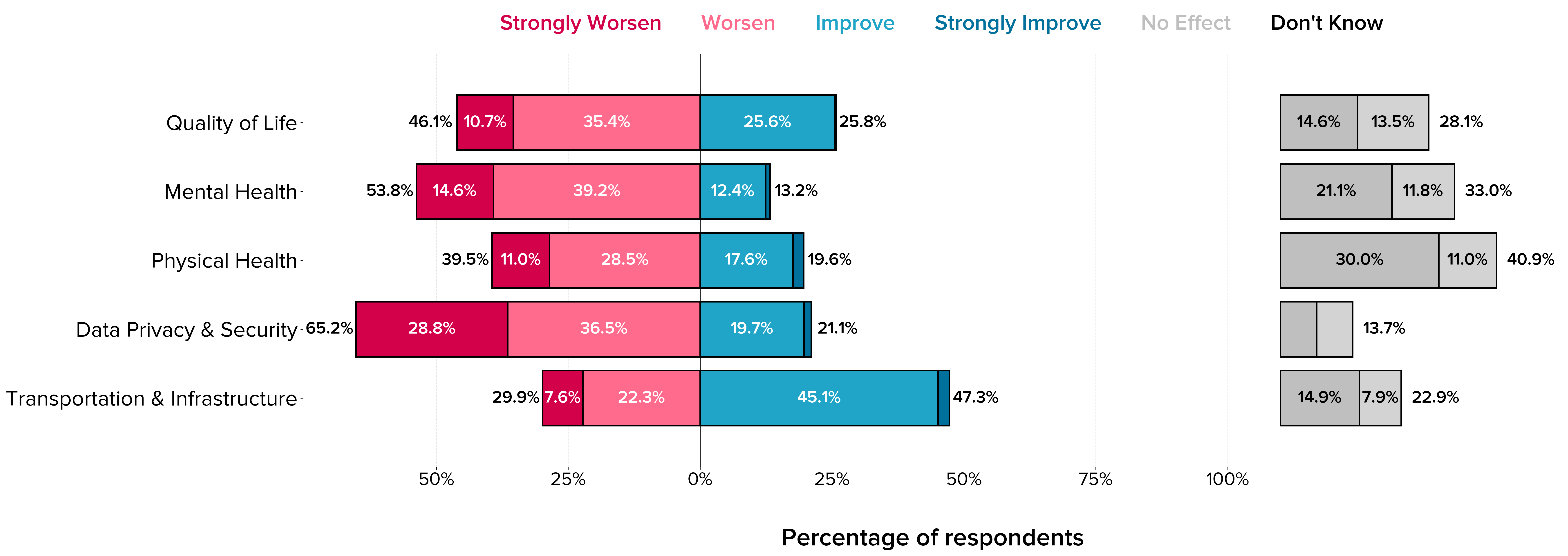}
    \caption{\textbf{Republican US officials' expectations of the local impacts of AI between 2025 and 2050.} The figure shows unweighted relative frequencies for QS2 across both survey waves.}
    
    \label{fig:QS2_Reps}
\end{figure*}

\begin{figure*}
    \centering
    \includegraphics[width=\textwidth,height=\textheight,keepaspectratio]{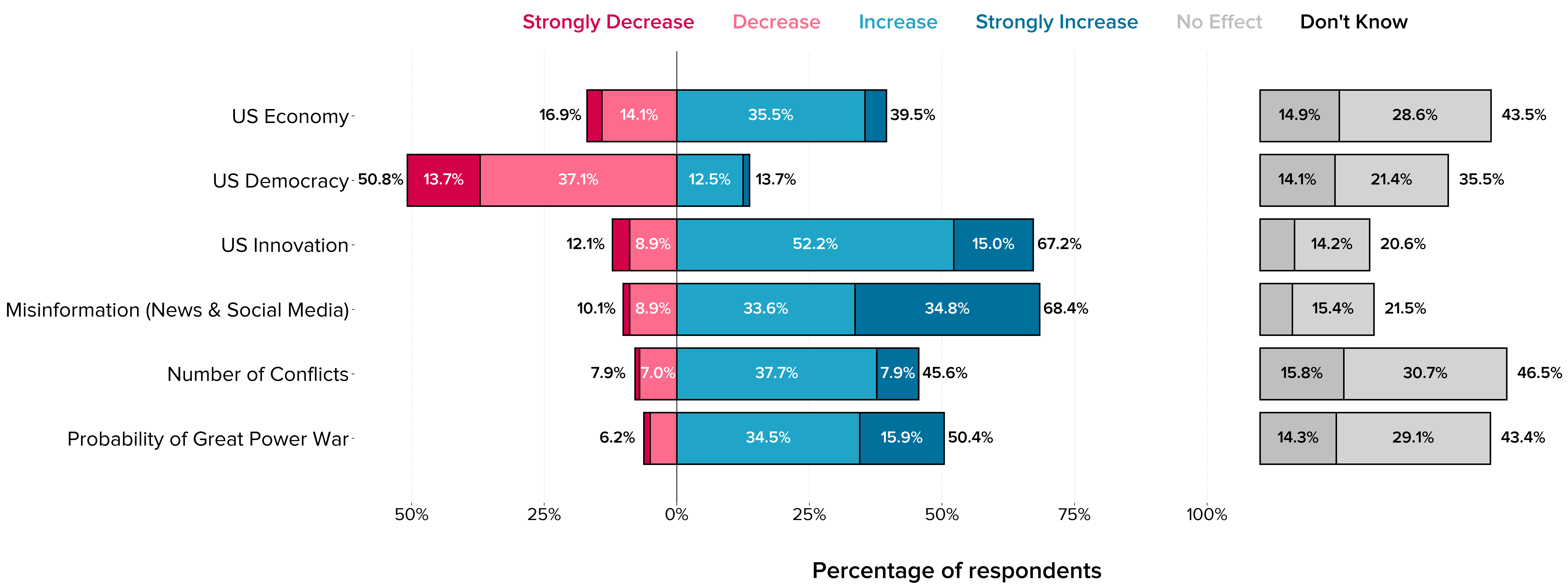}
    
    \caption{\textbf{Democratic US officials' expectations of the broad impacts of AI between 2025 and 2050.} The figure shows unweighted relative frequencies for QS3 across both survey waves.}
    
    \label{fig:QS3_Dems}
\end{figure*}

\begin{figure*}
    \centering
    \includegraphics[width=\textwidth,height=\textheight,keepaspectratio]{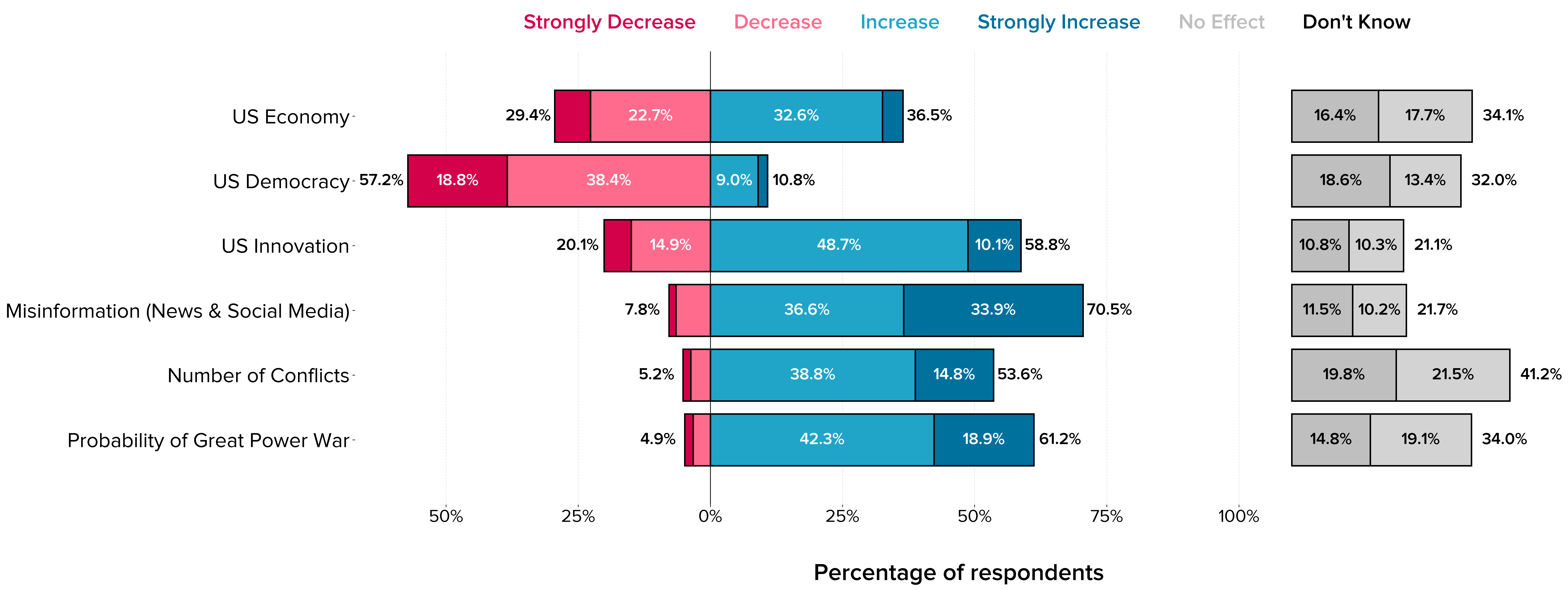}

    \caption{\textbf{Republican US officials' expectations of the broad impacts of AI between 2025 and 2050.} The figure shows unweighted relative frequencies for QS3 across both survey waves.}
    
    \label{fig:QS3_Reps}
\end{figure*}

\begin{figure*}
    \centering
    \includegraphics[width=\textwidth,height=\textheight,keepaspectratio]{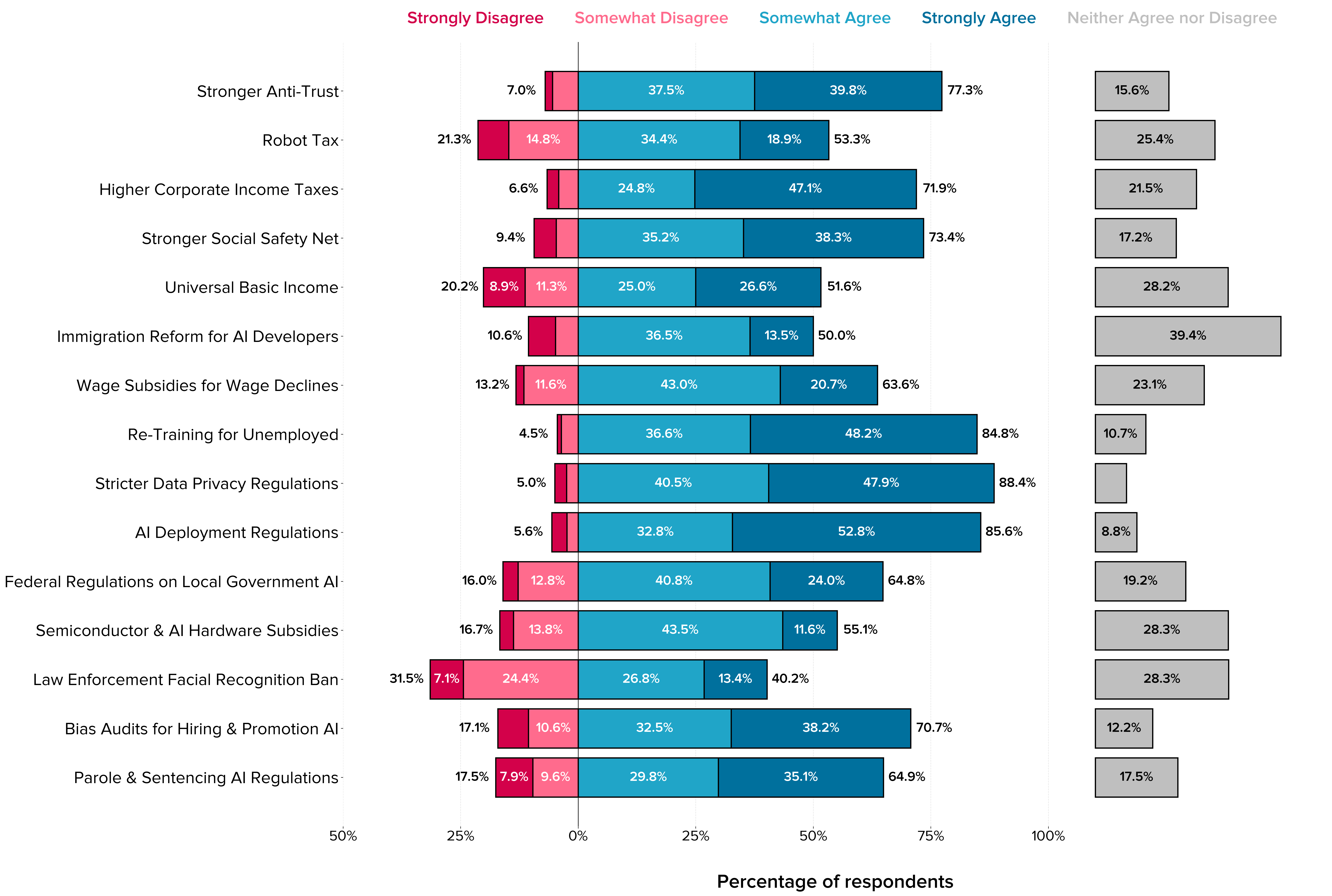}
    
    \caption{\textbf{Democratic US officials' views on what AI policies would be beneficial between 2025 and 2050.} The figure shows unweighted relative frequencies for QS4 across both survey waves.}
    
    \label{fig:QS4_Dems}
\end{figure*}

\begin{figure*}
    \centering
    \includegraphics[width=\textwidth,height=\textheight,keepaspectratio]{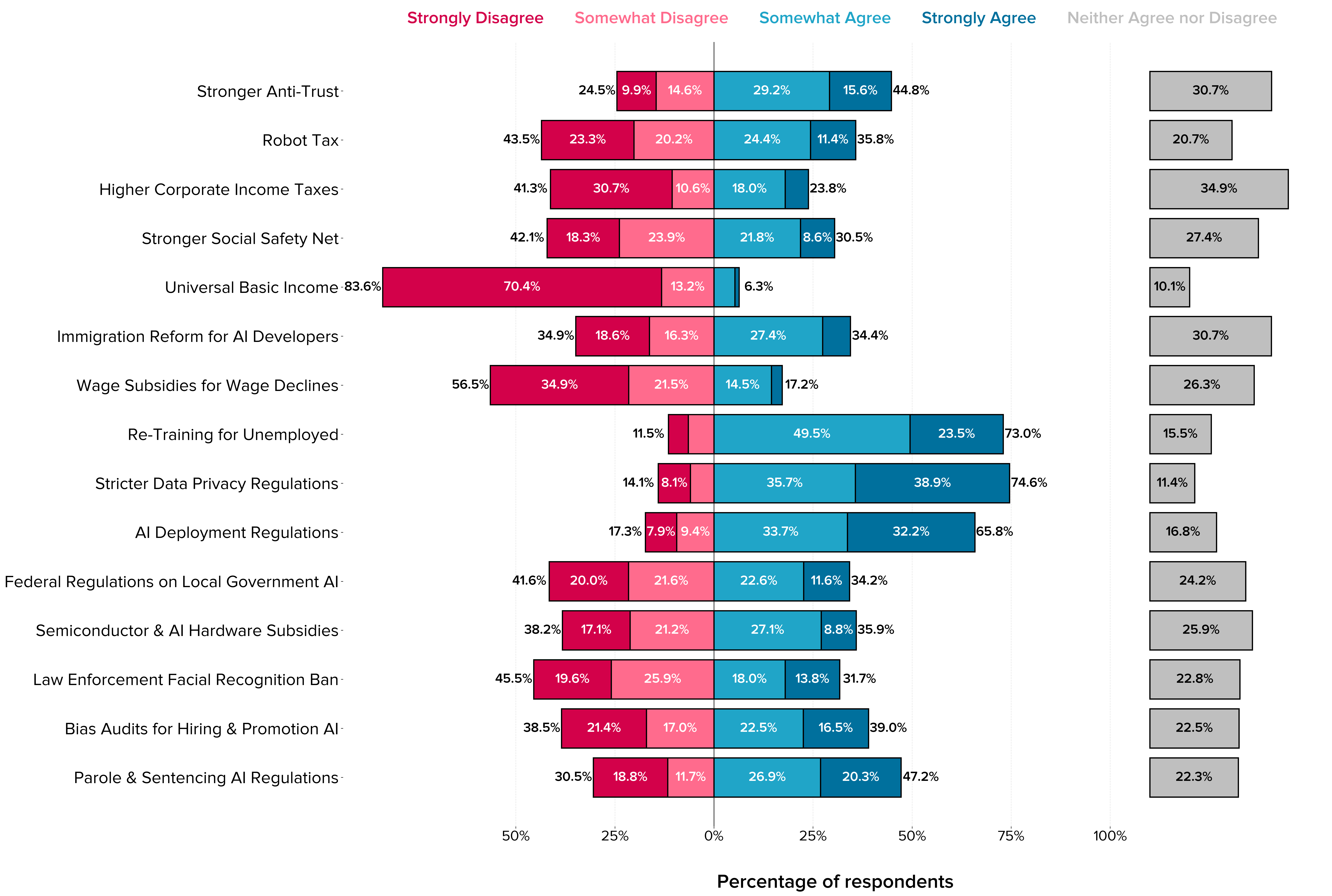}

    \caption{\textbf{Republican US officials' views on what AI policies would be beneficial between 2025 and 2050.} The figure shows unweighted relative frequencies for QS4 across both survey waves.}
        
    \label{fig:QS4_Reps}
\end{figure*}

\begin{figure*}
    \centering
    \includegraphics[width=\textwidth,height=\textheight,keepaspectratio]{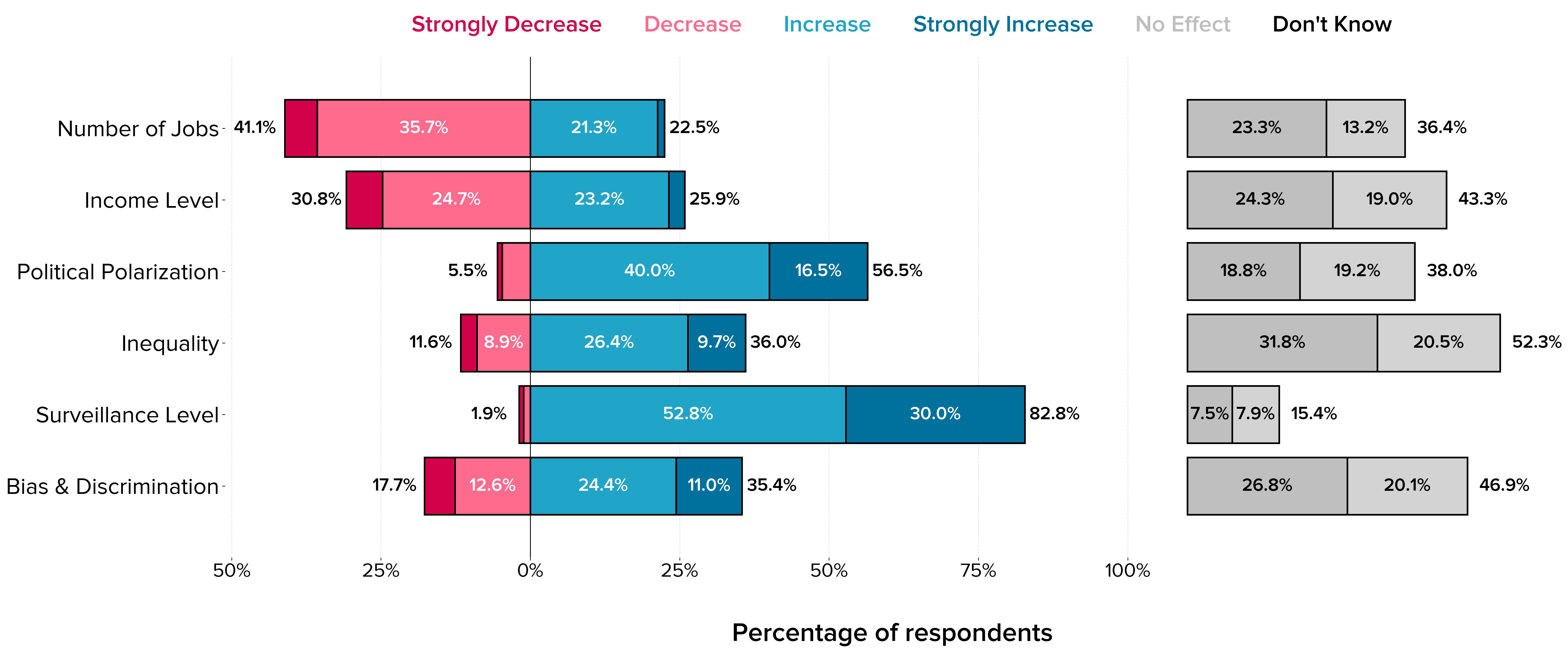}

    \caption{\textbf{Local US officials' expectations in 2022 of the local impacts of AI between 2025 and 2050.} The figure shows unweighted relative frequencies for QS1 for the 2022 wave only.}
    
    \label{fig:QS1_2022}
\end{figure*}

\begin{figure*}
    \centering
    \includegraphics[width=\textwidth,height=\textheight,keepaspectratio]{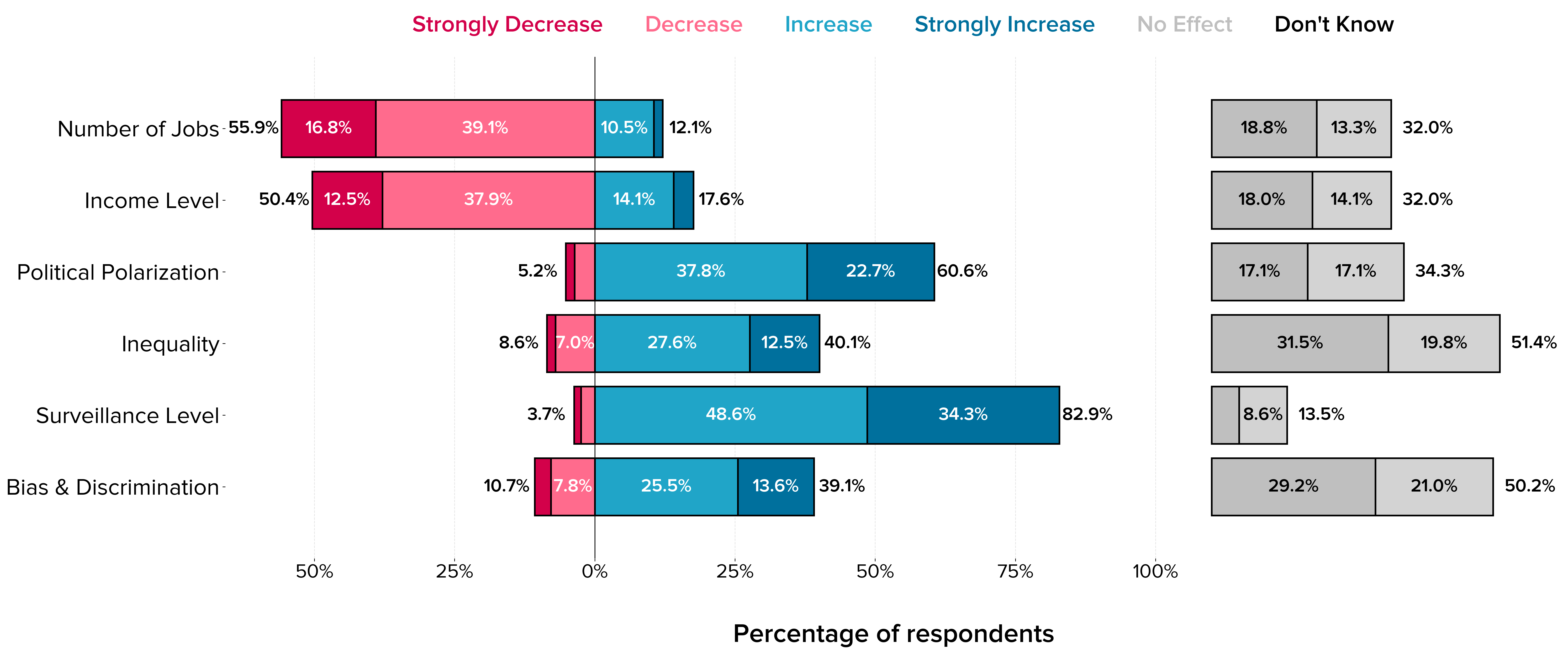}

    \caption{\textbf{Local US officials' expectations in 2023 of the local impacts of AI between 2025 and 2050.} The figure shows unweighted relative frequencies for QS1 for the 2023 wave only.}
    
    \label{fig:QS1_2023}
\end{figure*}

\begin{figure*}
    \centering
    \includegraphics[width=\textwidth,height=\textheight,keepaspectratio]{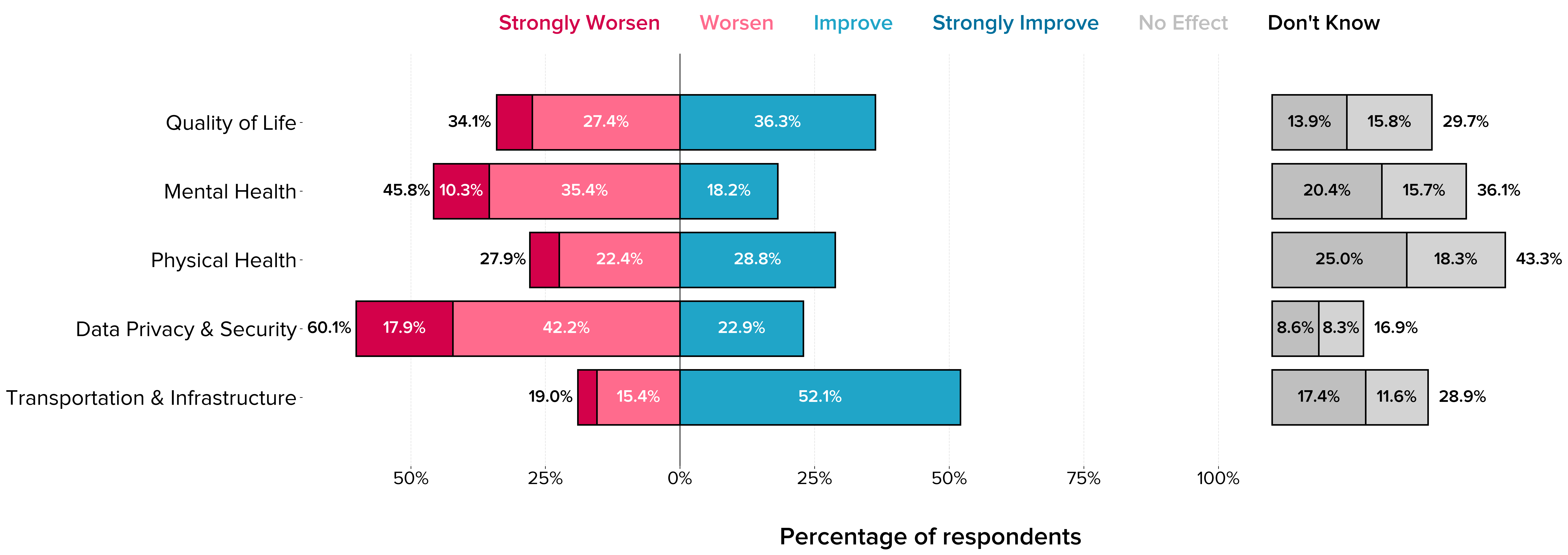}
    
    \caption{\textbf{Local US officials' expectations in 2022 of the local impacts of AI between 2025 and 2050.} The figure shows unweighted relative frequencies for QS2 for the 2022 wave only.}
    
    \label{fig:QS2_2022}
\end{figure*}

\begin{figure*}
    \centering
    \includegraphics[width=\textwidth,height=\textheight,keepaspectratio]{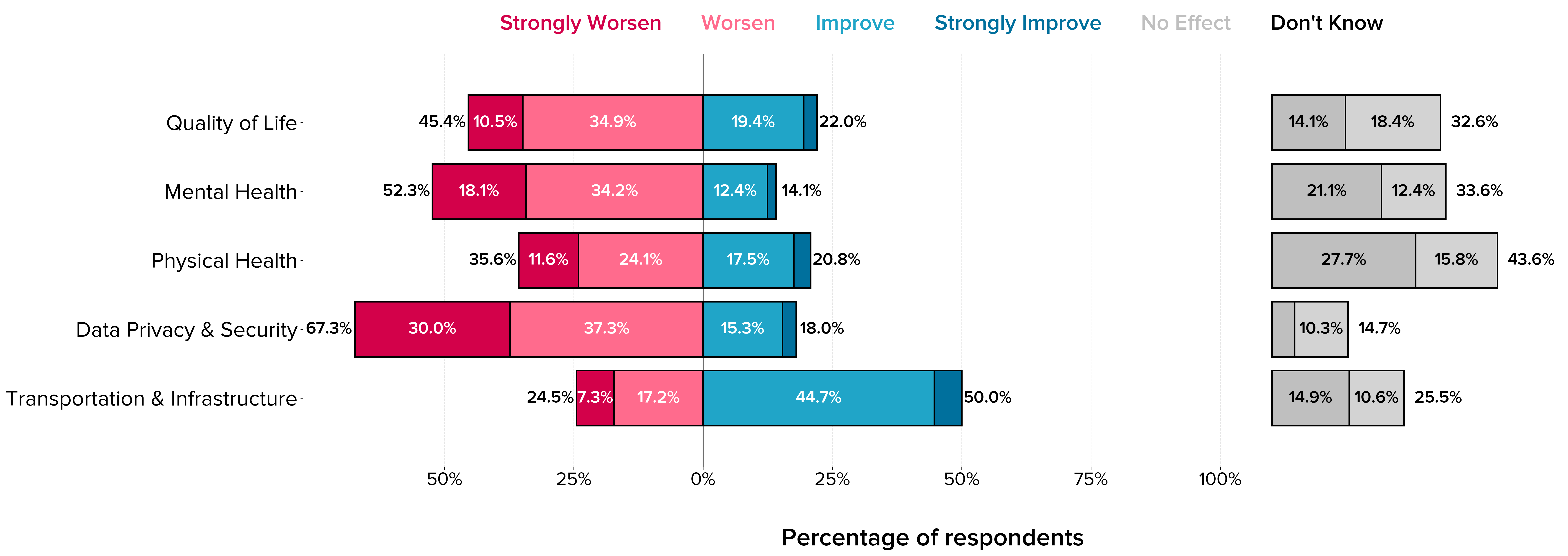}

    \caption{\textbf{Local US officials' expectations in 2023 of the local impacts of AI between 2025 and 2050.} The figure shows unweighted relative frequencies for QS2 for the 2023 wave only.}
    
    \label{fig:QS2_2023}
\end{figure*}

\begin{figure*}
    \centering
    \includegraphics[width=\textwidth,height=\textheight,keepaspectratio]{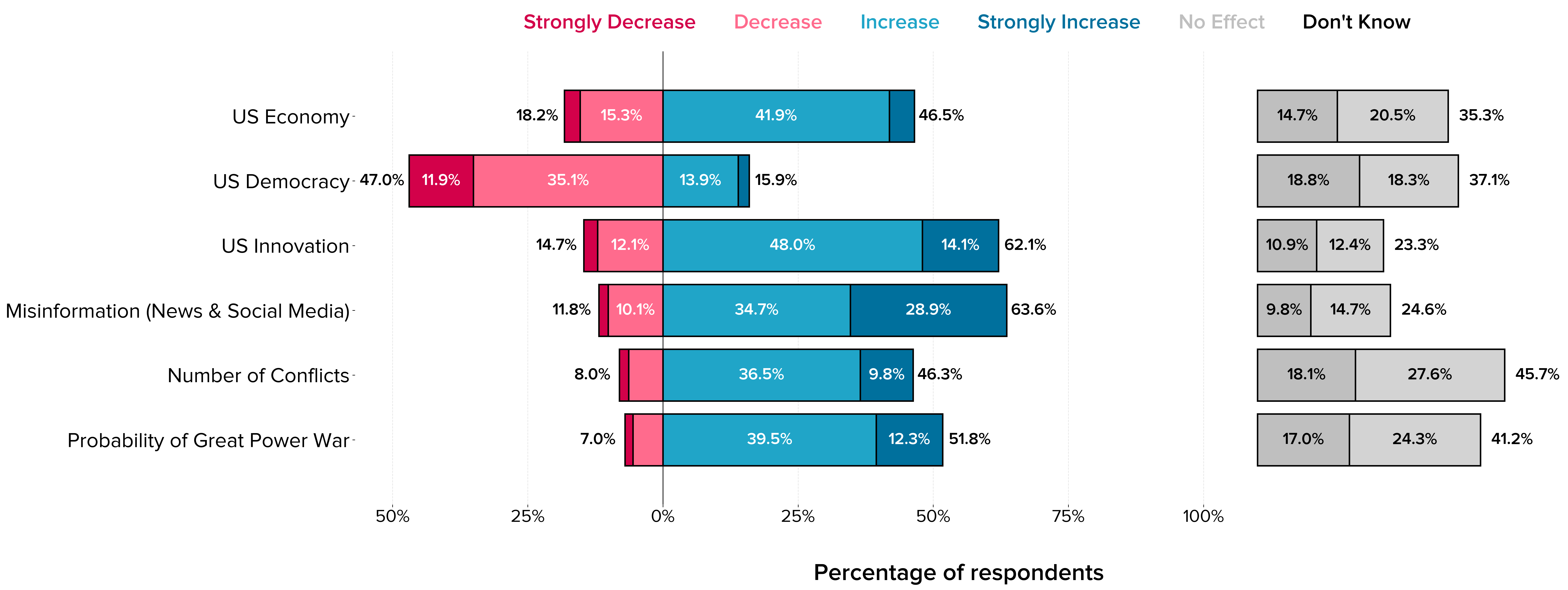}
    
    \caption{\textbf{Local US officials' expectations in 2022 of the local impacts of AI between 2025 and 2050.} The figure shows unweighted relative frequencies for QS3 for the 2022 wave only.}
    
    \label{fig:QS3_2022}
\end{figure*}

\begin{figure*}
    \centering
    \includegraphics[width=\textwidth,height=\textheight,keepaspectratio]{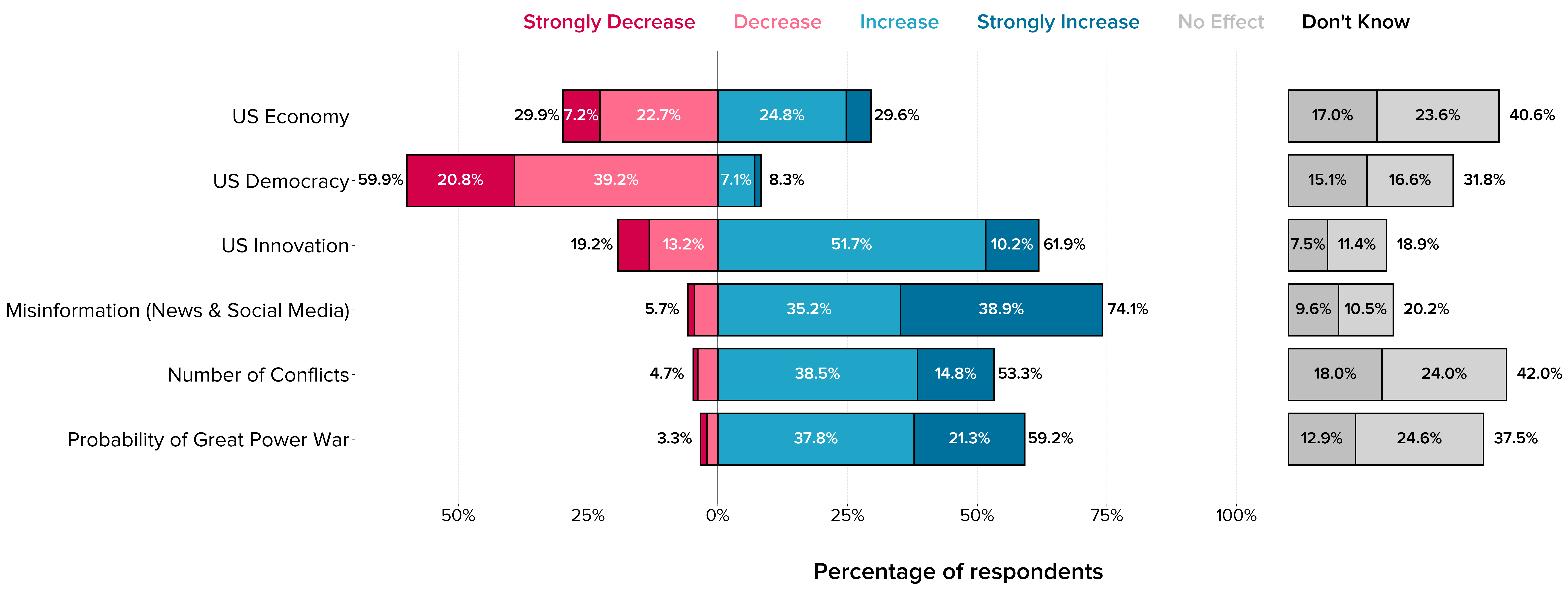}

    \caption{\textbf{Local US officials' expectations in 2023 of the local impacts of AI between 2025 and 2050.} The figure shows unweighted relative frequencies for QS3 for the 2023 wave only.}
    
    \label{fig:QS3_2024}
\end{figure*}

\begin{figure*}
    \centering
    \includegraphics[width=\textwidth,height=\textheight,keepaspectratio]{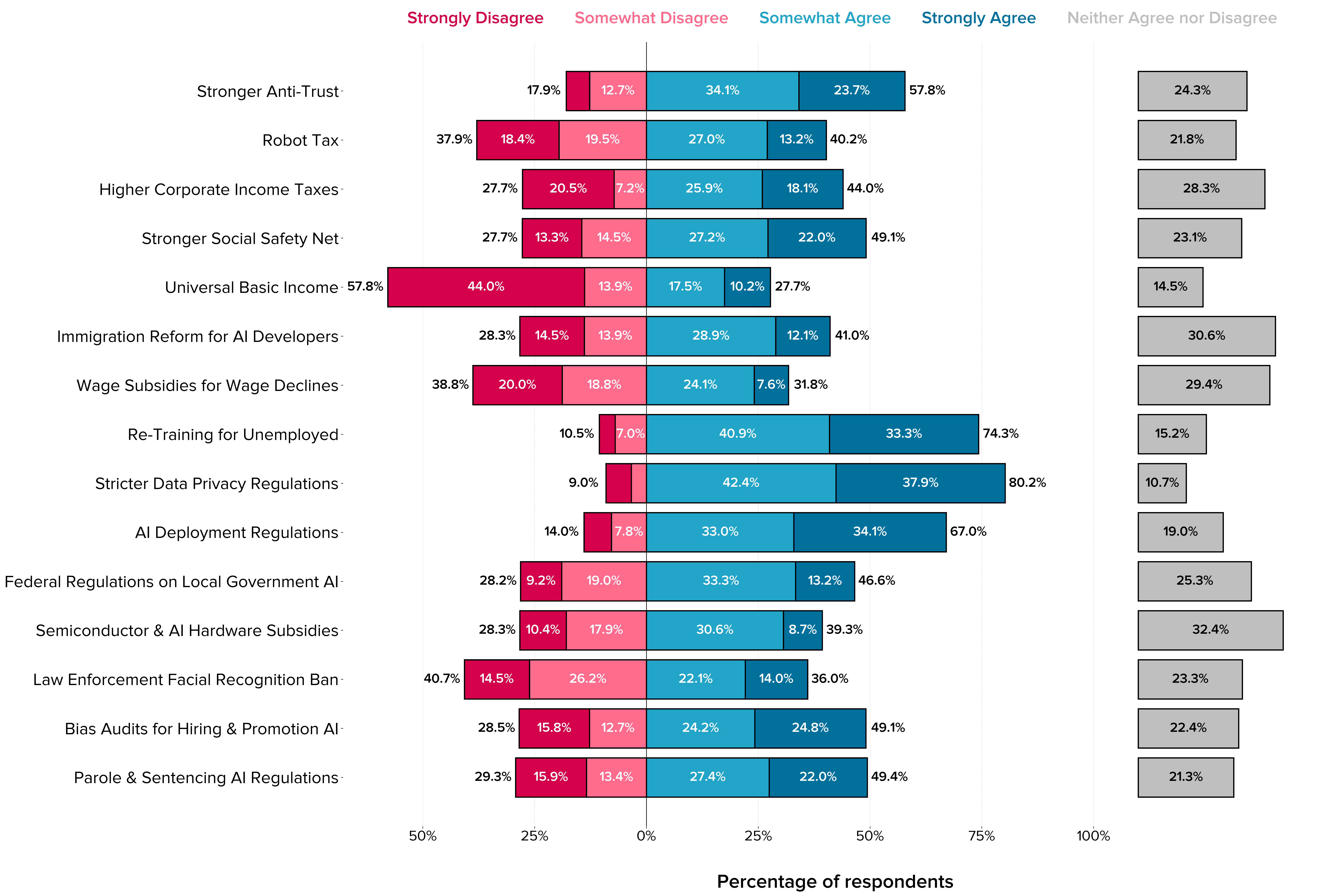}

    \caption{\textbf{Local US officials' views in 2022 on what AI policies would be beneficial between 2025 and 2050.} The figure shows unweighted relative frequencies for QS4 for the 2022 wave only.}
        
    \label{fig:QS4_2022}
\end{figure*}

\begin{figure*}
    \centering
    \includegraphics[width=\textwidth,height=\textheight,keepaspectratio]{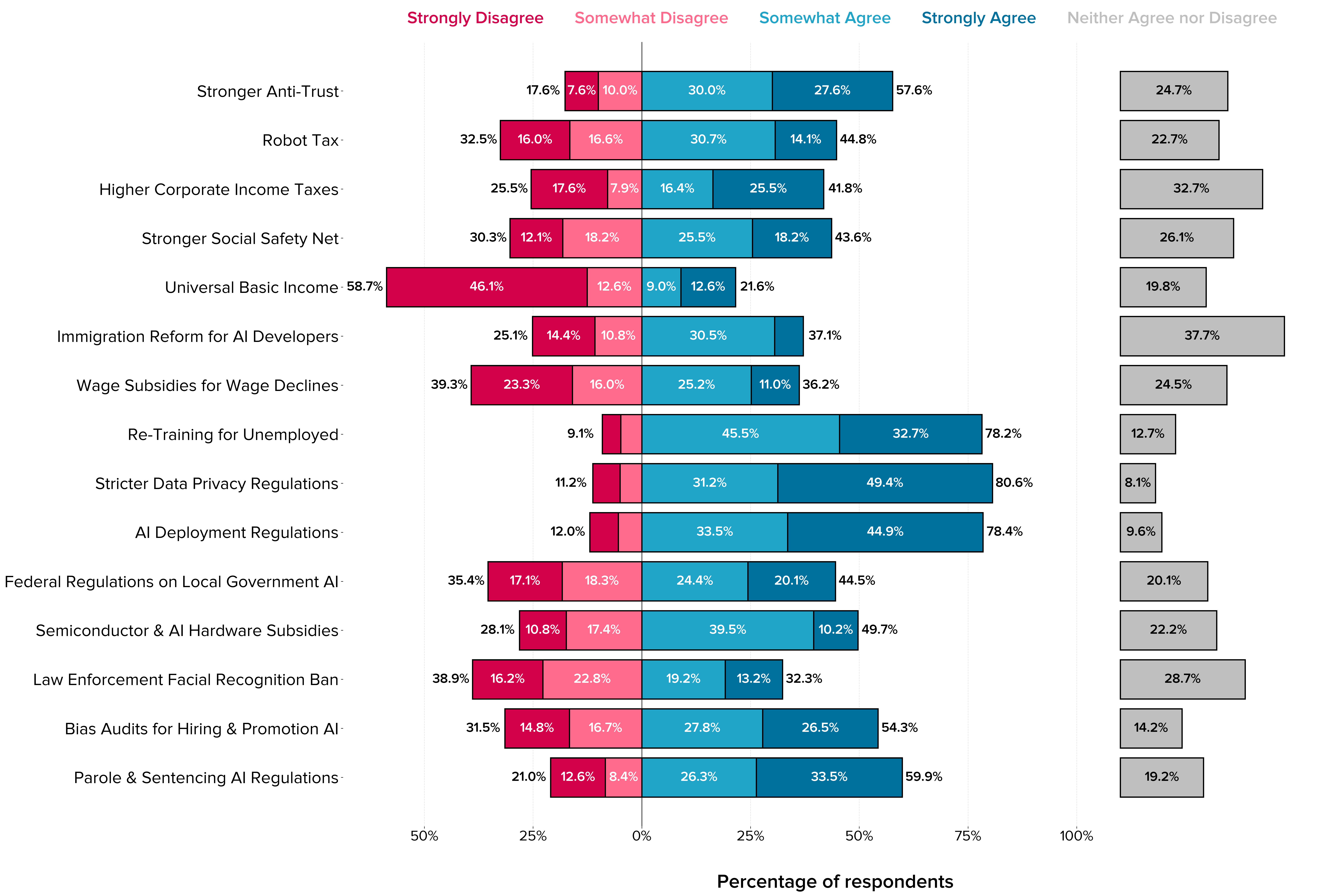}

    \caption{\textbf{Local US officials' views in 2023 on what AI policies would be beneficial between 2025 and 2050.} The figure shows unweighted relative frequencies for QS4 for the 2023 wave only.}
    
    \label{fig:QS4_2023}
\end{figure*}

Figures \ref{fig:QS1_Dems}--\ref{fig:QS4_2023} contain relative frequencies for QS1--4, segmented by political party and by year. Note that all figures display unweighted statistics across both survey waves. Labels for bars containing relative frequencies of less than 7.0\% are hidden for readability.

\clearpage

\paragraph*{S2 Variable definitions}
\label{supp:Variable_Defs}

The regression model is as follows:

\begin{equation*} 
    \begin{aligned}
        y_i = gender + age + edu + race + party + year_{2023} + party * year_{2023} \\
        + gov_{municipality} + gov_{county} + college + pop + Biden \\
        + \mathbbm{1}_{y_i \in \{Indices \backslash Index_{Policy \ Agreement}\}} * policy_{regai}
    \end{aligned}
\end{equation*}

where $y_i \in \{QS1-4, Q4.1, Indices\}$, $\mathbbm{1}_{y_i \in \{Indices \backslash Index_{Policy \ Agreement}\}}$ is an indicator variable for whether $y_i$ is any index except the policy agreement index, and $policy_{regai}$ is Q4.1 (which is excluded from the regression equation for the policy agreement index as it is an element of that index).

All variables are treated as continuous unless otherwise indicated. See \nameref{supp:Survey_Text} for the text of all questions.

The independent variables in our regression model are defined as follows (coding in parentheses):

\begin{itemize}
    \item $Gender$: self-identified gender of respondent
        \begin{itemize}
            \item Woman (0)
            \item Man (1)
            \item Other / self-describe (2) (removed from regression model due to too few responses)
        \end{itemize}
    \item $Age$: self-identified birth year cohort of respondent, treated as continuous variable by subtracting median of each bin from 2024
        \begin{itemize}
            \item 1920 or earlier (104)
            \item 1926 - 1930 (96)
            \item 1931 - 1935 (91)
            \item 1936 - 1940 (86)
            \item 1941 - 1945 (81)
            \item 1946 - 1950 (76)
            \item 1951 - 1955 (71)
            \item 1956 - 1960 (66)
            \item 1961 - 1965 (61)
            \item 1966 - 1970 (56)
            \item 1971 - 1975 (51)
            \item 1976 - 1980 (46)
            \item 1981 - 1985 (41)
            \item 1986 - 1990 (36)
            \item 1991 - 1995 (31)
            \item 1996 - 2000 (26)
            \item 2001 - 2005 (21)
            \item 2006 or later (18) ($n = 0$)
        \end{itemize}
    \item $Edu$: self-identified education level of respondent
        \begin{itemize}
            \item Less than high school' (0)
            \item High school graduate (1)
            \item Technical/trade school (2) 
            \item Some college (3)
            \item College graduate (4) 
            \item Some graduate school(5)
            \item Graduate degree (6)
        \end{itemize}
    \item $Race$: self-identified race of respondent
        \begin{itemize}
            \item White (0)
            \item Non-White (1)
        \end{itemize}
    \item $Party$: political party identification of respondent, collapsed from three different questions:  (1) A question asking respondent whether they think of themselves as a ``Democrat''; ``Republican''; ``Independent''; or ``Other party''. (2) A question asking whether respondent thinks of themselves as closer to the ``Democratic Party''; ``Republican Party''; or ``Neither''. (3) A question asking respondent whether they think of themselves as ``very conservative''; ``somewhat conservative''; ``moderate, middle of the road''; ``somewhat liberal''; ``very liberal''; or ``not sure''. (1) and (3) were asked of all respondents, and (2) was asked only to respondents who responded ``Independent'' or ``Other party'' to (1). Respondents who indicated in (2) that they ``think of [themselves] as closer to'' the Democratic or Republican parties were recoded as Democrats and Republicans, respectively. Then, respondents who ``think of [themselves] as'' very/somewhat liberal or as very/somewhat conservative were recoded as Democrats and Republicans, respectively. ``Other party'' responses to (1) were collapsed into the same category as ``independent''.
        \begin{itemize}
            \item Democrat (0)
            \item Independent (1)
            \item Republican (2)
        \end{itemize}
    \item $Year_{2023}$
        \begin{itemize}
            \item 2022 (0)
            \item 2023 (1)
        \end{itemize}
    \item $Party * Year_{2023}$: the interaction term between $party$ and $year_{2023}$.
    \item $Policy_{regai}$: response to Q4.1 (``Dreksler\_AI\_regulated'' in the documentation). Only included for indices that were not the policy agreement index, since the policy agreement index averaged responses to all variables including Q4.1.
\end{itemize}

The following control variables in our regression model are defined below:

\begin{itemize}
    \item $Census$: proportion of 25-years-or-older residents in respondent's geographic unit who have completed a 4-year, post-secondary degree. Data is from the 2015--2019 Five Year Data from the US Census American Community Survey, as compiled by IPUMS National Historical Geographic Information System (NHGIS). This variable is binned into terciles.
    \begin{itemize}
        \item First tercile (0)
        \item Second tercile (1)
        \item Third tercile (2)
    \end{itemize}
    \item $Pop$: total number of residents living in respondent's geographic unit. Data is from the 2015--2019 Five Year Data from the US Census American Community Survey, as compiled by IPUMS National Historical Geographic Information System (NHGIS). This variable is binned into terciles.
    \begin{itemize}
        \item First tercile (0)
        \item Second tercile (1)
        \item Third tercile (2)
    \end{itemize}
    \item $Biden$: proportion of the votes, by respondent's county, for Joe Biden in the 2020 Presidential election. Each sub-county government is matched to the relevant county in which it is contained. This variable is binned into terciles.
    \begin{itemize}
        \item First tercile (0)
        \item Second tercile (1)
        \item Third tercile (2)
    \end{itemize}
\end{itemize}

\paragraph*{S3 Survey text}
\label{supp:Survey_Text}

The text of the survey questionnaires are contained in the files `2022 Survey draft.pdf' and `2023 Survey draft.pdf.' in the OSF repository \cite{dreksler_2022-2023_2023}. The text of the substantive survey questions remained the same from 2022 to 2023, but note that our vendor (CivicPulse) made the following changes from 2022 to 2023:

\begin{itemize}
    \item Incomplete responses: our vendor collected incomplete responses in 2022 but not in 2023. We do not report statistics on or results from incomplete responses.
    \item Race: note that in the original data, the race variable is binary and is coded differently from 2022 to 2023.
    \item Weight computation: the weights used for our regression model were computed by CivicPulse for the pooled sample according to \cite{debell_computing_2009} using the $gender$, $pop$, and $Biden$ variables as defined in \nameref{supp:Variable_Defs} and in the documentation. 
    \item Census data included from 2022 to 2023: note that the 2023 dataset but not the 2022 dataset includes an $urban$ variable for Census data on the proportion of the urban population in respondent's Census unit. In the original 2023 dataset, the weights were the computed using $urban$, but  our regression weights were re-computed without this variable.
\end{itemize}

\paragraph*{S4 Indices definitions}
\label{supp:Indices_Defs}

We created respondent-level indices on which we fit our regression model. These indices are defined below using the variable definitions from \nameref{supp:Survey_Text} (original variable names in parentheses):

\begin{itemize}
    \item Policy agreement index: average of all policy questions (Q4.1--QS4)
    
    \item Positive impacts index: average of 
        \begin{itemize}
            \item Number of jobs (Local\_effects\_GR\_jobs)
            \item People's incomes (Local\_effects\_GR\_income)
            \item Quality of life (Local\_change\_GR\_qof)
            \item Mental health (Local\_change\_GR\_mh)
            \item Physical health (Local\_change\_GR\_ph)
            \item Data privacy and security (Local\_change\_GR\_data)
            \item Transportation and infrastructure (Local\_change\_GR\_transp)
            \item Size of the US economy (Broad\_effects\_GR\_econ)
            \item Strength of US democracy (Broad\_effects\_GR\_demo)
            \item Rates of innovation in the US (Broad\_effects\_GR\_innov)
        \end{itemize}
    
    \item Negative impacts index: average of
        \begin{itemize}
            \item Levels of political polarization (Local\_effects\_GR\_polar)
            \item Inequality (Local\_effects\_GR\_ineq)
            \item Levels of surveillance (Local\_effects\_GR\_surv)
            \item Bias and discrimination (Local\_effects\_GR\_discr)
            \item Amount of misinformation on US news and social media (Broad\_effects\_GR\_misinf)
            \item Number of conflicts and wars worldwide (Broad\_effects\_GR\_confl)
            \item Likelihood of a great power war (Broad\_effects\_GR\_war)
        \end{itemize}
    
    \item Economic impacts index: average of
        \begin{itemize}
            \item Number of jobs (Local\_effects\_GR\_jobs)
            \item People's incomes (Local\_effects\_GR\_income)
            \item Inequality (Local\_effects\_GR\_ineq)
            \item Size of the US economy (Broad\_effects\_GR\_econ)
        \end{itemize}
    
    \item Societal impacts index: average of
        \begin{itemize}
            \item Levels of surveillance (Local\_effects\_GR\_surv)
            \item Bias and discrimination (Local\_effects\_GR\_discr)
            \item Data privacy and security (Local\_change\_GR\_data)
            \item Amount of misinformation on US news and social media (Broad\_effects\_GR\_misinf)
        \end{itemize}
    
    \item Personal well-being \& community health index: average of
        \begin{itemize}
            \item Quality of life (Local\_change\_GR\_qof)
            \item Mental health (Local\_change\_GR\_mh)
            \item Physical health (Local\_change\_GR\_ph)
        \end{itemize}
    
    \item Progress and innovation impacts index: average of
        \begin{itemize}
            \item Transportation and infrastructure (Local\_change\_GR\_transp)
            \item Rates of innovation in the US (Broad\_effects\_GR\_innov)
        \end{itemize}
    
    \item Political impacts index: average of
        \begin{itemize}
            \item Levels of political polarization (Local\_effects\_GR\_polar)
            \item Strength of US democracy (Broad\_effects\_GR\_demo)
            \item Number of conflicts and wars worldwide (Broad\_effects\_GR\_confl)
            \item Likelihood of a great power war (Broad\_effects\_GR\_war)
        \end{itemize}
    
    \item Positive local community impacts index: average of
        \begin{itemize}
            \item Number of jobs (Local\_effects\_GR\_jobs)
            \item People's incomes (Local\_effects\_GR\_income)
            \item Quality of life (Local\_change\_GR\_qof)
            \item Mental health (Local\_change\_GR\_mh)
            \item Physical health (Local\_change\_GR\_ph)
            \item Data privacy and security (Local\_change\_GR\_data)
            \item Transportation and infrastructure (Local\_change\_GR\_transp)
        \end{itemize}
    
    \item Negative local community impacts index: average of
        \begin{itemize}
            \item Levels of political polarization (Local\_effects\_GR\_polar)
            \item Inequality (Local\_effects\_GR\_ineq)
            \item Levels of surveillance (Local\_effects\_GR\_surv)
            \item Bias and discrimination (Local\_effects\_GR\_discr)
        \end{itemize}
    
    \item Positive broad impacts index: average of
        \begin{itemize}
            \item Size of the US economy (Broad\_effects\_GR\_econ)
            \item Strength of US democracy (Broad\_effects\_GR\_demo)
            \item Rates of innovation in the US (Broad\_effects\_GR\_innov)
        \end{itemize}
    
    \item Negative broad impacts index: average of
        \begin{itemize}
            \item Amount of misinformation on US news and social media (Broad\_effects\_GR\_misinf)
            \item Number of conflicts and wars worldwide (Broad\_effects\_GR\_confl)
            \item Likelihood of a great power war (Broad\_effects\_GR\_war)
        \end{itemize}
\end{itemize}

\paragraph*{S5 Full Regression Results} \label{supp:Full_Regression_Results}

Tables \ref{tab:Results_regression_full_1-2}--\ref{tab:Results_regression_full_indices} contain the full results of the survey-weighted linear regression model as specified in \nameref{sec:Methods}, across both survey waves. P-values are Benjamini-Hochberg adjusted, and IDK responses are re-coded as neutral (0) where applicable. The results in this section are the full results of the truncated results presented in \nameref{supp:Full_Regression_Results}. 

\begin{figure*}
    \includegraphics[height=\textheight]{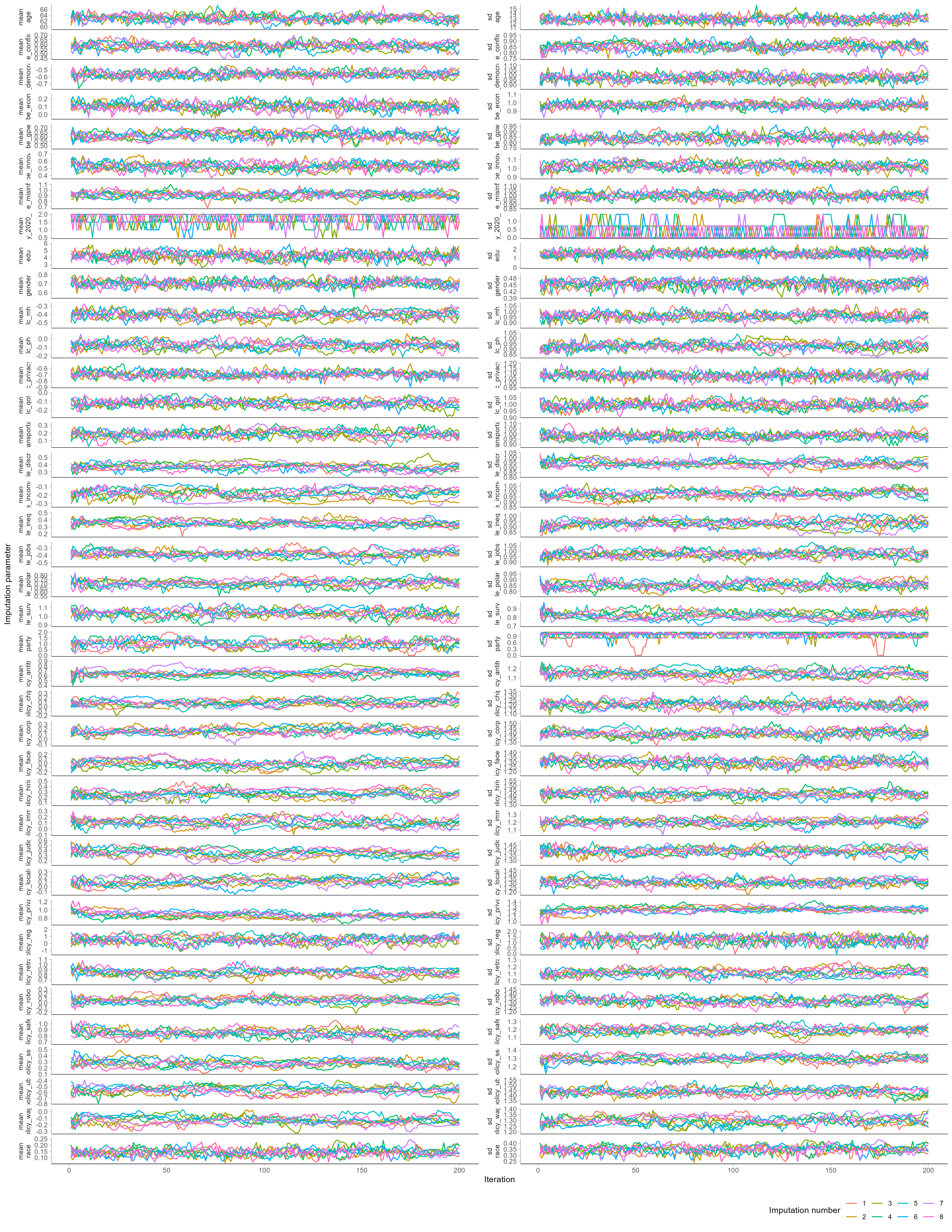}
    \caption{\textbf{Convergence Plots for 8 Randomly Selected MICE Imputations, Across 200 Iterations Each}}
    \label{fig:MICE_Convergence}
\end{figure*}

\begin{figure*}
    \includegraphics[height=\textheight]{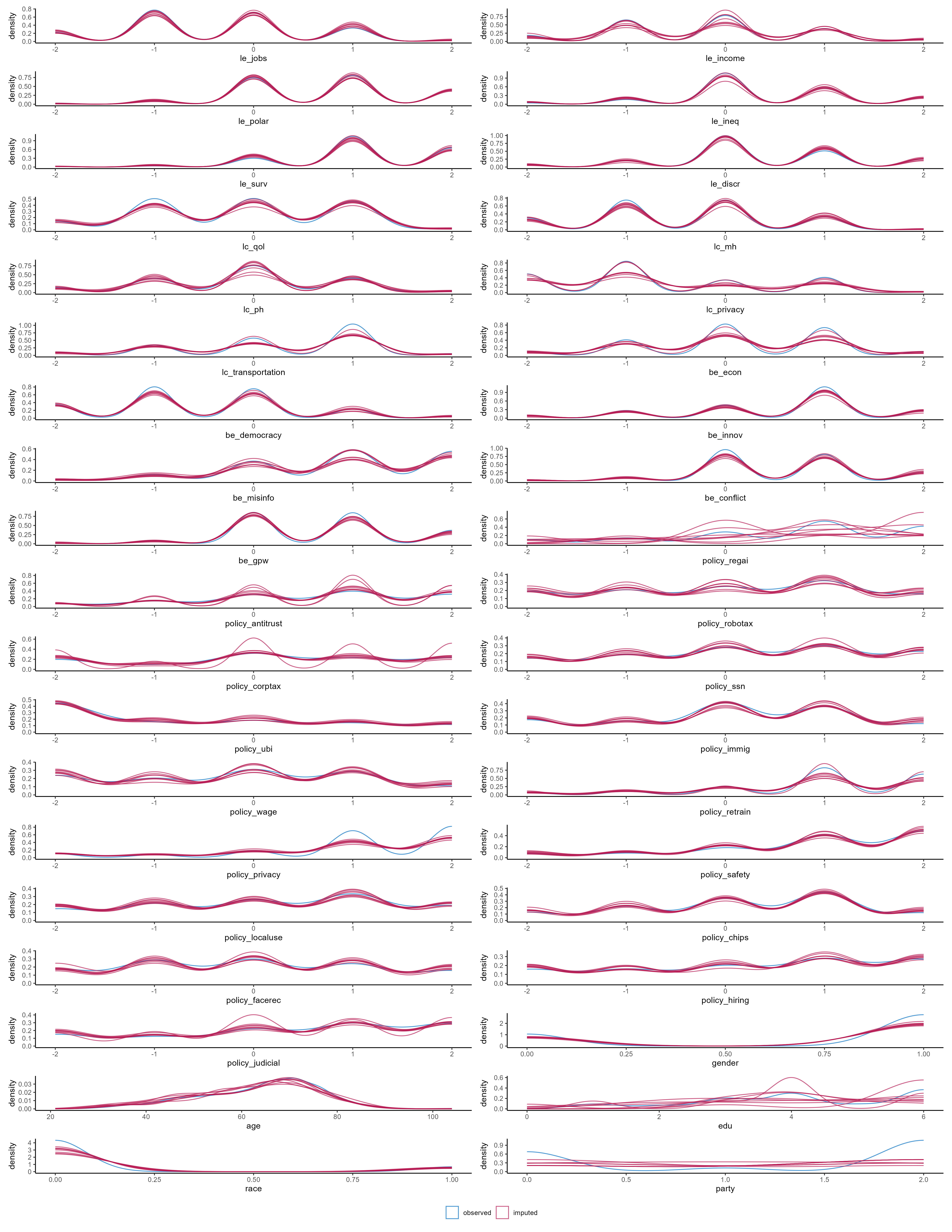}
    \caption{\textbf{Density Plots for 8 Randomly Selected MICE Imputations}}
    \label{fig:MICE_Density}
\end{figure*}

We also present figures to assess the fit of our imputations. Fig \ref{fig:MICE_Convergence} contains trace plots for imputed parameters across 200 iterations. We display only eight randomly selected imputed datasets for readability (our procedure imputed 120 datasets). Generally, convergence appears to occur for all parameters except the $Biden\_county\_2020$ and $party$ variables; however, missingness for these variables are extremely low since the first is Census-level data and the latter is combined from three different fields, so our results should not be implicated. Fig \ref{fig:MICE_Density} display density plots for the same eight imputed datasets layered on top of the original sample; for the vast majority of variables, the imputations look to map well onto the original distribution.
    
\nameref{supp:Alternative_Regression_Results} contains the full results for the same model specification but with IDK responses imputed where applicable.

\begin{sidewaystable}[phtb]
    \centering
    \small
    
    \begin{tabular}{lccccccc}
        \toprule
        & \multicolumn{7}{c}{\textbf{Covariate}} \\
        \textbf{Question}$^{\text{a}}$ & $Age$ & $Education$ & $Gender$$^{\text{b}}$ & $Race$$^{\text{c}}$ & $Party$$^{\text{d}}$ & $Year$$^{\text{e}}$ & $Party * Year$ \\
        
        \midrule
        \multicolumn{8}{l}{\textbf{QS1: Local Effects of AI}} \\
        
        \hspace{1em} Number of Jobs & 0.006 & -0.002 & 0.137 & 0.077 & -0.037 & -0.311 & -0.007 \\ 
         & (-0.001, 0.013) & (-0.057, 0.054) & (-0.045, 0.32) & (-0.168, 0.322) & (-0.144, 0.069) & (-0.537, -0.085) & (-0.149, 0.135) \\ 
        \hspace{1em} Income Level & 0.004 & -0.057 & -0.011 & -0.065 & -0.027 & -0.295 & 0.013 \\ 
         & (-0.003, 0.01) & (-0.11, -0.004) & (-0.194, 0.171) & (-0.321, 0.192) & (-0.136, 0.081) & (-0.518, -0.072) & (-0.131, 0.157) \\ 
        \hspace{1em} Political Polarization & -0.003 & 0.013 & -0.014 & -0.008 & -0.021 & 0.137 & -0.006 \\ 
         & (-0.008, 0.003) & (-0.034, 0.06) & (-0.182, 0.155) & (-0.237, 0.22) & (-0.122, 0.08) & (-0.079, 0.354) & (-0.14, 0.128) \\ 
        \hspace{1em} Inequality & 0.003 & 0.006 & -0.118 & 0.105 & -0.047 & 0.066 & 0.001 \\ 
         & (-0.003, 0.009) & (-0.042, 0.054) & (-0.31, 0.075) & (-0.146, 0.356) & (-0.152, 0.058) & (-0.162, 0.294) & (-0.137, 0.14) \\ 
        \hspace{1em} Surveillance Level & \textbf{-0.010\textsuperscript{**}} & \textbf{0.095\textsuperscript{***}} & 0.047 & 0.121 & 0.030 & -0.005 & 0.017 \\ 
         & \textbf{(-0.016, -0.005)} & \textbf{(0.051, 0.138)} & (-0.113, 0.207) & (-0.078, 0.32) & (-0.068, 0.128) & (-0.198, 0.189) & (-0.107, 0.141) \\ 
        \hspace{1em} Bias \& Discrimination & 0.007 & -0.012 & -0.199 & 0.161 & -0.001 & 0.194 & -0.021 \\ 
         & (-0.0, 0.013) & (-0.064, 0.04) & (-0.387, -0.012) & (-0.075, 0.398) & (-0.114, 0.112) & (-0.031, 0.42) & (-0.165, 0.124) \\

        \midrule
        \multicolumn{8}{l}{\textbf{QS2: Local Effects of AI}} \\
        
        \hspace{1em} Quality of Life & -0.003 & 0.054 & \textbf{0.263\textsuperscript{*}} & -0.161 & \textbf{-0.161\textsuperscript{*}} & -0.274 & 0.033 \\ 
         & (-0.009, 0.004) & (0.009, 0.099) & \textbf{(0.094, 0.432)} & (-0.396, 0.073) & \textbf{(-0.263, -0.058)} & (-0.486, -0.061) & (-0.106, 0.173) \\ 
        \hspace{1em} Mental Health & 0.003 & 0.049 & \textbf{0.313\textsuperscript{**}} & 0.111 & -0.137 & -0.220 & 0.066 \\ 
         & (-0.004, 0.009) & (0.005, 0.093) & \textbf{(0.149, 0.478)} & (-0.125, 0.347) & (-0.235, -0.039) & (-0.441, 0.0) & (-0.073, 0.205) \\ 
        \hspace{1em} Physical Health & 0.001 & 0.054 & 0.188 & 0.119 & \textbf{-0.181\textsuperscript{**}} & -0.159 & 0.050 \\ 
         & (-0.005, 0.008) & (0.005, 0.103) & (0.01, 0.365) & (-0.107, 0.345) & \textbf{(-0.278, -0.085)} & (-0.366, 0.047) & (-0.082, 0.183) \\ 
        \hspace{1em} Data Privacy \& & -0.002 & 0.042 & 0.091 & 0.246 & -0.014 & -0.077 & -0.105 \\ 
         \hspace{2em} Security & (-0.009, 0.006) & (-0.011, 0.095) & (-0.107, 0.288) & (-0.025, 0.518) & (-0.132, 0.104) & (-0.327, 0.173) & (-0.264, 0.053) \\ 
        \hspace{1em} Transportation \& & -0.001 & \textbf{0.121\textsuperscript{***}} & 0.168 & 0.023 & -0.084 & 0.092 & -0.068 \\ 
         \hspace{2em} Infrastructure & (-0.007, 0.005) & \textbf{(0.075, 0.167)} & (-0.012, 0.348) & (-0.194, 0.241) & (-0.18, 0.011) & (-0.103, 0.287) & (-0.202, 0.067) \\ 

        \bottomrule
    \end{tabular}

    \caption{\textbf{Full Results from Regression Analysis for QS1--2.}$^{\text{f}}$}

    \par 
    \raggedright \footnotesize
    * = $p < 0.05$, ** = $p < 0.01$, *** = $p < 0.001$. All statistically significant results in bold. \\
    $^{\text{a}}$ For each question, higher values represent belief that outcomes would increase. \\
    $^{\text{b}}$ 0 = woman and 1 = man. \\
    $^{\text{c}}$ 0 = white, 1 = non-white. \\
    $^{\text{d}}$ 0 = Democrat, 1 = independent or other party, and 2 = Republican. \\
    $^{\text{e}}$ 0 = 2022 and 1 = 2023. \\
    $^{\text{f}}$ Results are for the survey-weighted linear regression model as specified in \nameref{sec:Methods}, across both survey waves. P-values are Benjamini-Hochberg adjusted. IDK responses are re-coded as neutral (0). \\

    \label{tab:Results_regression_full_1-2}
\end{sidewaystable}

\begin{sidewaystable}[phtb]
    \centering
    \small
    
    \begin{tabular}{lccccccc}
        \toprule
        & \multicolumn{7}{c}{\textbf{Covariate}} \\
        \textbf{Question}$^{\text{a}}$ & $Age$ & $Education$ & $Gender$$^{\text{b}}$ & $Race$$^{\text{c}}$ & $Party$$^{\text{d}}$ & $Year$$^{\text{e}}$ & $Party * Year$ \\
        
        \midrule
        \multicolumn{8}{l}{\textbf{QS3: Broad Effects of AI}} \\

        \hspace{1em} US Economy & 0.007 & 0.019 & \textbf{0.246\textsuperscript{*}} & 0.157 & -0.104 & -0.269 & -0.031 \\ 
         & (0.001, 0.013) & (-0.028, 0.066) & \textbf{(0.077, 0.414)} & (-0.057, 0.37) & (-0.203, -0.005) & (-0.475, -0.063) & (-0.166, 0.103) \\ 
        \hspace{1em} US Democracy & 0.005 & -0.000 & 0.083 & 0.238 & -0.099 & -0.281 & 0.033 \\ 
         & (-0.001, 0.012) & (-0.047, 0.046) & (-0.083, 0.248) & (0.015, 0.461) & (-0.204, 0.006) & (-0.498, -0.065) & (-0.108, 0.173) \\ 
        \hspace{1em} US Innovation & -0.001 & 0.052 & 0.214 & 0.076 & -0.051 & 0.000 & -0.063 \\ 
         & (-0.008, 0.005) & (0.002, 0.101) & (0.032, 0.397) & (-0.151, 0.302) & (-0.158, 0.055) & (-0.22, 0.221) & (-0.209, 0.084) \\ 
        \hspace{1em} Misinformation & -0.008 & -0.014 & -0.061 & -0.040 & 0.043 & 0.298 & -0.045 \\ 
         \hspace{2em} (News \& Social Media) & (-0.014, -0.002) & (-0.06, 0.033) & (-0.234, 0.111) & (-0.289, 0.209) & (-0.072, 0.157) & (0.061, 0.535) & (-0.192, 0.103) \\ 
        \hspace{1em} Number of Conflicts & -0.003 & -0.048 & -0.153 & -0.025 & 0.029 & 0.105 & 0.019 \\ 
         & (-0.008, 0.002) & (-0.089, -0.008) & (-0.303, -0.002) & (-0.218, 0.167) & (-0.064, 0.123) & (-0.086, 0.297) & (-0.102, 0.14) \\ 
        \hspace{1em} Probability of & -0.006 & \textbf{-0.071\textsuperscript{*}} & -0.058 & -0.017 & -0.012 & 0.109 & 0.045 \\ 
         \hspace{1em} Great Power War & (-0.012, -0.001) & \textbf{(-0.113, -0.03)} & (-0.207, 0.09) & (-0.219, 0.185) & (-0.107, 0.082) & (-0.078, 0.295) & (-0.076, 0.167) \\ 
        
        \bottomrule
    \end{tabular}

    \caption{\textbf{Full Results from Regression Analysis for QS3.}$^{\text{f}}$}

    \par 
    \raggedright \footnotesize
    * = $p < 0.05$, ** = $p < 0.01$, *** = $p < 0.001$. All statistically significant results in bold. \\
    $^{\text{a}}$ For each question, higher values represent belief that outcomes would increase. \\
    $^{\text{b}}$ 0 = woman and 1 = man. \\
    $^{\text{c}}$ 0 = white, 1 = non-white. \\
    $^{\text{d}}$ 0 = Democrat, 1 = independent or other party, and 2 = Republican. \\
    $^{\text{e}}$ 0 = 2022 and 1 = 2023. \\
    $^{\text{f}}$ Results are for the survey-weighted linear regression model as specified in \nameref{sec:Methods}, across both survey waves. P-values are Benjamini-Hochberg adjusted. Respondents did not have the IDK option for QS3. \\

    \label{tab:Results_regression_full_3}
\end{sidewaystable}

\begin{sidewaystable}[phtb]
    \centering
    \small

    \begin{tabular}{lccccccc}
        \toprule
        & \multicolumn{7}{c}{\textbf{Covariate}} \\
        \textbf{Question}$^{\text{a}}$ & $Age$ & $Education$ & $Gender$$^{\text{b}}$ & $Race$$^{\text{c}}$ & $Party$$^{\text{d}}$ & $Year$$^{\text{e}}$ & $Party * Year$ \\
        
        \midrule
        \multicolumn{8}{l}{\textbf{QS4.1: General Support for AI Regulation}} \\

        \hspace{1em} Support AI Regulation & -0.003 & -0.007 & 0.056 & -0.265 & \textbf{-0.363\textsuperscript{***}} & \textbf{0.374\textsuperscript{**}} & 0.142 \\ 
         & (-0.009, 0.004) & (-0.057, 0.044) & (-0.126, 0.238) & (-0.505, -0.025) & \textbf{(-0.473, -0.252)} & \textbf{(0.178, 0.569)} & (-0.0, 0.284) \\ 

        \bottomrule
    \end{tabular}

    \caption{\textbf{Full Results from Regression Analysis for Q4.1.}$^{\text{f}}$}

    \par 
    \raggedright \footnotesize
    * = $p < 0.05$, ** = $p < 0.01$, *** = $p < 0.001$. All statistically significant results in bold. \\
    $^{\text{a}}$ For each question, higher values represent belief that outcomes would increase. \\
    $^{\text{b}}$ 0 = woman and 1 = man. \\
    $^{\text{c}}$ 0 = white, 1 = non-white. \\
    $^{\text{d}}$ 0 = Democrat, 1 = independent or other party, and 2 = Republican. \\
    $^{\text{e}}$ 0 = 2022 and 1 = 2023. \\
    $^{\text{f}}$ Results are for the survey-weighted linear regression model as specified in \nameref{sec:Methods}, across both survey waves. P-values are Benjamini-Hochberg adjusted. Respondents did not have the IDK option for QS4.1. \\

    \label{tab:Results_regression_full_4.1}
\end{sidewaystable}

\begin{sidewaystable}[phtb]
    \centering
    \footnotesize
    
    \begin{tabular}{lccccccc}
        \toprule
        & \multicolumn{7}{c}{\textbf{Covariate}} \\
        \textbf{Question}$^{\text{a}}$ & $Age$ & $Education$ & $Gender$$^{\text{b}}$ & $Race$$^{\text{c}}$ & $Party$$^{\text{d}}$ & $Year$$^{\text{e}}$ & $Party * Year$ \\
        
        \midrule
        \multicolumn{8}{l}{\textbf{QS4: Policy Support}} \\
        
        \hspace{1em} Stronger Anti-Trust & 0.004 & 0.033 & -0.087 & -0.234 & \textbf{-0.359\textsuperscript{***}} & -0.016 & 0.080 \\ 
         & (-0.005, 0.013) & (-0.034, 0.101) & (-0.319, 0.145) & (-0.546, 0.078) & \textbf{(-0.49, -0.228)} & (-0.272, 0.24) & (-0.086, 0.246) \\ 
        \hspace{1em} Robot Tax & -0.002 & 0.007 & \textbf{-0.552\textsuperscript{*}} & -0.296 & \textbf{-0.298\textsuperscript{*}} & 0.078 & 0.068 \\ 
         & (-0.014, 0.01) & (-0.078, 0.092) & \textbf{(-0.875, -0.229)} & (-0.686, 0.095) & \textbf{(-0.472, -0.123)} & (-0.259, 0.415) & (-0.14, 0.275) \\ 
        \hspace{1em} Higher Corporate & -0.011 & 0.003 & -0.257 & 0.112 & \textbf{-0.659\textsuperscript{***}} & 0.039 & -0.004 \\ 
         \hspace{2em} Income Taxes & (-0.021, -0.001) & (-0.076, 0.081) & (-0.539, 0.025) & (-0.256, 0.479) & \textbf{(-0.821, -0.497)} & (-0.268, 0.346) & (-0.197, 0.188) \\ 
        \hspace{1em} Stronger Social Safety Net & 0.009 & -0.016 & -0.146 & -0.053 & \textbf{-0.555\textsuperscript{***}} & 0.070 & 0.024 \\ 
         & (0.0, 0.019) & (-0.091, 0.06) & (-0.411, 0.119) & (-0.394, 0.288) & \textbf{(-0.699, -0.412)} & (-0.223, 0.363) & (-0.159, 0.207) \\ 
        \hspace{1em} Universal Basic Income & -0.013 & 0.086 & -0.087 & 0.428 & \textbf{-0.859\textsuperscript{***}} & -0.062 & 0.042 \\ 
         & (-0.022, -0.004) & (0.018, 0.154) & (-0.337, 0.163) & (0.037, 0.818) & \textbf{(-1.015, -0.703)} & (-0.397, 0.274) & (-0.138, 0.221) \\ 
        \hspace{1em} Immigration Reform for & 0.005 & -0.003 & 0.250 & -0.079 & \textbf{-0.317\textsuperscript{**}} & -0.132 & 0.062 \\ 
         \hspace{2em} AI Developers & (-0.004, 0.015) & (-0.078, 0.072) & (-0.004, 0.504) & (-0.421, 0.262) & \textbf{(-0.482, -0.151)} & (-0.492, 0.229) & (-0.142, 0.265) \\ 
        \hspace{1em} Wage Subsidies for & 0.001 & -0.039 & -0.296 & -0.157 & \textbf{-0.642\textsuperscript{***}} & 0.057 & -0.016 \\ 
         \hspace{2em} Wage Declines & (-0.009, 0.01) & (-0.113, 0.035) & (-0.534, -0.059) & (-0.493, 0.179) & \textbf{(-0.79, -0.494)} & (-0.234, 0.349) & (-0.194, 0.162) \\ 
        \hspace{1em} Re-Training for Unemployed & 0.003 & 0.014 & 0.134 & -0.079 & \textbf{-0.281\textsuperscript{**}} & -0.007 & 0.045 \\ 
         & (-0.006, 0.011) & (-0.051, 0.078) & (-0.097, 0.365) & (-0.393, 0.235) & \textbf{(-0.422, -0.139)} & (-0.279, 0.264) & (-0.122, 0.213) \\ 
        \hspace{1em} Stricter Data Privacy Regulations & -0.004 & \textbf{0.116\textsuperscript{**}} & -0.112 & -0.088 & -0.186 & 0.055 & 0.041 \\ 
         & (-0.013, 0.006) & \textbf{(0.049, 0.184)} & (-0.363, 0.139) & (-0.424, 0.249) & (-0.329, -0.043) & (-0.209, 0.32) & (-0.14, 0.223) \\ 
        \hspace{1em} AI Deployment Regulations & 0.001 & 0.039 & -0.208 & -0.282 & \textbf{-0.307\textsuperscript{***}} & 0.113 & 0.109 \\ 
         & (-0.01, 0.011) & (-0.037, 0.115) & (-0.452, 0.036) & (-0.638, 0.073) & \textbf{(-0.446, -0.168)} & (-0.14, 0.366) & (-0.065, 0.283) \\ 
        \hspace{1em} Federal Regulations on & -0.005 & 0.044 & -0.128 & 0.153 & \textbf{-0.320\textsuperscript{**}} & -0.130 & 0.071 \\ 
         \hspace{2em} Local Government AI & (-0.015, 0.004) & (-0.036, 0.123) & (-0.416, 0.16) & (-0.199, 0.505) & \textbf{(-0.491, -0.149)} & (-0.463, 0.203) & (-0.123, 0.265) \\ 
        \hspace{1em} Semiconductor \& & 0.008 & 0.004 & 0.104 & -0.172 & \textbf{-0.260\textsuperscript{*}} & 0.274 & -0.056 \\ 
         \hspace{2em} \& AI Hardware Subsidies & (-0.002, 0.018) & (-0.074, 0.083) & (-0.181, 0.389) & (-0.57, 0.227) & \textbf{(-0.419, -0.101)} & (-0.026, 0.574) & (-0.25, 0.138) \\ 
        \hspace{1em} Law Enforcement & -0.006 & 0.007 & -0.226 & 0.155 & -0.131 & -0.088 & 0.029 \\ 
         \hspace{2em} Facial Recognition Ban & (-0.017, 0.005) & (-0.076, 0.089) & (-0.538, 0.086) & (-0.257, 0.566) & (-0.314, 0.052) & (-0.438, 0.262) & (-0.182, 0.24) \\ 
        \hspace{1em} Bias Audits for Hiring \&  & -0.009 & 0.052 & -0.209 & -0.084 & \textbf{-0.386\textsuperscript{***}} & 0.030 & 0.075 \\ 
         \hspace{2em} Promotion AI & (-0.02, 0.002) & (-0.033, 0.137) & (-0.55, 0.133) & (-0.47, 0.302) & \textbf{(-0.561, -0.211)} & (-0.31, 0.369) & (-0.141, 0.29) \\ 
        \hspace{1em} Parole \& Sentencing & -0.011 & 0.084 & 0.177 & -0.008 & \textbf{-0.397\textsuperscript{***}} & 0.305 & 0.051 \\ 
         \hspace{2em} AI Regulations & (-0.024, 0.001) & (-0.006, 0.173) & (-0.134, 0.488) & (-0.408, 0.392) & \textbf{(-0.576, -0.219)} & (-0.052, 0.662) & (-0.164, 0.267) \\ 
         
        \bottomrule
    \end{tabular}
    \caption{\textbf{Full Results from Regression Analysis for QS4.}$^{\text{f}}$}

    \par 
    \raggedright \footnotesize
    * = $p < 0.05$, ** = $p < 0.01$, *** = $p < 0.001$. All statistically significant results in bold. \\
    $^{\text{a}}$ For each question, higher values represent belief that outcomes would increase. \\
    $^{\text{b}}$ 0 = woman and 1 = man. \\
    $^{\text{c}}$ 0 = white, 1 = non-white. \\
    $^{\text{d}}$ 0 = Democrat, 1 = independent or other party, and 2 = Republican. \\
    $^{\text{e}}$ 0 = 2022 and 1 = 2023. \\
    $^{\text{f}}$ Results are for the survey-weighted linear regression model as specified in \nameref{sec:Methods}, across both survey waves. P-values are Benjamini-Hochberg adjusted. Respondents did not have the IDK option for QS4. \\

    \label{tab:Results_regression_full_4}
\end{sidewaystable}

\begin{sidewaystable}[phtb]
    \centering
    \scriptsize
    
    \begin{tabular}{lcccccccc}
        \toprule
        & \multicolumn{8}{c}{\textbf{Covariate}} \\
        \textbf{Question}$^{\text{a}}$ & $Age$ & $Education$ & $Gender$$^{\text{b}}$ & $Race$$^{\text{c}}$ & $Party$$^{\text{d}}$ & $Year$$^{\text{e}}$ & $Party * Year$ & $Policy_{RegAI}$ \\
        \midrule
        \multicolumn{8}{l}{\textbf{Constructed Indices}} \\
        
        \hspace{1em} Policy Agreement & -0.002 & 0.019 & -0.092 & -0.082 & \textbf{-0.400\textsuperscript{***}} & 0.106 & 0.056 & -- \\
         & (-0.006, 0.002) & (-0.014, 0.051) & (-0.206, 0.021) & (-0.239, 0.075) & \textbf{(-0.470, -0.330)} & (-0.028, 0.239) & (-0.038, 0.150) & \\
        \hspace{1em} Positive Impacts (All) & 0.002 & 0.036 & \textbf{0.208\textsuperscript{**}} & 0.087 & -0.089 & -0.185 & -0.010 & 0.018 \\ 
         & (-0.002, 0.006) & (0.006, 0.066) & \textbf{(0.1, 0.317)} & (-0.057, 0.232) & (-0.156, -0.022) & (-0.321, -0.049) & (-0.1, 0.081) & (-0.026, 0.062) \\ 
        \hspace{1em} Negative Impacts (All) & -0.004 & -0.008 & -0.084 & 0.047 & 0.029 & 0.105 & 0.003 & 0.046 \\ 
         & (-0.007, -0.0) & (-0.037, 0.021) & (-0.194, 0.027) & (-0.09, 0.184) & (-0.041, 0.099) & (-0.033, 0.243) & (-0.086, 0.093) & (0.002, 0.09) \\ 
        \hspace{1em} Economic Impacts & 0.006 & -0.018 & 0.085 & 0.101 & -0.058 & -0.156 & -0.026 & 0.018 \\ 
         & (0.001, 0.011) & (-0.052, 0.017) & (-0.035, 0.205) & (-0.074, 0.275) & (-0.133, 0.018) & (-0.318, 0.006) & (-0.132, 0.079) & (-0.031, 0.068) \\ 
        \hspace{1em} Societal Impacts & -0.004 & 0.012 & -0.080 & 0.023 & 0.025 & 0.043 & -0.048 & 0.046 \\ 
         & (-0.009, 0.001) & (-0.024, 0.047) & (-0.209, 0.049) & (-0.161, 0.207) & (-0.058, 0.107) & (-0.122, 0.207) & (-0.158, 0.062) & (-0.004, 0.096) \\ 
        \hspace{1em} Personal Well-Being \& & 0.000 & 0.053 & \textbf{0.266\textsuperscript{**}} & 0.030 & \textbf{-0.174\textsuperscript{***}} & -0.236 & 0.063 & 0.003 \\ 
         \hspace{1em} Community Health Impacts & (-0.005, 0.006) & (0.015, 0.09) & \textbf{(0.122, 0.411)} & (-0.161, 0.221) & \textbf{(-0.256, -0.092)} & (-0.411, -0.061) & (-0.053, 0.179) & (-0.052, 0.058) \\ 
        \hspace{1em} Progress \& & -0.002 & \textbf{0.091\textsuperscript{***}} & 0.201 & 0.069 & -0.025 & 0.029 & -0.083 & 0.073 \\ 
         \hspace{2em} Innovation Impacts & (-0.007, 0.003) & \textbf{(0.052, 0.13)} & (0.058, 0.343) & (-0.118, 0.255) & (-0.111, 0.061) & (-0.147, 0.205) & (-0.202, 0.036) & (0.017, 0.129) \\ 
        \hspace{1em} Political Impacts & -0.003 & -0.040 & -0.024 & 0.036 & -0.019 & -0.058 & 0.050 & 0.023 \\ 
         & (-0.006, 0.001) & (-0.072, -0.009) & (-0.136, 0.087) & (-0.106, 0.178) & (-0.088, 0.050) & (-0.195, 0.079) & (-0.041, 0.141) & (-0.024, 0.070) \\ 
        \hspace{1em} Positive Impacts & 0.001 & 0.042 & \textbf{0.204\textsuperscript{*}} & 0.041 & -0.090 & -0.167 & -0.012 & 0.007 \\ 
         \hspace{2em} (Local Effects) & (-0.004, 0.005) & (0.009, 0.074) & \textbf{(0.083, 0.324)} & (-0.117, 0.200) & (-0.161, -0.019) & (-0.314, -0.020) & (-0.110, 0.086) & (-0.040, 0.053) \\ 
        \hspace{1em} Negative Impacts & -0.001 & 0.022 & -0.070 & 0.109 & 0.015 & 0.044 & 0.001 & 0.055 \\ 
         \hspace{2em} (Local Effects) & (-0.006, 0.003) & (-0.014, 0.057) & (-0.199, 0.060) & (-0.059, 0.277) & (-0.066, 0.097) & (-0.118, 0.207) & (-0.107, 0.109) & (0.002, 0.107) \\ 
        \hspace{1em} Positive Impacts & 0.004 & 0.021 & \textbf{0.195\textsuperscript{*}} & 0.193 & -0.077 & -0.201 & -0.021 & 0.038 \\ 
         \hspace{2em} (Broad Effects) & (-0.001, 0.009) & (-0.019, 0.062) & \textbf{(0.059, 0.332)} & (0.007, 0.379) & (-0.165, 0.010) & (-0.378, -0.024) & (-0.140, 0.098) & (-0.020, 0.096) \\ 
        \hspace{1em} Negative Impacts & -0.006 & -0.045 & -0.093 & -0.038 & 0.029 & 0.167 & -0.002 & 0.025 \\ 
         \hspace{2em} (Broad Effects) & (-0.010, -0.001) & (-0.080, -0.010) & (-0.223, 0.038) & (-0.205, 0.128) & (-0.057, 0.115) & (0.007, 0.328) & (-0.109, 0.105) & (-0.028, 0.079) \\ 

        \bottomrule
    \end{tabular}
    \caption{\textbf{Full Results from Regression Analysis for Constructed Indices.}$^{\text{f}}$}

    \par 
    \raggedright \footnotesize
    * = $p < 0.05$, ** = $p < 0.01$, *** = $p < 0.001$. All statistically significant results in bold. \\
    $^{\text{a}}$ For each question, higher values represent belief that outcomes would increase. \\
    $^{\text{b}}$ 0 = woman and 1 = man. \\
    $^{\text{c}}$ 0 = white, 1 = non-white. \\
    $^{\text{d}}$ 0 = Democrat, 1 = independent or other party, and 2 = Republican. \\
    $^{\text{e}}$ 0 = 2022 and 1 = 2023. \\
    $^{\text{f}}$ Results are for the survey-weighted linear regression model as specified in \nameref{sec:Methods}, across both survey waves. P-values are Benjamini-Hochberg adjusted. IDK responses are re-coded as neutral (0). Index definitions are contained in \nameref{supp:Indices_Defs}. \\

    \label{tab:Results_regression_full_indices}
\end{sidewaystable}

\clearpage

\paragraph*{S6 Alternative Regression Results}
\label{supp:Alternative_Regression_Results}

Tables \ref{tab:Results_regression_imputed_1-2}--\ref{tab:Results_regression_imputed_indices} contain the full results the full results of the survey-weighted linear regression model as specified in \nameref{sec:Methods}, across both survey waves, but with IDK responses imputed where applicable rather than re-coded as neutral (0) as is done for the results in \nameref{subsubsec:Results_Regression_Demographic}, \nameref{subsubsec:Results_Regression_Year},  and \nameref{supp:Full_Regression_Results}. P-values are Benjamini-Hochberg adjusted.

All coefficients that were statistically significant at $p < 0.05$ when IDKs are re-coded as neutral are also significant when IDKs are imputed. In addition, the model fit on IDK-imputed data sees 21 more statistically significant coefficients than the same model fit on IDKs recoded as neutral (54 vs. 33 coefficients).

\begin{sidewaystable}[phtb]
    \centering
    \small
    
    \begin{tabular}{lccccccc}
        \toprule
        & \multicolumn{7}{c}{\textbf{Covariate}} \\
        \textbf{Question}$^{\text{a}}$ & $Age$ & $Education$ & $Gender$$^{\text{b}}$ & $Race$$^{\text{c}}$ & $Party$$^{\text{d}}$ & $Year$$^{\text{e}}$ & $Party * Year$ \\
        
        \midrule
        \multicolumn{6}{l}{\textbf{QS1: Local Effects of AI}} \\
        
        \hspace{1em} Number of Jobs & 0.005 & 0.004 & 0.204 & 0.064 & -0.046 & \textbf{-0.387\textsuperscript{*}} & 0.017 \\ 
         & (-0.002, 0.013) & (-0.056, 0.064) & (0.005, 0.403) & (-0.19, 0.319) & (-0.163, 0.071) & \textbf{(-0.626, -0.147)} & (-0.137, 0.171) \\ 
        \hspace{1em} Income Level & 0.006 & -0.068 & 0.057 & -0.037 & -0.039 & -0.357 & 0.036 \\ 
         & (-0.002, 0.014) & (-0.13, -0.007) & (-0.169, 0.284) & (-0.31, 0.235) & (-0.175, 0.097) & (-0.623, -0.091) & (-0.132, 0.204) \\ 
        \hspace{1em} Political Polarization & -0.001 & 0.012 & -0.163 & 0.055 & -0.006 & 0.165 & -0.040 \\ 
         & (-0.008, 0.006) & (-0.041, 0.064) & (-0.35, 0.024) & (-0.186, 0.296) & (-0.115, 0.103) & (-0.057, 0.387) & (-0.183, 0.102) \\ 
        \hspace{1em} Inequality & 0.004 & -0.001 & -0.240 & 0.190 & -0.060 & 0.112 & -0.007 \\ 
         & (-0.003, 0.011) & (-0.058, 0.056) & (-0.45, -0.03) & (-0.093, 0.472) & (-0.183, 0.064) & (-0.135, 0.358) & (-0.162, 0.148) \\ 
        \hspace{1em} Surveillance Level & \textbf{-0.008\textsuperscript{*}} & \textbf{0.066\textsuperscript{*}} & -0.102 & 0.109 & 0.025 & 0.031 & -0.005 \\ 
         & \textbf{(-0.013, -0.002)} & \textbf{(0.024, 0.108)} & (-0.263, 0.059) & (-0.076, 0.295) & (-0.072, 0.121) & (-0.161, 0.222) & (-0.127, 0.118) \\ 
        \hspace{1em} Bias \& Discrimination & 0.007 & -0.023 & -0.282 & 0.167 & 0.002 & 0.246 & -0.046 \\ 
         & (-0.0, 0.015) & (-0.083, 0.036) & (-0.503, -0.062) & (-0.121, 0.454) & (-0.128, 0.131) & (-0.026, 0.518) & (-0.212, 0.12) \\

        \midrule
        \multicolumn{6}{l}{\textbf{QS2: Local Effects of AI}} \\
        
        \hspace{1em} Quality of Life & -0.003 & 0.060 & \textbf{0.352\textsuperscript{**}} & -0.148 & \textbf{-0.210\textsuperscript{**}} & \textbf{-0.392\textsuperscript{*}} & 0.077 \\ 
         & (-0.011, 0.004) & (0.006, 0.113) & \textbf{(0.156, 0.548)} & (-0.415, 0.120) & \textbf{(-0.320, -0.100)} & \textbf{(-0.637, -0.146)} & (-0.076, 0.230) \\ 
        \hspace{1em} Mental Health & 0.001 & 0.064 & \textbf{0.396\textsuperscript{***}} & 0.135 & \textbf{-0.173\textsuperscript{*}} & -0.288 & 0.107 \\ 
         & (-0.006, 0.008) & (0.015, 0.113) & \textbf{(0.219, 0.574)} & (-0.113, 0.383) & \textbf{(-0.278, -0.068)} & (-0.528, -0.048) & (-0.040, 0.255) \\ 
        \hspace{1em} Physical Health & 0.003 & \textbf{0.073\textsuperscript{*}} & 0.210 & 0.133 & \textbf{-0.213\textsuperscript{**}} & -0.259 & 0.108 \\ 
         & (-0.004, 0.010) & \textbf{(0.020, 0.126)} & (0.017, 0.403) & (-0.132, 0.399) & \textbf{(-0.323, -0.103)} & (-0.495, -0.022) & (-0.042, 0.258) \\ 
        \hspace{1em} Data Privacy \& & -0.004 & 0.057 & 0.149 & 0.206 & -0.054 & -0.194 & -0.044 \\ 
         \hspace{2em} Security & (-0.012, 0.004) & (0.004, 0.11) & (-0.067, 0.365) & (-0.088, 0.5) & (-0.181, 0.073) & (-0.456, 0.067) & (-0.209, 0.121) \\ 
        \hspace{1em} Transportation \& & -0.002 & \textbf{0.151\textsuperscript{***}} & 0.197 & 0.077 & -0.112 & 0.010 & -0.019 \\ 
         \hspace{2em} Infrastructure & (-0.009, 0.004) & \textbf{(0.099, 0.203)} & (-0.003, 0.396) & (-0.171, 0.325) & (-0.215, -0.01) & (-0.213, 0.232) & (-0.164, 0.126) \\ 

        \bottomrule
    \end{tabular}
    \caption{\textbf{Full Results from Regression Analysis for QS1--2.}$^{\text{f}}$}

    \par 
    \raggedright \footnotesize
    * = $p < 0.05$, ** = $p < 0.01$, *** = $p < 0.001$. All statistically significant results in bold. \\
    $^{\text{a}}$ For each question, higher values represent belief that outcomes would increase. \\
    $^{\text{b}}$ 0 = woman and 1 = man. \\
    $^{\text{c}}$ 0 = white, 1 = non-white. \\
    $^{\text{d}}$ 0 = Democrat, 1 = independent or other party, and 2 = Republican. \\
    $^{\text{e}}$ 0 = 2022 and 1 = 2023. \\
    $^{\text{f}}$ Results are for the survey-weighted linear regression model as specified in \nameref{sec:Methods}, across both survey waves. P-values are Benjamini-Hochberg adjusted. IDK responses are imputed with MICE. \\

    \label{tab:Results_regression_imputed_1-2}
\end{sidewaystable}

\begin{sidewaystable}[phtb]
    \centering
    \small

    \begin{tabular}{lccccccc}
        \toprule
        & \multicolumn{7}{c}{\textbf{Covariate}} \\
        \textbf{Question}$^{\text{a}}$ & $Age$ & $Education$ & $Gender$$^{\text{b}}$ & $Race$$^{\text{c}}$ & $Party$$^{\text{d}}$ & $Year$$^{\text{e}}$ & $Party * Year$ \\
        
        \midrule
        \multicolumn{6}{l}{\textbf{QS3: Broad Effects of AI}} \\
        
         \hspace{1em} US Economy & 0.007 & 0.031 & 0.279 & 0.193 & -0.141 & \textbf{-0.381\textsuperscript{*}} & -0.008 \\ 
         & (-0.001, 0.015) & (-0.023, 0.086) & (0.059, 0.499) & (-0.065, 0.450) & (-0.253, -0.030) & \textbf{(-0.629, -0.133)} & (-0.168, 0.153) \\ 
        \hspace{1em} US Democracy & 0.005 & 0.009 & 0.172 & 0.225 & -0.109 & \textbf{-0.338\textsuperscript{*}} & 0.060 \\ 
         & (-0.003, 0.012) & (-0.043, 0.060) & (-0.026, 0.369) & (-0.034, 0.484) & (-0.230, 0.012) & \textbf{(-0.584, -0.092)} & (-0.091, 0.212) \\ 
        \hspace{1em} US Innovation & 0.002 & 0.056 & 0.199 & 0.082 & -0.087 & -0.053 & -0.049 \\ 
         & (-0.006, 0.009) & (-0.000, 0.112) & (-0.011, 0.408) & (-0.171, 0.335) & (-0.203, 0.029) & (-0.291, 0.185) & (-0.207, 0.110) \\ 
        \hspace{1em} Misinformation & -0.007 & -0.006 & -0.159 & -0.027 & 0.051 & \textbf{0.348\textsuperscript{*}} & -0.083 \\ 
         \hspace{2em} (News \& Social Media) & (-0.014, -0.001) & (-0.057, 0.045) & (-0.350, 0.031) & (-0.275, 0.221) & (-0.071, 0.173) & \textbf{(0.102, 0.594)} & (-0.236, 0.070) \\ 
        \hspace{1em} Number of Conflicts & -0.002 & -0.040 & \textbf{-0.296\textsuperscript{**}} & 0.012 & 0.059 & 0.207 & -0.041 \\ 
         & (-0.008, 0.005) & (-0.091, 0.011) & \textbf{(-0.465, -0.127)} & (-0.205, 0.228) & (-0.048, 0.166) & (-0.018, 0.432) & (-0.178, 0.096) \\ 
        \hspace{1em} Probability of & -0.006 & \textbf{-0.074\textsuperscript{*}} & -0.141 & 0.068 & 0.007 & 0.282 & -0.036 \\ 
         \hspace{2em} Great Power War & (-0.012, 0.001) & \textbf{(-0.121, -0.027)} & (-0.316, 0.034) & (-0.168, 0.304) & (-0.103, 0.117) & (0.065, 0.499) & (-0.173, 0.101) \\ 
        
        \bottomrule
    \end{tabular}
    \caption{\textbf{Full Results from Regression Analysis for QS3.}$^{\text{f}}$}

    \par 
    \raggedright \footnotesize
    * = $p < 0.05$, ** = $p < 0.01$, *** = $p < 0.001$. All statistically significant results in bold. \\
    $^{\text{a}}$ For each question, higher values represent belief that outcomes would increase. \\
    $^{\text{b}}$ 0 = woman and 1 = man. \\
    $^{\text{c}}$ 0 = white, 1 = non-white. \\
    $^{\text{d}}$ 0 = Democrat, 1 = independent or other party, and 2 = Republican. \\
    $^{\text{e}}$ 0 = 2022 and 1 = 2023. \\
    $^{\text{f}}$ Results are for the survey-weighted linear regression model as specified in \nameref{sec:Methods}, across both survey waves. P-values are Benjamini-Hochberg adjusted. Respondents did not have the IDK option for QS3. \\

    \label{tab:Results_regression_imputed_3}
\end{sidewaystable}

\begin{sidewaystable}[phtb]
    \centering
    \small
    
    \begin{tabular}{lccccccc}
        \toprule
        & \multicolumn{7}{c}{\textbf{Covariate}} \\
        \textbf{Question}$^{\text{a}}$ & $Age$ & $Education$ & $Gender$$^{\text{b}}$ & $Race$$^{\text{c}}$ & $Party$$^{\text{d}}$ & $Year$$^{\text{e}}$ & $Party * Year$ \\
        
        \midrule
        \multicolumn{6}{l}{\textbf{QS4.1: General Support for AI Regulation}} \\
         \hspace{1em} Support AI Regulation & -0.003 & -0.006 & 0.056 & -0.259 & \textbf{-0.361\textsuperscript{***}} & \textbf{0.376\textsuperscript{**}} & 0.141 \\ 
          & (-0.009, 0.004) & (-0.057, 0.045) & (-0.125, 0.237) & (-0.499, -0.018) & \textbf{(-0.472, -0.250)} & \textbf{(0.180, 0.572)} & (-0.001, 0.284) \\ 
         
        \bottomrule
    \end{tabular}
    \caption{\textbf{Full Results from Regression Analysis for Q4.1.}$^{\text{f}}$}

    \par 
    \raggedright \footnotesize
    * = $p < 0.05$, ** = $p < 0.01$, *** = $p < 0.001$. All statistically significant results in bold. \\
    $^{\text{a}}$ For each question, higher values represent belief that outcomes would increase. \\
    $^{\text{b}}$ 0 = woman and 1 = man. \\
    $^{\text{c}}$ 0 = white, 1 = non-white. \\
    $^{\text{d}}$ 0 = Democrat, 1 = independent or other party, and 2 = Republican. \\
    $^{\text{e}}$ 0 = 2022 and 1 = 2023. \\
    $^{\text{f}}$ Results are for the survey-weighted linear regression model as specified in \nameref{sec:Methods}, across both survey waves. P-values are Benjamini-Hochberg adjusted. Respondents did not have the IDK option for QS4.1. \\
    
    \label{tab:Results_regression_imputed_4.1}
\end{sidewaystable}

\begin{sidewaystable}[phtb]
    \centering
    \footnotesize
    
    \begin{tabular}{lccccccc}
        \toprule
        & \multicolumn{7}{c}{\textbf{Covariate}} \\
        \textbf{Question}$^{\text{a}}$ & $Age$ & $Education$ & $Gender$$^{\text{b}}$ & $Race$$^{\text{c}}$ & $Party$$^{\text{d}}$ & $Year$$^{\text{e}}$ & $Party * Year$ \\
        
        \midrule
        \multicolumn{6}{l}{\textbf{QS4: Policy Support}} \\
        
        \hspace{1em} Stronger Anti-Trust & 0.004 & 0.036 & -0.046 & -0.279 & \textbf{-0.352\textsuperscript{***}} & -0.012 & 0.071 \\ 
         & (-0.005, 0.013) & (-0.031, 0.104) & (-0.316, 0.224) & (-0.609, 0.052) & \textbf{(-0.488, -0.217)} & (-0.271, 0.248) & (-0.091, 0.233) \\ 
        \hspace{1em} Robot Tax & -0.004 & 0.000 & \textbf{-0.589\textsuperscript{**}} & -0.271 & \textbf{-0.275\textsuperscript{*}} & 0.109 & 0.048 \\ 
         & (-0.017, 0.009) & (-0.084, 0.085) & \textbf{(-0.889, -0.288)} & (-0.679, 0.138) & \textbf{(-0.441, -0.109)} & (-0.228, 0.446) & (-0.156, 0.253) \\ 
        \hspace{1em} Higher Corporate & -0.010 & 0.002 & -0.246 & 0.113 & \textbf{-0.671\textsuperscript{***}} & 0.064 & -0.019 \\ 
         \hspace{2em} Income Taxes & (-0.020, -0.000) & (-0.083, 0.086) & (-0.523, 0.031) & (-0.247, 0.473) & \textbf{(-0.831, -0.512)} & (-0.232, 0.360) & (-0.216, 0.177) \\ 
        \hspace{1em} Stronger Social Safety Net & 0.009 & -0.008 & -0.148 & -0.053 & \textbf{-0.560\textsuperscript{***}} & 0.049 & 0.038 \\ 
         & (-0.000, 0.019) & (-0.086, 0.069) & (-0.410, 0.115) & (-0.367, 0.260) & \textbf{(-0.707, -0.414)} & (-0.247, 0.345) & (-0.148, 0.224) \\ 
        \hspace{1em} Universal Basic Income & \textbf{-0.014\textsuperscript{*}} & 0.094 & -0.110 & 0.389 & \textbf{-0.850\textsuperscript{***}} & -0.037 & 0.035 \\ 
         & \textbf{(-0.023, -0.005)} & (0.022, 0.166) & (-0.361, 0.141) & (0.006, 0.771) & \textbf{(-1.010, -0.690)} & (-0.351, 0.276) & (-0.142, 0.213) \\ 
        \hspace{1em} Immigration Reform for & 0.005 & 0.001 & 0.208 & -0.073 & \textbf{-0.319\textsuperscript{**}} & -0.187 & 0.079 \\ 
         \hspace{2em} AI Developers & (-0.005, 0.014) & (-0.074, 0.076) & (-0.074, 0.491) & (-0.427, 0.281) & \textbf{(-0.485, -0.153)} & (-0.527, 0.154) & (-0.111, 0.268) \\ 
        \hspace{1em} Wage Subsidies for Wage Declines & 0.000 & -0.036 & -0.292 & -0.168 & \textbf{-0.648\textsuperscript{***}} & 0.074 & -0.020 \\ 
         & (-0.009, 0.010) & (-0.116, 0.044) & (-0.560, -0.025) & (-0.491, 0.155) & \textbf{(-0.803, -0.494)} & (-0.224, 0.373) & (-0.198, 0.159) \\ 
        \hspace{1em} Re-Training for Unemployed & 0.002 & 0.022 & 0.154 & -0.050 & \textbf{-0.272\textsuperscript{**}} & 0.006 & 0.055 \\ 
         & (-0.008, 0.012) & (-0.046, 0.091) & (-0.103, 0.411) & (-0.358, 0.257) & \textbf{(-0.407, -0.137)} & (-0.258, 0.270) & (-0.113, 0.222) \\ 
        \hspace{1em} Stricter Data Privacy Regulations & -0.004 & \textbf{0.113\textsuperscript{*}} & -0.099 & -0.058 & -0.188 & 0.080 & 0.038 \\ 
         & (-0.014, 0.007) & \textbf{(0.042, 0.184)} & (-0.334, 0.136) & (-0.414, 0.297) & (-0.335, -0.041) & (-0.203, 0.364) & (-0.142, 0.217) \\ 
        \hspace{1em} AI Deployment Regulations & 0.000 & 0.041 & -0.187 & -0.262 & \textbf{-0.298\textsuperscript{**}} & 0.116 & 0.105 \\ 
         & (-0.009, 0.010) & (-0.033, 0.115) & (-0.437, 0.063) & (-0.602, 0.077) & \textbf{(-0.451, -0.146)} & (-0.147, 0.379) & (-0.071, 0.281) \\ 
        \hspace{1em} Federal Regulations on & -0.005 & 0.043 & -0.120 & 0.135 & \textbf{-0.320\textsuperscript{**}} & -0.109 & 0.076 \\ 
         \hspace{2em} Local Government AI & (-0.016, 0.006) & (-0.044, 0.131) & (-0.397, 0.157) & (-0.226, 0.496) & \textbf{(-0.487, -0.153)} & (-0.426, 0.209) & (-0.122, 0.275) \\ 
        \hspace{1em} Semiconductor \& AI & 0.008 & 0.000 & 0.083 & -0.133 & \textbf{-0.265\textsuperscript{*}} & 0.236 & -0.048 \\ 
         \hspace{2em} Hardware Subsidies & (-0.002, 0.017) & (-0.086, 0.086) & (-0.187, 0.352) & (-0.549, 0.283) & \textbf{(-0.425, -0.104)} & (-0.060, 0.531) & (-0.234, 0.138) \\ 
        \hspace{1em} Law Enforcement& -0.007 & 0.012 & -0.249 & 0.120 & -0.117 & -0.051 & 0.001 \\ 
         \hspace{2em} Facial Recognition Ban & (-0.018, 0.004) & (-0.070, 0.094) & (-0.574, 0.076) & (-0.296, 0.536) & (-0.302, 0.067) & (-0.400, 0.299) & (-0.210, 0.212) \\ 
        \hspace{1em} Bias Audits for Hiring \& & -0.009 & 0.057 & -0.237 & -0.028 & \textbf{-0.389\textsuperscript{**}} & 0.021 & 0.084 \\ 
         \hspace{2em} Promotion AI & (-0.020, 0.003) & (-0.031, 0.146) & (-0.555, 0.081) & (-0.394, 0.338) & \textbf{(-0.576, -0.203)} & (-0.328, 0.371) & (-0.129, 0.298) \\ 
        \hspace{1em} Parole \& Sentencing & -0.011 & 0.088 & 0.169 & -0.019 & \textbf{-0.391\textsuperscript{***}} & 0.275 & 0.059 \\ 
         \hspace{2em} AI Regulations & (-0.023, 0.001) & (0.002, 0.174) & (-0.161, 0.499) & (-0.445, 0.407) & \textbf{(-0.574, -0.208)} & (-0.062, 0.612) & (-0.151, 0.269) \\ 

        \bottomrule
    \end{tabular}
    \caption{\textbf{Full Results from Regression Analysis for QS4.}$^{\text{f}}$}

    \par 
    \raggedright \footnotesize
    * = $p < 0.05$, ** = $p < 0.01$, *** = $p < 0.001$. All statistically significant results in bold. \\
    $^{\text{a}}$ For each question, higher values represent belief that outcomes would increase. \\
    $^{\text{b}}$ 0 = woman and 1 = man. \\
    $^{\text{c}}$ 0 = white, 1 = non-white. \\
    $^{\text{d}}$ 0 = Democrat, 1 = independent or other party, and 2 = Republican. \\
    $^{\text{e}}$ 0 = 2022 and 1 = 2023. \\
    $^{\text{f}}$ Results are for the survey-weighted linear regression model as specified in \nameref{sec:Methods}, across both survey waves. P-values are Benjamini-Hochberg adjusted. Respondents did not have the IDK option for QS4. \\

    \label{tab:Results_regression_imputed_4}
\end{sidewaystable}

\begin{sidewaystable}[phtb]
    \centering
    \scriptsize
    
    \begin{tabular}{lcccccccc}
        \toprule
        & \multicolumn{8}{c}{\textbf{Covariate}} \\
        \textbf{Question}$^{\text{a}}$ & $Age$ & $Education$ & $Gender$$^{\text{b}}$ & $Race$$^{\text{c}}$ & $Party$$^{\text{d}}$ & $Year$$^{\text{e}}$ & $Party * Year$ & $Policy_{RegAI}$ \\
        \midrule
        \multicolumn{9}{l}{\textbf{Constructed Indices}} \\

        \hspace{1em} Policy Agreement & -0.002 & 0.019 & -0.098 & -0.083 & \textbf{-0.399\textsuperscript{***}} & 0.106 & 0.056 & -- \\ 
        & (-0.006, 0.002) & (-0.014, 0.052) & (-0.212, 0.016) & (-0.237, 0.072) & \textbf{(-0.469, -0.33)} & (-0.027, 0.239) & (-0.038, 0.15) & \\ 
        \hspace{1em} Positive Impacts (All) & 0.002 & \textbf{0.052\textsuperscript{*}} & \textbf{0.235\textsuperscript{**}} & 0.104 & \textbf{-0.114\textsuperscript{*}} & \textbf{-0.275\textsuperscript{*}} & 0.037 & 0.005 \\ 
         & (-0.003, 0.007) & \textbf{(0.018, 0.086)} & \textbf{(0.104, 0.366)} & (-0.068, 0.276) & \textbf{(-0.192, -0.036)} & \textbf{(-0.441, -0.108)} & (-0.071, 0.144) & (-0.046, 0.055) \\ 
        \hspace{1em} Negative Impacts (All) & -0.003 & -0.012 & \textbf{-0.194\textsuperscript{*}} & 0.061 & 0.051 & 0.201 & -0.051 & 0.050 \\ 
         & (-0.007, 0.001) & (-0.043, 0.02) & \textbf{(-0.318, -0.07)} & (-0.089, 0.211) & (-0.028, 0.13) & (0.048, 0.355) & (-0.151, 0.049) & (0.0, 0.101) \\ 
        \hspace{1em} Economic Impacts & 0.007 & -0.012 & 0.105 & 0.109 & -0.071 & -0.237 & -0.005 & 0.012 \\ 
         & (0.001, 0.012) & (-0.051, 0.028) & (-0.04, 0.249) & (-0.085, 0.303) & (-0.158, 0.016) & (-0.426, -0.048) & (-0.127, 0.118) & (-0.045, 0.069) \\ 
        \hspace{1em} Societal Impacts & -0.003 & 0.013 & -0.147 & 0.014 & 0.015 & 0.090 & -0.067 & 0.038 \\ 
         & (-0.008, 0.002) & (-0.025, 0.051) & (-0.285, -0.009) & (-0.186, 0.214) & (-0.075, 0.104) & (-0.09, 0.27) & (-0.187, 0.053) & (-0.016, 0.092) \\ 
        \hspace{1em} Personal Well-Being \& & 0.000 & \textbf{0.066\textsuperscript{*}} & \textbf{0.319\textsuperscript{**}} & 0.040 & \textbf{-0.203\textsuperscript{***}} & \textbf{-0.336\textsuperscript{*}} & 0.117 & 0.003 \\ 
         \hspace{2em} \& Community Health Impacts & (-0.006, 0.006) & \textbf{(0.025, 0.107)} & \textbf{(0.159, 0.479)} & (-0.178, 0.258) & \textbf{(-0.295, -0.111)} & \textbf{(-0.54, -0.131)} & (-0.015, 0.249) & (-0.056, 0.062) \\ 
        \hspace{1em} Progress \& & -0.001 & \textbf{0.105\textsuperscript{***}} & 0.209 & 0.128 & -0.057 & -0.031 & -0.053 & 0.063 \\ 
         \hspace{2em} Innovation Impacts & (-0.006, 0.004) & \textbf{(0.061, 0.148)} & (0.04, 0.377) & (-0.074, 0.329) & (-0.149, 0.035) & (-0.228, 0.165) & (-0.184, 0.077) & (0.001, 0.126) \\ 
        \hspace{1em} Political Impacts & -0.002 & -0.047 & -0.090 & 0.045 & -0.020 & -0.064 & 0.037 & 0.024 \\ 
         & (-0.006, 0.003) & (-0.082, -0.012) & (-0.220, 0.040) & (-0.114, 0.204) & (-0.098, 0.057) & (-0.224, 0.096) & (-0.068, 0.141) & (-0.028, 0.075) \\ 
        \hspace{1em} Positive Impacts & 0.001 & \textbf{0.057\textsuperscript{*}} & \textbf{0.248\textsuperscript{**}} & 0.048 & \textbf{-0.113\textsuperscript{*}} & \textbf{-0.246\textsuperscript{*}} & 0.030 & 0.000 \\ 
         \hspace{2em} (Local Effects) & (-0.004, 0.006) & \textbf{(0.020, 0.093)} & \textbf{(0.111, 0.385)} & (-0.137, 0.234) & \textbf{(-0.193, -0.033)} & \textbf{(-0.420, -0.071)} & (-0.082, 0.143) & (-0.051, 0.052) \\ 
        \hspace{1em} Negative Impacts & -0.000 & 0.016 & \textbf{-0.191\textsuperscript{*}} & 0.121 & 0.032 & 0.119 & -0.049 & 0.062 \\ 
         \hspace{2em} (Local Effects) & (-0.005, 0.005) & (-0.025, 0.056) & \textbf{(-0.332, -0.050)} & (-0.058, 0.300) & (-0.057, 0.121) & (-0.056, 0.294) & (-0.167, 0.069) & (0.004, 0.119) \\ 
        \hspace{1em} Positive Impacts & 0.004 & 0.035 & \textbf{0.223\textsuperscript{*}} & 0.209 & -0.107 & \textbf{-0.289\textsuperscript{*}} & 0.009 & 0.031 \\ 
         \hspace{2em} (Broad Effects) & (-0.002, 0.010) & (-0.010, 0.080) & \textbf{(0.058, 0.389)} & (0.005, 0.414) & (-0.205, -0.009) & \textbf{(-0.495, -0.084)} & (-0.127, 0.145) & (-0.033, 0.095) \\ 
        \hspace{1em} Negative Impacts & -0.005 & -0.037 & \textbf{-0.205\textsuperscript{*}} & -0.003 & 0.044 & \textbf{0.290\textsuperscript{*}} & -0.065 & 0.025 \\ 
         \hspace{2em} (Broad Effects) & (-0.010, 0.000) & (-0.076, 0.001) & \textbf{(-0.347, -0.063)} & (-0.188, 0.181) & (-0.053, 0.140) & \textbf{(0.106, 0.474)} & (-0.184, 0.053) & (-0.033, 0.082) \\ 

        \bottomrule
    \end{tabular}
    \caption{\textbf{Full Results from Regression Analysis for Constructed Indices.}$^{\text{f}}$}

    \par 
    \raggedright \footnotesize
    * = $p < 0.05$, ** = $p < 0.01$, *** = $p < 0.001$. All statistically significant results in bold. \\
    $^{\text{a}}$ For each question, higher values represent belief that outcomes would increase. \\
    $^{\text{b}}$ 0 = woman and 1 = man. \\
    $^{\text{c}}$ 0 = white, 1 = non-white. \\
    $^{\text{d}}$ 0 = Democrat, 1 = independent or other party, and 2 = Republican. \\
    $^{\text{e}}$ 0 = 2022 and 1 = 2023. \\
    $^{\text{f}}$ Results are for the survey-weighted linear regression model as specified in \nameref{sec:Methods}, across both survey waves. P-values are Benjamini-Hochberg adjusted. IDK responses are imputed with MICE. Index definitions are contained in \nameref{supp:Indices_Defs}. \\

    \label{tab:Results_regression_imputed_indices}
\end{sidewaystable}

\clearpage

\paragraph*{S7 Survey experiment}
\label{supp:Survey_Experiment}

We included a survey experiment to test whether respondents’ beliefs on the effects of AI would changed based on a prompt about the long-term effects of AI. Respondents were randomly assigned to receive the prompt below:

\begin{quote}
    Think about children born in your community today. According to current average life expectancy rates, some of them might be alive in the year 2100. 
\end{quote}

All respondents were then presented with Q4.2: ``Do you think that AI will have an overall positive or negative effect on the US from now until 2100?'' Respondents were then asked to explain their response in a free-text field. The response scale was a 5-point Likert scale ranging from very negative (-2) to very positive (2); respondents were permitted to respond with the IDK option.

\begin{figure}[!hptb] 
    \centering
    \includegraphics[width=\textwidth,height=\textheight,keepaspectratio]{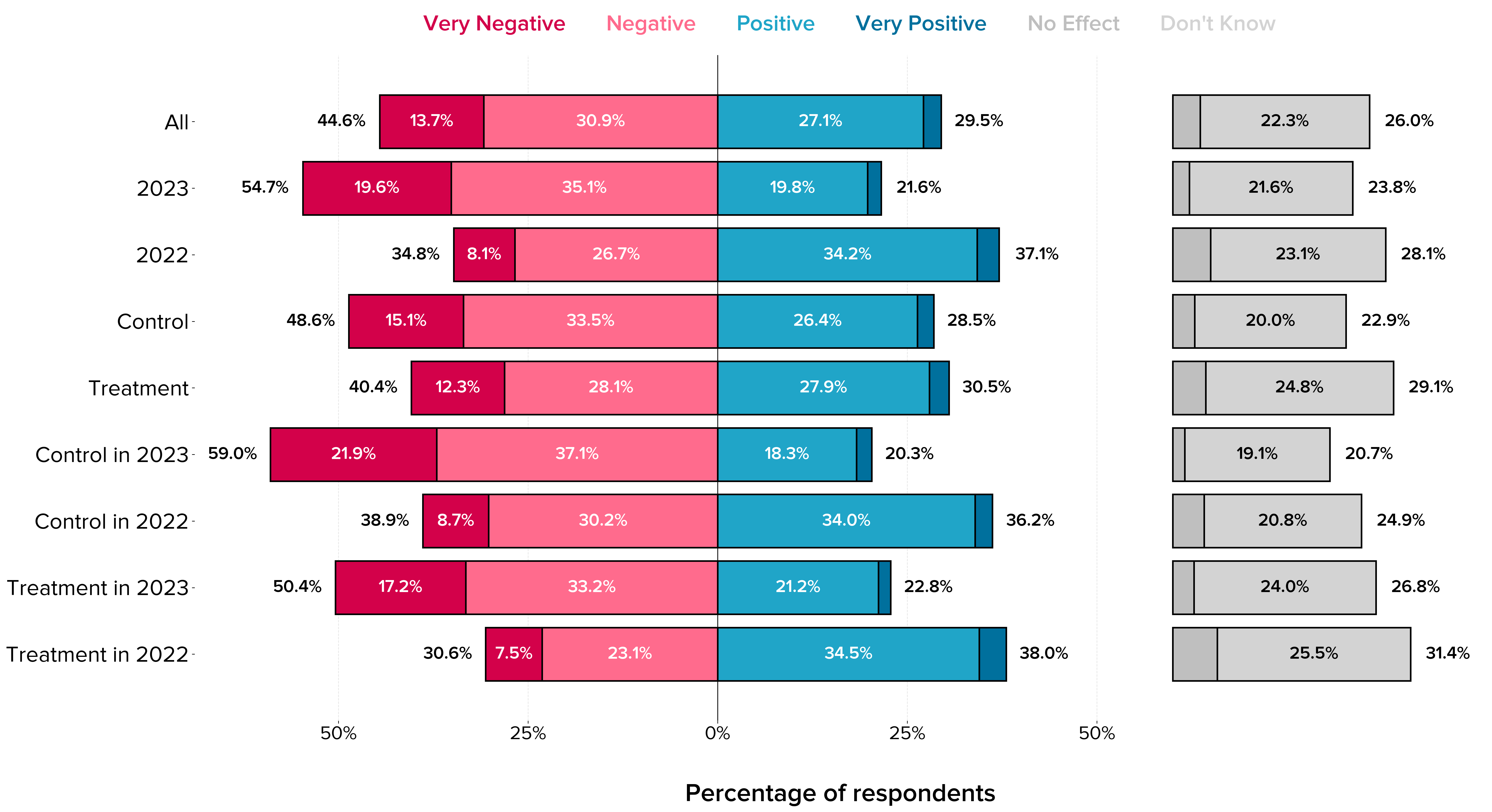}
    \caption{\textbf{Local US officials' responses to the question ``Do you think that AI will have an overall positive or negative effect on the US from now until 2100?'', with overall support and segments by treatment group and year.}$^{\text{a}}$}

    \par\raggedright \footnotesize
    $^{\text{a}}$ The figure shows unweighted relative frequencies across both survey waves. \\
    
    \label{fig:Results_Q4.2}
\end{figure}

Relative frequencies for responses to Q4.2 are shown in Table \ref{fig:Results_Q4.2}.

We estimated the average treatment effect (ATE) of the experimental treatment above using the following survey-weighted linear regression model:

\begin{equation*} 
    \begin{aligned}
        y = \mathbbm{1}_{Treatment} + gender + age + edu + race + party + year_{2023} + party * year_{2023} \\
        + gov_{municipality} + gov_{county} + college + pop + Biden \\
    \end{aligned}
\end{equation*}

where $y$ is the response to Q4.2, $\mathbbm{1}_{Treatment}$ is an indicator variable for whether the respondent was treated, and the other variables are as defined in \nameref{supp:Variable_Defs}. As with the main regression results presented in \nameref{supp:Full_Regression_Results} and \nameref{supp:Alternative_Regression_Results}, we estimate the ATE on both 1) an imputed dataset where the IDK values are set to neutral and not imputed, and 2) an imputed dataset where the IDK values are treated as missing and imputed.

As is shown in \ref{tab:ATE_Results}, the experimental treatment was not found to have a statistically significant effect at the 5\% level. These results were not adjusted for multiple comparisons.

\begin{table*}[htbp]
    \centering
    \small
    
    \begin{tabular}{lcccc}
        \toprule
         & ATE & Standard Error & P-Value & 95\% CI \\
        \midrule

        Neutral-Coded IDKs & 0.120 & 0.078 & 0.127 & (-0.034, 0.273) \\
        Imputed IDKs & 0.126 & 0.078 & 0.107 & (-0.027, 0.0230) \\

        \bottomrule
    \end{tabular}
    \caption{ATE Regression Results for Experimental Treatment (Q4.2)}

    \label{tab:ATE_Results}
\end{table*}

\paragraph*{S8 Sample representativeness}
\label{supp:samplerepresentativeness}

Same representativeness statistics relative to Census statistics are presented below in Tables \ref{tab:sample_representativeness_2022} and \ref{tab:sample_representativeness_2023}.

\begin{table}[ht]
    \centering
    \small
    \caption{Sample representativeness of county, municipality, and township officials in the 2022 survey wave}
    \begin{tabular}{llccc}
        \toprule
        \textbf{Category} & \textbf{Area Characteristics} & \textbf{Sample Median} & \textbf{Population Median} \\
        \hline
        \multirow[t]{4}{*}{County Officials} 
        & Proportion Urban            & 0.52 & 0.40 \\
        & Proportion College-educated & 0.21 & 0.19 \\
        & Population Size             & 42,360 & 25,750 \\
        & Democratic Vote Share 2020  & 0.36 & 0.30 \\
        \hline
        \multirow[t]{4}{*}{Municipality Officials} 
        & Proportion Urban            & 0.99 & 0.98 \\
        & Proportion College-educated & 0.25 & 0.21 \\
        & Population Size             & 5,030 & 4,180 \\
        & Democratic Vote Share 2020  & 0.42 & 0.40 \\
        \hline
        \multirow[t]{4}{*}{Township Officials} 
        & Proportion Urban            & 0.09 & 0.01 \\
        & Proportion College-educated & 0.27 & 0.22 \\
        & Population Size             & 3,750 & 2,680 \\
        & Democratic Vote Share 2020  & 0.48 & 0.39 \\
        \bottomrule
    \end{tabular}
    \label{tab:sample_representativeness_2022}
\end{table}

\begin{table}[ht]
    \centering
    \small
    \caption{Sample representativeness of county, municipality, and township officials in the 2023 survey wave}
    \begin{tabular}{llcc}
        \toprule
        \textbf{Category} & \textbf{Area Characteristics} & \textbf{Sample Median} & \textbf{Population Median} \\
        \midrule
        \multirow[t]{3}{*}{County Officials} 
        & Proportion College-educated & 0.22 & 0.19 \\
        & Population Size             & 42,360 & 25,750 \\
        & Democratic Vote Share 2020  & 0.35 & 0.30 \\
        \midrule
        \multirow[t]{3}{*}{Municipality Officials} 
        & Proportion College-educated & 0.25 & 0.21 \\
        & Population Size             & 5,080 & 4,180 \\
        & Democratic Vote Share 2020  & 0.43 & 0.40 \\
        \midrule
        \multirow[t]{3}{*}{Township Officials} 
        & Proportion College-educated & 0.27 & 0.22 \\
        & Population Size             & 3,900 & 2,680 \\
        & Democratic Vote Share 2020  & 0.48 & 0.39 \\
        \bottomrule
    \end{tabular}
    \label{tab:sample_representativeness_2023}
\end{table}

\clearpage

\section*{Ethics statement}
This study was approved by the Syracuse University Institutional Review Board (\#22-045). Informed consent was collected by CivicPulse at the start of the surveys, ensuring only respondents who agreed to participate entered the sample. The authors report no conflicts of interest. The data collection was funded by a grant from the Long-Term Future Fund.

\section*{Acknowledgments}
We would like to thank Markus Anderljung for feedback on the survey draft and CivicPulse for fielding the survey and generating the weights.

\nolinenumbers

\bibliography{references}

\end{document}